\newif\ifOneCol

\ifOneCol
    \documentclass[draftcls,onecolumn,11pt]{IEEEtran} 
\else
	\documentclass[journal]{IEEEtran}                    
\fi

\usepackage{enumerate}
\usepackage{verbatim}

\usepackage{cite}

\usepackage{graphicx}
\ifCLASSINFOpdf
\else
\fi

\usepackage[export]{adjustbox}%
\usepackage[font=small]{caption}
\usepackage{amsmath}
\usepackage[cmintegrals]{newtxmath}
\usepackage{multirow,makecell}
\usepackage{array}
\usepackage{etaremune}
\usepackage{amssymb}
\usepackage{booktabs}
\usepackage{colortbl}

\ifCLASSOPTIONcompsoc
  \usepackage[caption=false,font=normalsize,labelfont=sf,textfont=sf]{subfig}
\else
  \usepackage[caption=false,font=footnotesize]{subfig}
\fi

\usepackage{soul}

\usepackage[table]{xcolor}

\begin{document}
\bibliographystyle{IEEEtran}

\title{A Survey of Molecular Communication in\\ Cell Biology: Establishing a New Hierarchy for \\Interdisciplinary Applications}
%
%
%

\author{
\thanks{{Manuscript draft. Funding information here.} (Dadi Bi and Apostolos Almpanis are co-first authors.) (Corresponding author: Yansha Deng.)}
{Dadi Bi},~\IEEEmembership{Student Member,~IEEE}, {Apostolos Almpanis},~\IEEEmembership{Student Member,~IEEE}, \\{Adam Noel},~\IEEEmembership{Member,~IEEE}, {Yansha Deng},~\IEEEmembership{Member,~IEEE}, and {Robert Schober,~\IEEEmembership{Fellow,~IEEE}}
\thanks{D. Bi and Y. Deng are with the Department of Engineering, King's College London, London WC2R 2LS, U.K. (email: \{dadi.bi, yansha.deng\}@kcl.ac.uk).}
\thanks{A. Almpanis and A. Noel are with the School of Engineering, University of Warwick, Coventry CV4 7AL, U.K. (email: \{tolis.almpanis, adam.noel\}@warwick.ac.uk).}%
\thanks{R. Schober is with the Institute for Digital Communications, Friedrich-Alexander-Universit\"at Erlangen-N\"urnberg, 91058 Erlangen, Germany (e-mail: robert.schober@fau.de).}%
}

\maketitle

\begin{abstract}
Molecular communication (MC) engineering is inspired by the use of chemical signals as information carriers in cell biology. The biological nature of chemical signaling makes MC a promising methodology for interdisciplinary applications requiring communication between cells and other microscale devices. However, since the life sciences and communications engineering fields have distinct approaches to formulating and solving research problems, the mismatch between them can hinder the translation of research results and impede the development and implementation of interdisciplinary solutions. To bridge this gap, this survey proposes a novel communication hierarchy for MC signaling in cell biology and maps phenomena, contributions, and problems to the hierarchy. The hierarchy includes: 1) the physical propagation of cell signaling at the Physical Signal Propagation level; 2) the  generation,  reception, and biochemical pathways of molecular signals at the Physical and Chemical Signal Interaction level; 3) the quantification of physical signals, including macroscale observation and control methods, and conversion of signals to information at the Signal-Data Interface level; 4) the interpretation of information in cell signals and the realization of synthetic systems to store, process, and communicate molecular signals at the Local Data Abstraction level; and 5) applications relying on communication with MC signals at the Application level. To further demonstrate the proposed hierarchy, it is applied to case studies on quorum sensing, neuronal signaling, and communication via DNA. Finally, several open problems are identified for each level and the integration of multiple levels. The proposed hierarchy provides language for communication engineers to study and interface with biological systems, and also helps biologists to understand how communications engineering concepts can be exploited to interpret, control, and manipulate signaling in cell biology.
\end{abstract}


%
\IEEEpeerreviewmaketitle

\section{Introduction}

\IEEEPARstart{L}{ike} human beings, cells have their own ``social activities'' and are in constant communication with each other. One way they achieve this is by continuously sensing, receiving, and interpreting extracellular signaling molecules, and then coordinating their behaviors in response. This form of information exchange, termed molecular communication (MC), is a biologically-inspired communication paradigm, where information is exchanged via chemical signals \cite{Akyildiz2008,Nakano2013c}. The basic concepts and the architecture of MC were initially proposed and described to the research community in 2005 \cite{Suda2005,Hiyama2005}. After empirical work aimed to validate the feasibility of MC, this novel field has been primarily occupied and developed by the theoretical communications research community \cite{Nakano2012}.

Significant progress has been made over the last decade with a flourish of activity to understand the biophysical characteristics of molecule propagation using tools and mechanisms from communication engineering. The focus of channel modeling research has spanned from basic Brownian motion \cite{Eckford2007a}  to molecular transport with fluid flow \cite{Bicen2013} and active propagation that relies on energy sources, such as molecular motors \cite{Chahibi2016} and bacterial chemotaxis
\cite{Suda2019}. 
The interactions between information molecules and the receiver have been extensively studied for passive reception \cite{Noel2014} and full absorption \cite{yilmaz2014}, and recent works have modeled receiver-side reaction kinetics more precisely, e.g., reversible adsorption \cite{Deng2016} and ligand-binding \cite{arman2016tnb}.
 While many works have been based on transmission using simple on-off keying modulation \cite{Mahfuz2010}, more sophisticated modulation and coding schemes have been developed for molecular transmission with higher data rates and improved communication reliability \cite{Kuran2012,Shih2013}. Accompanying MC system design has been information-theoretical research to quantify the fundamental limits of molecular signaling, i.e., the communication capacity \cite{Nakano2013c}. 
In addition to  theoretical research, experimental research on MC has sought to validate theoretical models and provide pathways towards applications, both at microscale \cite{Akyildiz2012,Krishnaswamy2013,Nakano2014,Tuccitto2018,Grebenstein2019a} and macroscale \cite{Farsad2013a,Giannoukos2018}. More details on channel modeling, modulation and coding, communication capacity, physical design, and biological building blocks can be found in recent surveys \cite{Jamali2019b,Farsad2016,Akyildiz2019a,Kuscu2019a,Soldner2020}, respectively. 

With the ultimate goal of enabling  practical and  paradigm-shifting applications, such as  disease diagnosis, drug delivery, and health monitoring, the MC community has sought exploitation in cross-disciplinary research. For example, for disease diagnosis, evaluating the capacity of the brain to encode and retrieve memories could reveal the dysfunction and loss of synaptic communication due to Alzheimer's and other neurodegenerative diseases \cite{Akan2017}. For drug delivery, MC theory has been applied to characterize the transport of drug particles in blood vessels with the aim to optimize the drug injection rate while reducing its side effects \cite{Chahibi2013}. For health monitoring, MC could coordinate the movement of intra-body nanoscale sensors to collect health data, which could be further transmitted to external devices via micro-to-macro interfaces for real-time monitoring \cite{Nakano2014,Nakano2019}. 
Additional MC applications are identified in the surveys \cite{Chude-Okonkwo2017a,Akyildiz2019}.
To lead towards successful implementation of MC for the aforementioned applications, both synthetic biology and microfluidics have been regarded as promising tools for the design, test, and manufacture of microscale MC systems. Synthetic biology offers tools to engineer MC transceivers with modulation and coding functionalities via  genetic circuits  \cite{Unluturk2015,Marcone2018}. Microfluidics provides microscale experimental platforms to flexibly manipulate and control molecular transport to realize MC functionalities with high performance and reagent economy \cite{DeLeo2013,Krishnaswamy2013}.

Clearly, MC research incorporates elements from different scientific communities and this survey seeks to bring them closer together. MC theory can provide valuable insights for both man-made and natural systems. However, life scientists and engineers tend to have quite different ways of thinking and employ different language. In biology, the typical way to consider a natural (or synthetic) system is to describe its parts in appropriate detail and how these parts integrate to provide a functional system from start to finish. A representative example of this approach is cell signaling with G-protein-coupled receptors (which we describe in further detail in Section~\ref{sec_reception}); this system is conventionally described as a chain of events from receptor activation to the subsequent cell response that is experimentally observed \cite{Birnbaumer1990}. Another example is the \textit{Wnt} signaling pathway, which is a highly-conserved system present in all animals that regulates many important processes like cell proliferation, differentiation, and cell survival \cite{Klaus2008}. Wnt signaling is typically depicted as a series of molecule-release incidents that can trigger a response in neighboring cells and determine their fate. The trend in biology to detail functional components is consistent with the discipline's focus on understanding biological mechanisms (especially when a number of components are unknown) and controlling biological systems to maximize production yield (e.g., from a bioreactor).

On the other hand, it is common in engineering to design systems with a more modular approach; the different subsystems and their theoretical limits are studied and tested in isolation. For example, in a synthetic communication system, the physical channel through which information propagates is distinct from the encoding and decoding techniques that are applied, and these can be described separately, although they must be combined to implement a functioning system. This way of thinking is evident in \cite{Nakano2014b} where different levels for MC systems are described as part of a hierarchy. Despite the progress that has been made within the MC research community, translation of results to enable the desired interdisciplinary applications has been limited, in part due to the mismatch of different perspectives and the distinct methods that each community uses to formulate research problems.

Biological systems tend to be difficult to study, both due to their complexity and also because technology is not always sufficient, so parts of a natural system might remain unknown until technological developments enable observation. Thus, there is a tendency in life sciences to sequester complicated systems into small manageable parts, with the risk of losing higher-level interactions. Quorum sensing is an informative example, where individual microbes were being studied for decades, but only recently came the realization that communication via quorum sensing between microorganisms is of fundamental importance for the coordinated behaviors that we observe (e.g., biofilms, virulence) \cite{Kolter2006}. Systems biology tries to enforce a more holistic view of natural systems and to exploit concepts in biology that originated in other disciplines, including engineering \cite{Alon2006}. The tools commonly employed include big data and network motif analysis, i.e., the study of individual biological systems in engineering terms (e.g., biological circuits as logic gates \cite{Anderson2006}). However, the systems biology approach is still decidedly biologically-focused, and there can be a benefit to studying systems biology problems from a more structured engineering perspective. Inspired by systems biology, we wish to highlight the evident relevance of communication theory to signaling in microscale biological systems. A holistic view of natural and synthetic biological systems from a modular communication-centred perspective is a missing link that would help bridge contributions in MC to biological applications.

\subsection{State-of-the-Art in Communication Hierarchies}
\label{sec_hi}

The notion of a communication hierarchy is commonly found in the design of communication networks. By standardizing the role of each layer, the layers can be designed in isolation without compromising the functionality of the system. We seek such a communication hierarchy for signaling in cell biology (whether natural or engineered), \emph{not} to facilitate communication network design or to map existing layering approaches to cell biology, but rather to enable an interdisciplinary understanding of natural and synthetic MC systems. Nevertheless, existing approaches provide a useful reference against which we can compare our approach.

The formal communication standard within the MC community is IEEE 1906.1, ``Recommended Practice for Nanoscale and Molecular Communication Framework''; see \cite{Bush2015}. It includes a definition for a nanoscale communication system that maps to the basic communication elements (i.e., transmitter, receiver, medium, message carrier, and message). However, it does not specify a particular protocol for communication. Interestingly, it explicitly \emph{excludes} purely natural systems by specifying that at least one component must be synthetic. Another standardization effort is the Molecular Communications Markup Language (MolComML), which specifies the essential components of MC systems for making different simulation platforms more comparable and cross-compatible \cite{Felicetti2018}.

For telecommunication systems, the Open Systems Interconnection model (OSI) \cite{Zimmermann1980} and Transmission Control Protocol/Internet Protocol (TCP/IP) \cite{Cerf1974} are popular frameworks, and both were designed for interoperability in heterogeneous digital communication networks. Entities that are at the same layer in different communicating devices interact via a protocol designed for that layer. Tasks managed by the layers include interaction with programs that need to communicate, establishing connections between devices, transmission error control, and the physical transmission of bits over the communication channel. Incidentally, both of these frameworks have already been considered for the design of MC systems.

Works proposing protocols for MC systems include \cite{Nakano2014b,Felicetti2014a}. In \cite{Nakano2014b}, the authors present a layered architecture that is inspired by both OSI and TCP/IP. Preliminary descriptions of this architecture were presented in \cite{Nakano2012,Nakano2013c}. The layers comprise an application layer, a molecular transport layer, a molecular network layer, a molecular link layer, and a physical layer (comprised of signaling and bio-nanomachine sublayers). These layers map to those in TCP/IP, and facilitate the design of synthetic (though bio-inspired) communication systems. Similarly, the authors of \cite{Felicetti2014a} also present an architecture based on TCP/IP, with the goal of operating a synthetic communication system over a range of tens of microns ($\mu$m). In particular, the protocol specifies how to establish connections between devices, how to reliably transfer data, and how the receiver can control the transmission rate. The protocols in \cite{Nakano2014b,Felicetti2014a} are designed to establish digital communications and assume that the system designer has full control over the specification of the communicating devices. 

A roadmap for the development of synthetic biological MC systems is proposed in \cite{Soldner2020}. It involves five stages and illustrates the steps to facilitate MC-enabled commercial applications. The main purpose of the roadmap is to help define the scope of \cite{Soldner2020} (i.e., transmitter and receiver building blocks for different signaling molecules) as opposed to establishing a communication hierarchy.

\subsection{Contribution Summary}

The intended audience for this survey are those who are interested in how communications engineering concepts emerge and apply to understanding and controlling signaling in cell biology. This includes members of the communication engineering community who may not be familiar with MC or cell biology, and researchers in synthetic biology and bio-engineering who may not be familiar with communication systems and networks. Ultimately, this survey is written to build and support a bridge between these distinct domains by linking them together and identifying opportunities for interdisciplinary collaboration.

\begin{figure}[!t]
	\centering
	\includegraphics[width=2in]{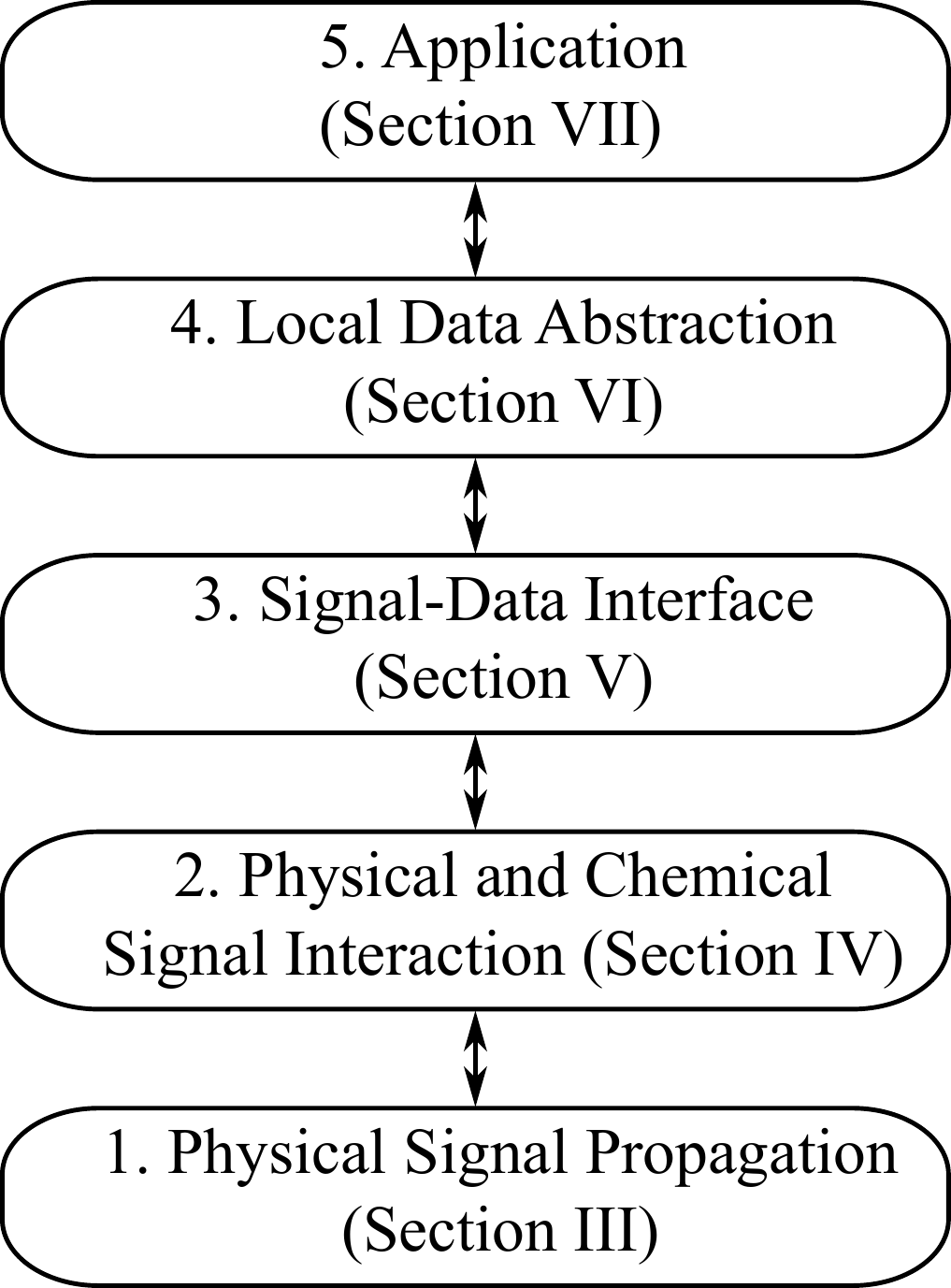}
	\caption{The proposed communication level hierarchy for MC signaling in cell biology that is also used to organize this paper.}
	\label{fig_hierarchy}
\end{figure}

To facilitate our objectives, this survey introduces a communication level hierarchy for microscale biological systems. Our perspective in the design of a communication level hierarchy for signaling in cell biology
is primarily \emph{not} to create a protocol and build synthetic communication systems. Instead, we seek to map our hierarchy and communications concepts directly to biological behavior. Thus, our approach will help engineers and biologists understand communication and signal processing (including computation and control) in cell biology, while providing language for synthetic biology and new opportunities to interface with biological systems. Furthermore, to emphasize that we are not designing a formal communication technology protocol, we refer to the tiers in our proposed hierarchy as ``\emph{levels}'' instead of ``layers''. The levels, also summarized in Fig.~\ref{fig_hierarchy}, are: 1) \textbf{Physical Signal Propagation}; 2) \textbf{Physical and Chemical Signal Interaction}; 3) \textbf{Signal-Data Interface}; 4) \textbf{Local Data Abstraction}; and 5) \textbf{Application}. Within the context of this hierarchy, this survey makes the following contributions:
\begin{enumerate}
    \item We map the communication processes of cell biology signaling to the levels of the proposed hierarchy. This includes macroscale interactions (i.e., experimental observation and control) with the microscale biological systems. 
    \item We provide biological case studies on quorum sensing, neuronal signaling, and communication via DNA, that map to all levels of the proposed hierarchy.
    \item We link contributions in the MC engineering domain with applications in biology and synthetic biology. This enables us to identify many opportunities for interdisciplinary contributions that advance understanding and control of signaling in cell biology.
\end{enumerate}

There are several challenges faced by this survey to effectively serve an audience with these diverse backgrounds, i.e., communications engineering and cell biology. First and foremost is the disparity in background knowledge. Although much of this survey describes cell biology and cellular subsystems, we do not expect a reader with a communications background to be familiar with these topics. Thus, we have sought for the cell biology in this survey to be self-contained, and we frequently cite \cite{Alberts2015} as a model background reference. We have also sought for the communications theory in this survey to be self-contained. However, a reader with a biology background and no foundation in communication systems may find it helpful to refer to a fundamental text in wireless communications (such as \cite{Goldsmith}). Additional background on the corresponding mathematics and signal processing can be found in \cite{Lathi}. Furthermore, we include glossaries of biological and communications terms in Tables~\ref{appendix:Biological_terms} and \ref{appendix:Communication_terms}, respectively, in the Appendix.

The second challenge for this survey is one of language. Different domains have distinct ways of articulating research problems and presenting results. This makes it difficult to recognize when research groups from different fields are seeking answers to the same question, or when an answer has already been obtained but from a different perspective. We hope that this survey and its proposed hierarchy is a useful guide for expanding a reader's language for interpreting results from both communications engineering and bio-engineering.

The third challenge for this survey is one of research focus. As we have already established, contributions in MC from the communication engineering community have focused on the design of new communication systems, whereas the subject of biology is concerned with understanding existing systems. Nevertheless, we intend for the hierarchy in this survey to identify ample common ground.

\subsection{Survey Organization}

\begin{figure*}[!t]
	\centering
	\includegraphics[width=0.66\textwidth]{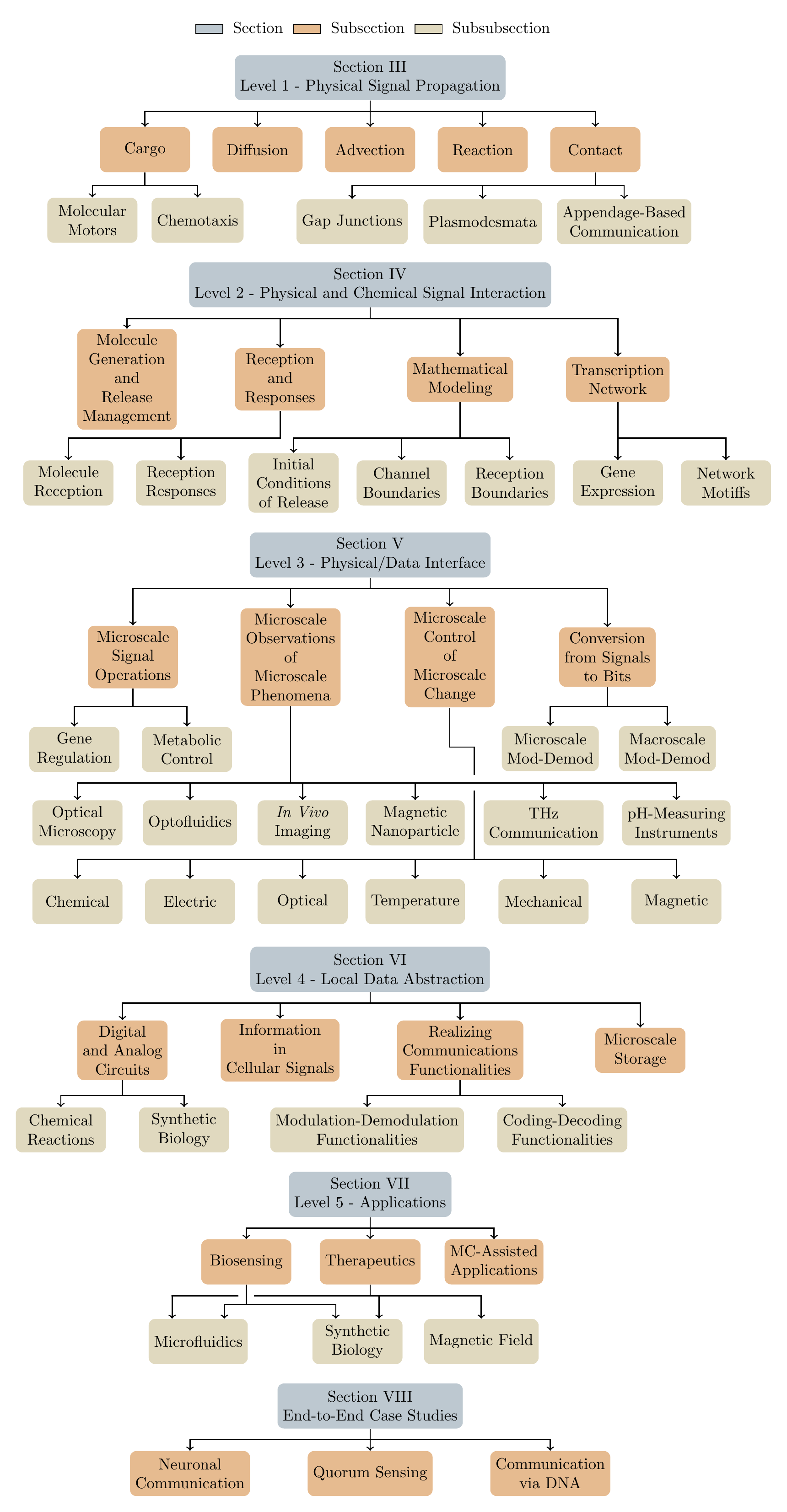}
	\caption{Organization and content of Sections III-VIII of this survey. Sections are shown in blue, subsections in orange, and subsubsections in beige.}
	\label{table_outline}
\end{figure*}

The remainder of this survey is organized as follows. We provide an overview of the proposed communication level hierarchy for cell biology in Section~\ref{sec_hierarchy}. We also present a general definition of a communicating device so that our discussions of communication between devices are in the appropriate context.

Sections~\ref{sec_physical} to \ref{sec_applications} sequentially detail the communication levels in a bottom-up approach, such that we traverse the levels in Fig.~\ref{fig_hierarchy} from Level 1 to Level 5. A graphical summary of the content of these levels is provided in Fig.~\ref{table_outline}. We start in Section~\ref{sec_physical} with the underlying fundamental concepts for the physical propagation of molecules (Level 1). We summarize the mathematical modeling of diffusion-based propagation but we also detail other mechanisms including cargo transport and contact-based signaling.

In Section~\ref{sec_interaction}, we discuss the biochemical and biophysical processes for devices to generate and receive signals (Level 2), including signaling pathways and physical responses. This section discusses initial and boundary conditions that are associated with molecule propagation. It also considers the biochemical signaling pathways associated with gene expression and corresponding transcription networks.

Section~\ref{sec_interface} addresses the mathematical quantification of physical signals and how they are observed and controlled (Level 3). We discuss microscale signal operations in terms of gene regulation and metabolic control, then review experimental methods for observing and controlling microscale molecular signals. We complete the section by mapping quantified physical signals to information bits.

Section~\ref{sec_data} focuses on the meaning of information in cell biology signaling (Level 4). We discuss the information contained in cellular signals, the realization of signal processing units with chemical reactions and synthetic biology, and the development of synthetic communication functionalities. We also describe DNA as a potential mechanism for microscale storage.

In Section~\ref{sec_applications}, we reach the top of the hierarchy and discuss applications (Level 5). We focus on biosensing and therapeutics as exemplary applications relying on the integration of natural and engineered cell biology systems.

Section~\ref{sec_case} presents three end-to-end case studies that span all of the levels of the proposed hierarchy. We discuss bacterial quorum sensing, neuronal signaling, and communication via DNA. We map each case study to all of the communication levels, thus demonstrating the flexibility of our approach.

Section~\ref{sec_open} provides a selection of open interdisciplinary research problems that can be identified and formulated using the proposed hierarchy. This includes problems that map to particular levels or the integration of multiple levels, as well as questions that are formulated by applying the hierarchy to the end-to-end case studies.

We conclude our survey in Section~\ref{sec_conclusion}. We re-iterate the intent of the hierarchy as a bridge to understanding and controlling communication in cell biology. We also emphasize key open research problems that are introduced throughout the work and for which we hope our framework will help to develop solutions.

\subsection{Comparison with Existing MC Surveys}


In the past 12 years, there have been several surveys focused on MC. To differentiate the scope of our survey with those of other surveys, in Table \ref{table_survey} we compare and summarize the differences of existing surveys according to the structure of our survey. This format also emphasizes our perspective and new contributions.

\begin{table*}[!t]
	\caption{Comparison of MC Surveys.}
	\label{table_survey}
	\centering	
	{\renewcommand{\arraystretch}{1}
		\scalebox{0.717}
		{
	\begin{tabular}{c|c||c|c|c|c|c|c|c|c|c|c|c|c|c|c|c|c|c|c|c}
		\hline
		\multicolumn{2}{c||}{Reference}
		&\cite{Akyildiz2008}
		&\cite{Nakano2012}
		&\cite{Darchini2013}
		&\cite{Nakano2014b}
		&\cite{Farsad2016}
		&\cite{Felicetti2016}
		&\cite{Akan2017}
		&\cite{Chahibi2017}
		&\cite{Chude-Okonkwo2017a}
		&\cite{Carmean2018}
		&\cite{Jamali2019b}
		&\cite{Akyildiz2019a}
		&\cite{Kuscu2019a}
		&\cite{Nakano2019}
		&\cite{Nguyen2019}
		&\cite{Kim2019b}
		&\cite{Jornet2019}
		&\cite{Soldner2020} 
		&This paper\\ \hline
		\multicolumn{2}{c||}{Year}
		&2008
		&2012
		&2013
		&2014
		&2016
		&2016
		&2016
		&2017
		&2017
		&2018
		&2019
		&2019
		&2019
		&2019
		&2019
		&2019
		&2019
		&2020
		&-
		\\ \hline \hline
	 	\multirow{9}{*}{\makecell[c]{Section III:\\Level 1\\Physical\\Signal\\Propagation}}
	 	&{\makecell{Diffusion-Based\\Propagation}}
	 	&            
	 	&\checkmark  
	 	&            
	 	&\checkmark  
	 	&\checkmark  
	 	&            
	 	&\checkmark  
	 	&\checkmark  
	 	&            
	 	&            
	 	&\checkmark  
	 	&\checkmark  
	 	&\checkmark  
	 	&\checkmark  
	 	&  
	 	&  
	 	&  
	 	&  
	 	&\checkmark  
	 	\\ \cline{2-21}
	 	&{\makecell{Advection-Diffusion-\\Based Propagation}}
	    &            
	 	&\checkmark  
	 	&            
	 	&  
	 	&\checkmark  
	 	&            
	 	&  
	 	&\checkmark  
	 	&\checkmark            
	 	&            
	 	&\checkmark  
	 	&\checkmark  
	 	&\checkmark  
	 	&  
	 	&  
	 	&  
	 	&  
	 	&  
	 	&\checkmark  
	 	\\ \cline{2-21}
	 	&{\makecell{Advection-Diffusion-\\Reaction-Based\\Propagation}} 
	 	&            
	 	&\checkmark  
	 	&            
	 	&  
	 	&  
	 	&            
	 	&  
	 	&  %
	 	&            
	 	&            
	 	&\checkmark  
	 	&\checkmark  
	 	&\checkmark  
	 	&  
	 	&  
	 	&  
	 	&  
	 	&  
	 	&\checkmark  
	 	\\ \cline{2-21}
	 	&{\makecell{Cargo-Based\\Propagation}} 
	 	&\checkmark            
	 	&  
	 	&\checkmark            
	 	&  
	 	&\checkmark  
	 	&            
	 	&\checkmark  
	 	&\checkmark  
	 	&            
	 	&            
	 	&\checkmark  
	 	&\checkmark  
	 	&\checkmark  
	 	&  
	 	&  
	 	&  
	 	&  
	 	&  
	 	&\checkmark  
	 	\\ \cline{2-21}
	 	&{\makecell{Contact-Based\\Propagation}} 
	    &            
	 	&  
	 	&\checkmark            
	 	&  
	 	&\checkmark  
	 	&            
	 	&\checkmark  
	 	&  
	 	&            
	 	&            
	 	&\checkmark  
	 	&\checkmark  
	 	&  
	 	&  
	 	&  
	 	&  
	 	&  
	 	&  
	 	&\checkmark  
	 	\\ \hline
 		\multirow{9}{*}{\makecell[c]{Section IV:\\Level 2\\Device\\Interface}}
 		&{\makecell{Molecule Generation\\and Release Management}}
 		&\checkmark            
 		&  
 		&            
 		&\checkmark  
 		&\checkmark  
 		&            
 		&  
 		&\checkmark  
 		&\checkmark            
 		&            
 		&\checkmark  
 		&\checkmark  
 		&\checkmark  
 		&  
 		&  
 		&  
 		&  
 		&\checkmark  
 		&\checkmark  
 		\\ \cline{2-21} 
 		&{\makecell{Molecule Reception\\and Responses}}
 		&\checkmark            
 		&  
 		&            
 		&  
 		&\checkmark  
 		&            
 		&  
 		&\checkmark  
 		&\checkmark            
 		&            
 		&\checkmark  
 		&\checkmark  
 		&\checkmark  
 		&  
 		&  
 		&  
 		&  
 		&\checkmark  
 		&\checkmark  
 		\\ \cline{2-21}
 		&{\makecell{Mathematical Modeling\\of Emission, Propagation,\\and Reception}}
 		&            
 		&  
 		&            
 		&  
 		&  
 		&            
 		&  
 		&  
 		&\checkmark            
 		&            
 		&\checkmark  
 		&\checkmark  
 		&  
 		&  
 		&  
 		&  
 		&  
 		&  
 		&\checkmark  
 		\\ \cline{2-21}
 		&{\makecell{Biochemical\\Signaling Pathway:\\Transcription Network}}
 		&            
 		&  
 		&            
 		&  
 		&  
 		&            
 		&  
 		&  
 		&            
 		&            
 		&  
 		&  
 		&  
 		&  
 		&  
 		&  
 		&  
 		&  
 		&\checkmark  
 		\\ \hline
 		\multirow{7}{*}{\makecell[c]{Section V:\\Level 3\\ Physical/Data\\Interface}}
 		&{\makecell{Microscale Signal\\Operations}}  
 		&            
 		&  
 		&            
 		&\checkmark  
 		&  
 		&            
 		&  
 		&  
 		&            
 		&\checkmark            
 		&  
 		&  
 		&  
 		&  
 		&\checkmark  
 		&  
 		&  
 		&  
 		&\checkmark  
 		\\ \cline{2-21}
 		&{\makecell{Macroscale Observations\\of Microscale Phenomena}}
 		&            
 		&  
 		&            
 		&  
 		&  
 		&            
 		&  
 		&  
 		&            
 		&            
 		&  
 		&  
 		&  
 		&  
 		&  
 		&\checkmark  
 		&\checkmark  
 		&\checkmark  
 		&\checkmark  
 		\\ \cline{2-21}
 		&{\makecell{Macroscale Control\\of Microscale Change}} 
 		&            
 		&  
 		&            
 		&  
 		&\checkmark  
 		&\checkmark            
 		&  
 		&\checkmark  
 		&            
 		&            
 		&  
 		&  
 		&  
 		&  
 		&  
 		&\checkmark  
 		&\checkmark  
 		&\checkmark  
 		&\checkmark  
 		\\ \cline{2-21}
 		&{\makecell{Conversion from\\Signals to Bits}}
 		&            
 		&  
 		&\checkmark            
 		&\checkmark  
 		&\checkmark  
 		&            
 		&  
 		&  
 		&\checkmark            
 		&\checkmark            
 		&  
 		&  
 		&\checkmark  
 		&  
 		&  
 		&\checkmark  
 		&  
 		&  
 		&\checkmark  
 		\\ \hline
 		\multirow{7}{*}{\makecell[c]{Section VI:\\Level 4\\Local Data}}
 		&{\makecell{Information in\\ Cellular Signals}} 
 		&            
 		&\checkmark  
 		&\checkmark            
 		&\checkmark  
 		&\checkmark  
 		&            
 		&  
 		&  
 		&\checkmark            
 		&\checkmark            
 		&  
 		&\checkmark  
 		&\checkmark  
 		&  
 		&  
 		&  
 		&  
 		&  
 		&\checkmark  
 		\\ \cline{2-21}
 		&{\makecell{Digital and\\Analog Circuits}} 
 		&            
 		&  
 		&            
 		&  
 		&  
 		&            
 		&  
 		&  
 		&            
 		&            
 		&  
 		&  
 		&  
 		&  
 		&\checkmark  
 		&  
 		&  
 		&  
 		&\checkmark  
 		\\ \cline{2-21}
 		&{\makecell{Realizing\\Communication\\Functionalities}}
 		&            
 		&  
 		&            
 		&  
 		&  
 		&            
 		&  
 		&  
 		&            
 		&            
 		&  
 		&  
 		&  
 		&  
 		&  
 		&  
 		&  
 		&  
 		&\checkmark  
 		\\ \cline{2-21} 
 		&{\makecell{Microscale\\Storage}}
 		&            
 		&  
 		&            
 		&  
 		&  
 		&            
 		&  
 		&  
 		&            
 		&\checkmark            
 		&  
 		&  
 		&  
 		&  
 		&  
 		&  
 		&  
 		&  
 		&\checkmark  
 		\\ \hline 
 		\multirow{6}{*}{\makecell[c]{Section VII:\\Level 5\\Application}} 		
 		&\multirow{2}{*}{{Biosensing}}
 		& 		            
 		&   
 		&            
 		&   
 		&   
 		&            
 		&   
 		&   
 		&             
 		&             
 		&   
 		&   
 		&   
 		&   
 		&   
 		&  
 		&   
 		&   
 		&\multirow{2}{*}{{\checkmark}}   
 		\\ 		
 		&
 		&   
 		&   
 		&   
 		&   
 		&   
 		&   
 		&   
 		&   
 		&   
 		&   
 		&   
 		&   
 		&   
 		&   
 		&   
 		&   
 		&   
 		&   
 		&
 		\\ \cline{2-21}       
 		&\multirow{2}{*}{{Therapeutics}}
 		&             
 		&  
 		&            
 		&   
 		&   
 		&{\multirow{2}{*}{{\checkmark}}}             
 		&   
 		&{\multirow{2}{*}{{\checkmark}}}    
 		&{\multirow{2}{*}{{\checkmark}}}              
 		&            
 		&  
 		&  
 		&  
 		&   
 		&  
 		&  
 		&  
 		&   
 		&\multirow{2}{*}{{\checkmark}}  
 		\\ 
 		 		&
 		&   
 		&   
 		&   
 		&   
 		&   
 		&   
 		&   
 		&   
 		&   
 		&   
 		&   
 		&   
 		&   
 		&   
 		&   
 		&   
 		&   
 		&   
 		&
 		\\ \cline{2-21}  
 		&\multirow{2}{*}{\makecell{MC-Assisted Application}}
 		&\multirow{2}{*}{{\checkmark}}            
 		&\multirow{2}{*}{{\checkmark}}  
 		&            
 		&\multirow{2}{*}{{\checkmark}}  
 		&\multirow{2}{*}{{\checkmark}}  
 		&\multirow{2}{*}{{\checkmark}}            
 		&\multirow{2}{*}{{\checkmark}}  
 		&\multirow{2}{*}{{\checkmark}}  
 		&\multirow{2}{*}{{\checkmark}}            
 		&            
 		&  
 		&  
 		&\multirow{2}{*}{{\checkmark}}  
 		&\multirow{2}{*}{{\checkmark}}  
 		&  
 		&  
 		&  
 		&\multirow{2}{*}{{\checkmark}}  
 		&\multirow{2}{*}{{\checkmark}}   
 		\\ 
 		&
 		&   
 		&   
 		&   
 		&   
 		&   
 		&   
 		&   
 		&   
 		&   
 		&   
 		&   
 		&   
 		&   
 		&   
 		&   
 		&   
 		&   
 		&   
 		&
 		\\ \hline
	\end{tabular}
	}}
\end{table*}

\section{Communication Hierarchy for Signaling in Cell Biology}
\label{sec_hierarchy}

In this section, we elaborate on our broad definition of communicating devices as we understand them in the context of cell biology. We then present our proposed communication hierarchy for signaling in cell biology. We briefly discuss each of the levels and compare them with the layers in existing communication protocols. The hierarchy is then used throughout the rest of this survey to articulate the different stages of a communication process between devices. 

\subsection{Defining Communication Devices}

A minimum requirement for any communication system is that there must be \emph{communicating devices}. A device can act as a \emph{transmitter} if it is a source of information and has a mechanism to translate that information into a physical signal for other devices to detect. A device can act as a \emph{receiver} if it needs to detect a signal from a transmitter and recover the embedded information. Of course, a single device can behave as both a transmitter and a receiver, i.e., as a \emph{transceiver}. Throughout this survey, we use the term ``device'' instead of ``transceiver'' to emphasize that the communicating devices can have diverse functions in addition to communication, and when relevant these functions are integrated within our proposed hierarchy. As we will see, one key difference between a fully-engineered communication system and a living cell system is that the transfer of information is explicitly present in the design of the engineered communication system, whereas it may only be implicitly represented in the biological system. For example, cells in a bacterial colony do not always directly communicate in the sense of information transfer between one specific cell and another specific intended target cell. Nevertheless, the release and detection of cellular signals, and the corresponding responses to those signals, indicate that communication is in fact taking place.

Due to the unstructured yet complex nature of biological communication, there is some flexibility in how we define a communicating device. In this survey, we take a very general approach so that we can identify devices as appropriate in the respective context. Thus, examples of devices include:
\begin{enumerate}
    \item Organelles and large macromolecules that engage in intracellular signaling. Depending on their functionality and mobility, macromolecules could also represent the signaling molecules between other devices.
    \item Individual cells in intercellular signaling networks. Again, individual cells could also represent the signal or its transport mechanism, depending on the context. For example, synthetic bacteria that move via chemotaxis have been proposed to carry information via plasmids in an artificial nanoscale network \cite{Unluturk2017}. Another example can be found in the animal immune system, where sub-populations of T-cells can act as intermediates for information about pathogens \cite{Lanzavecchia2000}.
    \item A colony of cells, including tissues and multi-cellular organisms, as an aggregation of individual cells in inter-population or inter-species communication.
    \item Experimental equipment or other synthetic means to introduce a signal or observe signals in a cell biology environment.
\end{enumerate}

An important detail is that any of these devices could be natural or synthetic (e.g., genetically-modified cells, synthetic macromolecules, microscale robots, etc.). Even though some devices listed above can be much larger than individual cells (e.g., multi-cellular organisms, cellular tissues), we still include them within our framework if they have identifiable signaling with microscale devices (including individual cells that are part of the larger device). Nevertheless, we often find it most convenient throughout this survey to consider a communicating device to be an individual cell.

\subsection{Our Proposed Hierarchy}

\begin{figure}[!t]
	\centering
	\includegraphics[width=3.3in]{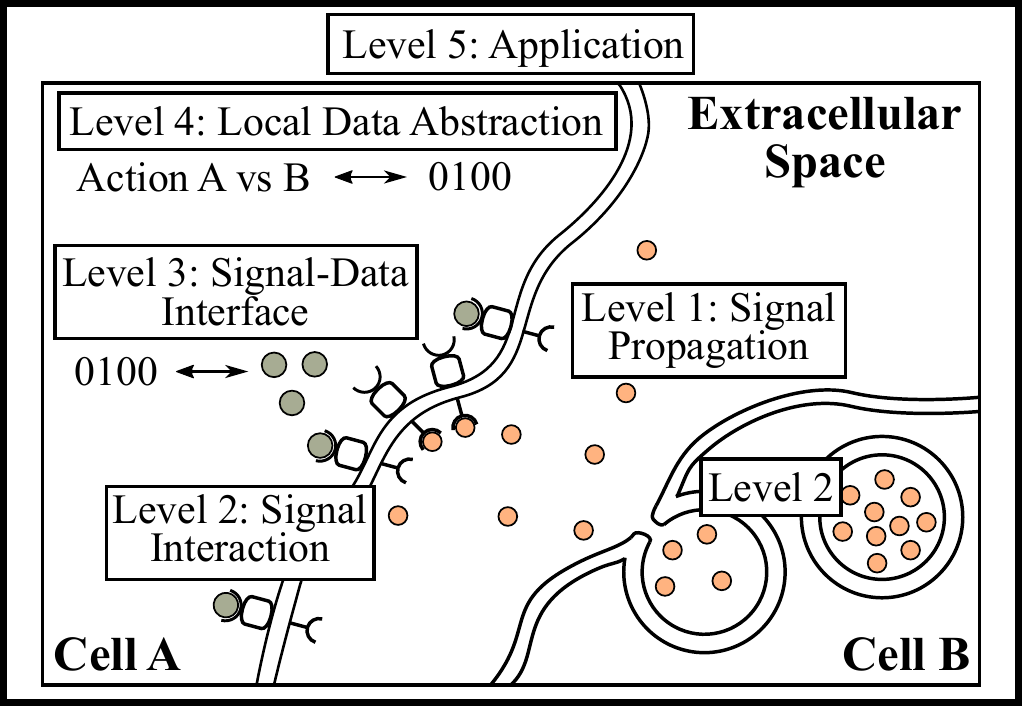}
	\caption{The proposed communication level hierarchy applied to an example cellular signaling environment. The communicating devices are two cells separated by extracellular space. In this scenario, Level 1 describes the signal propagation across the extracellular space. Levels 2-4 describe the biochemical pathways, signal-data interface, and data abstraction within the individual cells, respectively. Level 5 describes the overall behavior requiring communication between the two cells.}
	\label{fig_hierarchy_overlay}
\end{figure}

We summarize our proposed hierarchy in Fig.~\ref{fig_hierarchy} and apply it to a sample cellular signaling scenario in Fig.~\ref{fig_hierarchy_overlay}. The levels of the hierarchy are as follows:
\begin{enumerate}
    \item \textbf{Physical Signal Propagation} -- how molecules are transported between communicating devices, e.g., via diffusion, fluid flow, or contact-based means. This level is \emph{not} defined within devices themselves, but directly connects devices that are communicating; e.g., the two cells in Fig.~\ref{fig_hierarchy_overlay}.
    \item \textbf{Physical and Chemical Signal Interaction} -- how the physical signal is generated at a transmitting device and sampled at a receiving device, e.g., the stimulation of generation, release, and binding of the molecules. This also includes the biochemical signaling pathways that process molecular signals.
    \item \textbf{Signal-Data Interface} -- how physical signals are mathematically quantified, observed, and controlled. This includes the conversion of data between its mathematical representation and its physical form, i.e., modulation and demodulation in communication networks.
    \item \textbf{Local Data Abstraction} -- the meaning of quantified data at a local device, e.g., timing information, quantization of concentration, or the genetic information in a strand of DNA. This includes the information theoretic limits of molecular signals. We classify the field of bioinformatics to be primarily at this level \cite{Compeau2018}. The level also includes encoding and decoding in communication networks.
    \item \textbf{Application} -- the top-level behavior that is relying on communication. This could be entirely within a biological context, e.g., differentiation of cells in a multi-cellular organism or symbiosis between different species, or within a mixed synthetic and biological context, e.g., disease detection by medical sensors.
\end{enumerate}

We can compare our hierarchy with the TCP/IP communication protocol, which has similar concepts but several key differences. For clarity, we make a direct comparison with the TCP-based protocol proposed in \cite{Nakano2014b}. In \cite{Nakano2014b}, the application layer is an interface for applications to access the communications functionality; in our hierarchy, the application level \emph{is} the application itself. In \cite{Nakano2014b}, the transport, network, and link layers provide operations that are critical to the operation of a synthetic communication network; our focus is on biological systems, which in general do not implement these specific operations, and so they are not directly represented in our approach. In particular, we focus on the mathematical representation of the information (through the local data abstraction level) rather than the management of that information (as provided by the aforementioned layers and which is also the focus of \cite{Felicetti2014a}).

In \cite{Nakano2014b}, the signaling sublayer includes modulation and demodulation, transmission and sampling, and signal propagation. We separate these three critical tasks into distinct levels, i.e., Levels 1--3, because of the diversity of implementations at each of these levels and their interoperability for different biological and engineering applications. Finally, \cite{Nakano2014b} has a bio-nanomachine sublayer to define routine operations for the ongoing maintenance of a device; we do not focus on such behavior so we do not define a corresponding level, though we do integrate life-preserving tasks within the hierarchy (e.g., gene expression in Level 2; gene regulation and cell metabolism in Level 3).

Important advantages of our approach are its flexibility and scalability, which are both useful for the study of and interaction with propagating signals in cell biology. If we tried to map a natural communication system to a structured communication protocol, then we would observe many instances where components are missing, e.g., the network layer is defined for packet management in \cite{Nakano2014b} but such a function is not evident in cell-based diffusive signaling. This is in part because the quantity of information that is communicated in biological systems is often relatively small (on the order of a few bits; e.g., see \cite{Ruiz2018}) when compared with the objectives of synthetic MC systems. Important exceptions are DNA (as there are millions of base pairs -- 2 bits each -- in the genomes of most organisms; see \cite[Fig.~1.32]{Alberts2015}) and the aggregate information stored in the brain (as the human brain has about $10^{11}$ neurons and $10^{14}$ synaptic connections between them \cite[Ch.~11]{Alberts2015}). Our approach enables us to be more holistic and flexible in our mapping by identifying multiple and potentially overlapping hierarchies within a single environment without concern for how the hierarchies would all map and adhere to a single uniform communication protocol. Thus, we can simultaneously characterize internal communication within a cell, signaling between cells, and experimental observation of individual cells or an entire population, which is not practical with a formal communication protocol. There can also be asymmetry in the communicating devices, and some may not even exhibit all of the levels, so long as they are joined by some physical propagation channel. For example, chains of motor neurons propagate action potentials that convey the signal for muscle contraction, but the neurons do not need to understand the information that is being relayed. Additionally, the newly-discovered telocytes can act as intermediates for the flow of information between different types of cells \cite{Cretoiu2014,Cretoiu2017}.

\begin{table*}[!t]
    \rowcolors{4}{gray!15}{white}
	\renewcommand{\arraystretch}{1.6}
	\caption{Summary of Physical Signal Propagation}.
	\label{table_propagation}
	\centering
	\begin{tabular}{ l l l c}
		\toprule
		\multicolumn{1}{c}{Propagation Mechanism} & \multicolumn{1}{c}{Example} & \multicolumn{1}{c}{Speed} & \multicolumn{1}{c}{Reference}\\
		\midrule
		\multirow{3}{*}{\hfill Diffusion-Based Propagation}
		& Calcium ions & Diffusion coefficient: $5.3\times 10^{-10}$ m$^2$/s & \cite{Donahue1987} \\
		& Pheromones & Diffusion coefficient: $6.49\times 10^{-7}$ m$^2$/s & \cite{Bossert} \\	
		& \textit{lac} repressor protein & Diffusion coefficient: $10^{-14} - 5 \times 10^{-12}$ m$^2$/s & \cite{Tempestini2018}\\
		Advection-Diffusion-Based Propagation & Human skin capillary& Mean velocity: $3.5\time 10^{-4} - 9.5 \times 10^{-4}$m/s & \cite{Fagrell1977}\\
		Advection-Diffusion-Reaction-Based Propagation & Antibody-antigen interaction & Diffusion coefficient: $\approx 10^{-10}$ m$^2$/s & \cite{Gervais}\\
		Cargo-Based Propagation& Kinesin propagation & Mean velocity: $\approx 4.5 \times 10^{-6}$ m/s & \cite{Turner1996}\\
		Contact-Based Communication & Gap junctions & Mean velocity: $\approx 5 \times 10^{-1}$ m/s & \cite{Cole1988}\\
		\bottomrule
	\end{tabular}
\end{table*}

\section{Level 1 - Physical Signal Propagation}
\label{sec_physical}

A fundamental characteristic of any communication network is how information propagates between the devices. In this section, we survey the means by which MC signals physically propagate. These correspond to Level 1 of our proposed communication hierarchy (see Figs.~\ref{fig_hierarchy} and \ref{fig_hierarchy_overlay}). We focus on diffusion-based phenomena (Sections~\ref{Diffusion} to \ref{Convection-Diffusion-Reaction}) because they are prevalent at the microscale and they have received significant attention within the MC engineering community \cite{Jamali2019b}. 
We include mathematical descriptions for diffusion, which integrate with the mathematical characterizations of initial and boundary conditions (for Level 2 in Section~\ref{sec_interaction}) to determine the corresponding channel response. In addition, we describe cargo-based transport with motors and chemotaxis (Section~\ref{sec_cargo}), as well as contact-based propagation including gap junctions and plasmodesmata (Section~\ref{sec_contact}).
A summary of the propagation mechanisms that we discuss, including representative molecules for each mechanism, is also provided in Table~\ref{table_propagation}.

It is helpful to have a sense of the scope in diversity of molecules and mechanisms that are used in cell signaling. In biological systems, regardless of physical scale but especially in plants and animals, there is a tremendous variety of molecules with distinct characteristics (e.g., size, shape, electrical charge) that act as messengers between cells or whole organisms, in addition to the biochemical machinery to support them. For example, extracellular vesicles have been found to carry more than 19 distinct proteins that are believed to be involved in signaling between animal cells \cite{Tkach2018}. A search through the \textit{in silico Human Surfaceome} database\footnote{Publicly available at: http://wlab.ethz.ch/surfaceome/.} \cite{Bausch-Fluck2018} reveals 1201 surface receptor proteins and 137 ligands. Considering that not all molecules or receptors are known, that other communication modalities (e.g., ion channels, gases, contact signaling) are equally or more important depending on the cell type, and that there is significant cross-talk between modalities, these examples give a glimpse of the scope in diversity and complexity that characterizes communication in natural systems. Nevertheless, many of the molecules propagate according to one (or more) of the approaches surveyed in this section. Examples include the diverse cell signaling processes using ions (e.g., Ca$^{2+}$, K$^+$) that convey a different message depending on the type of the recipient cell, microbial quorum sensing where organisms secrete their own variants of molecules as communication signals, or pheromones carrying messages over long distances \cite{Wyatt2003}.

Furthermore, in contrast to fully-engineered systems where there is usually an effort to standardize components and minimize signal variability, natural systems (and those that are synthetically derived from natural systems) can express variation within populations of the same species or cells of the same type. This variation, which can include strength of gene expression or response to external stimuli \cite{Plener2015,Grote2015}, leads to variability in signal density, duration, and timing. Environmental factors also influence the propagation of any molecular signal (e.g., temperature, pH, presence of flow). Last but not least is the fact that competition for resources or between predators and prey has led to an arms race between different organisms where survival depends on the successful detection of or interference with each other's signals (e.g., ``eavesdropping'' in bacteria \cite{Rebolleda-Gomez2019}; discussed in Section~\ref{sec_quorum}). With signal interference and also the inherent stochasticity of molecule release and propagation, cell signaling is inherently very noisy and these features must be taken into consideration in order to understand a natural system's reliability or to design a synthetic MC-based system.

\subsection{Diffusion-Based Propagation}
\label{Diffusion}
Diffusion refers to the random walk, namely Brownian motion, of molecules in a medium arising from the molecules' thermal energy \cite{Einstein1956}. It is a simple and efficient movement paradigm without a need for infrastructure or external energy sources. Therefore, there are many examples found in nature, including calcium signaling among cells \cite{Nakano2005}, pheromonal communication among animals \cite{Bossert}, and propagation of DNA binding molecules over a DNA segment \cite{berg1993random}.

\begin{figure}[!t]
	\centering
	\includegraphics[width=2.7in]{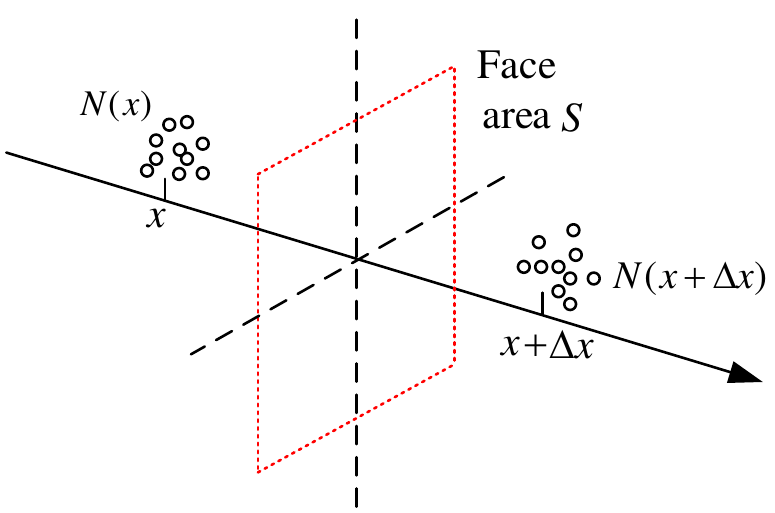}
	\caption{A macroscopic one dimensional (1D) random walk model where the small circles represent molecules and move along the $x$-axis. The dotted red square is in the plane that is orthogonal to the $x$-axis.}
	\label{f_fick_law}
\end{figure}

The mathematics of Brownian motion are often modeled using Fick's laws of diffusion. As a conceptual example, we find it useful to describe Fick's first law of diffusion from first principles using the macroscopic approach presented in \cite[Ch.~2]{berg1993random}. We consider the simplified case shown in Fig. \ref{f_fick_law}, where molecules move one step at a time along only one axis with a displacement step $\Delta x$ and a time step $\Delta T$. 
We assume that each molecule walks independently and the probabilities of moving forward and backward are both $1/2$. Let $N(x)$ denote the number of molecules at position $x$ and time $t$. During the time interval $[t,t+\Delta T]$, we expect that half of the molecules at $x$ will move to $x+\Delta x$ and traverse the normal face that is orthogonal to the axis and located at $(x+\Delta x)/2$. At the same time, we expect that half of the molecules at $x+\Delta x$ will cross the face in the opposite direction. Hence, the net \textit{expected} number of molecules coming to $x+\Delta x$ will be $\frac{1}{2}[N(x)-N(x+\Delta x)]$. Dividing by the face area $S$ and time step $\Delta T$, the net flux $J_{\text{Diff}}$ crossing the face by diffusion is 
\begin{equation}
\label{net_flux3}
J_{\text{Diff}}{\big|_{1\mathrm{D}}}=-\frac{1}{2{\Delta T}}\frac{[N(x+\Delta x)-N(x)]}{{S}}.
\end{equation}

If we further multiply the right-hand side of \eqref{net_flux3} by ${{\Delta x}^2}/{{\Delta x}^2}$, then it becomes
\begin{equation}
    \label{net_flux3_1}
    \begin{aligned}
        J_{\text{Diff}}{\big|_{1\mathrm{D}}}&=-\frac{1}{{\Delta x}}\frac{{\Delta x}^2}{2{\Delta T}}\frac{[N(x+\Delta x)-N(x)]}{{S}{\Delta x}}\\
        &=-\frac{1}{{\Delta x}}\frac{{\Delta x}^2}{2{\Delta T}}[C(x+\Delta x)-C(x)],
    \end{aligned} 
\end{equation}
where $C(x+\Delta x)=N(x+\Delta x)/({S}{\Delta x})$ and $C(x)=N(x)/({S}{\Delta x})$ are the molecular concentrations at locations $x+\Delta x$ and $x$, respectively.
By considering $\Delta x \to 0$ and defining the \textit{diffusion coefficient} $D={{\Delta x}^2}/{(2\Delta T)}$, we arrive at Fick's first law in 1D sapce \cite[eq.~(2.1)]{berg1993random}, i.e.,
\begin{equation}
\label{fick_1st_law_1D}
J_{\text{Diff}}{\big|_{1\mathrm{D}}}=-D\frac{\partial C(x,t)}{\partial x}.
\end{equation}

Correspondingly, Fick's first law in three dimensional (3D) space is 
\begin{equation}
\label{fick_1st_law_3D}
J_{\text{Diff}}\big|_{3\mathrm{D}}=-D\frac{\partial C(\textbf{d},t)}{\partial x},
\end{equation}
where vector $\textbf{d}=[x,y,z]$ specifies the molecule position.

Fick's first law describes the relationship between the diffusion flux and the concentration gradient. The value of the diffusion coefficient $D$ determines how fast a certain type of molecule moves. In general, $D$ is dependent upon the environment (e.g., temperature, viscosity) as well as the molecule size and shape. For example, in a given environment, smaller molecules tend to diffuse faster. However, even when a molecule's diffusion coefficient is on the order of $1000\,\mu$m$^2$/s (a relatively large value), it is estimated that it would take nearly half a millisecond for such a molecule to travel over $1$\,$\mu$m (the width of a bacterium) \cite{berg1993random}, demonstrating that diffusion alone is quite a slow process.

The impact of diffusion on concentration change with respect to time can be described by Fick's second law as \cite[eq.~(2.5)]{berg1993random}
\begin{equation}
	\label{fick_2st_law}
	\frac{\partial C(\textbf{d},t)}{\partial t}=D{\nabla}^2 C(\textbf{d},t),
\end{equation}
where ${\nabla}^2$ is the Laplace operator. Solutions to \eqref{fick_2st_law} can be obtained under different initial and boundary conditions, depending on the diffusion environment. We discuss examples of initial and boundary conditions in greater detail in Section~\ref{sec:ICs_BCs}.

\subsection{Advection-Diffusion-Based Propagation}
\label{Convection-Diffusion}
The diffusion process can be accelerated by introducing additional phenomena. In particular, molecule transport can be assisted by two physical mechanisms: 1) force-induced drift, and 2) advection, i.e., bulk flow. Force-induced drift is caused by applying an external force directly to the particles rather than the fluid containing them. Examples include applying a magnetic field to magnetic nanoparticles, an electrical field to charged particles, and a gravitational force to particles with sufficient mass \cite{Jamali2019b}. Advection refers to molecule transport assisted by bulk movement of the entire fluid, including the molecules of interest. Examples include endocrine signaling in blood vessels and the manipulation of fluids in microfluidic channels (we elaborate on applications using microfluidics in Section \ref{sec_applications}).
Here, we focus on advection and in the following, we present a mathematical framework to approximate molecular transport assisted by advection.
 
Analogous to diffusion, the advection process also results in a flux of concentration crossing the surface of a given region. It has been shown that the concentration flux caused by advection is simply a concentration shift over time; thus the flux $J_{\text{Adv}}$ with \textit{local} velocity $\textbf{u}$ can be described by  \cite{Hundsdorfer2013a}
\begin{equation}
J_{\text{Adv}}=\textbf{u}C.
\label{eq9}
\end{equation}

The temporal change in concentration is jointly determined by the diffusion flux and the advection flux, and can be expressed as {\cite[eq.~(4.3)]{Kirby2010}}
\begin{equation}
\frac{\partial}{{\partial}t}\int_{V} CdV=-\int_{S}(J_{\text{Diff}}+J_{\text{Adv}})\cdot \textbf{n}dS,
\label{eq10}
\end{equation}
where $V$ is the volume of a given region with differential element $dV$, $S$ is the surface of the volume with differential element $dS$, and $\textbf{n}$ is a unit outward normal vector. Substituting \eqref{fick_1st_law_3D} and \eqref{eq9} into \eqref{eq10}, and  applying the divergence theorem, we obtain the advection-diffusion equation in differential form as \cite{Bruus2008} 
\begin{equation}
\frac{\partial 
C(\textbf{d},t)}{{\partial}t}= D{\nabla}^2 C(\textbf{d},t)-\textbf{u} \cdot \nabla C(\textbf{d},t),
\label{eq11}
\end{equation}
where $\nabla$ is the Nabla operator.

It is clear from \eqref{eq11} that the flow properties, such as the velocity $\textbf{u}$, have an impact on the distribution of the molecule concentration. Several dimensionless numbers have been defined to classify and characterize the transport behavior of fluids. Two important examples are the Reynolds number and P\'{e}clet number, which we describe in the following.

\subsubsection{Reynolds number}
The Reynolds number (Re) is defined as {\cite[eq.~(2.39)]{Bruus2008}}
\begin{equation}
	\label{re_number}
	\text{Re}=\frac{{\uprho}u_{\text{eff}}d}{{\mu}},
\end{equation} 
where $\uprho$ is the flow density, $u_\text{eff}$ is the mean velocity, and $\mu$ is the fluid dynamic viscosity. $d$ is the characteristic length scale, and for flows in a pipe or tube, it becomes the hydraulic diameter of the pipe {\cite[eq.~(3)]{Bicen2013}}.
The Reynolds number determines whether the flow is in the \textit{laminar} regime or the \textit{turbulent} regime. In the laminar regime, the Reynolds number is normally less than $2300$, and regular streamline flow patterns can be experimentally observed. In contrast, in the turbulent regime, the Reynolds number is larger than $2300$, and a single stable flow pattern cannot be observed in practice. For microscale channels (whether synthetic or blood vessels), the Reynolds number is frequently less than $20$; thus, laminar flows are often assumed \cite{Kumar2010a}. For example, it has been demonstrated that most blood vessels (except the aorta with Re $\in$ $[1200,4500]$) are laminar \cite{Back,Jensen2018}. Based on this, the authors of \cite{Chahibi2013} derived a time-varying drug delivery concentration profile based on the advection-diffusion equation in \eqref{eq11}. This work provided an initial understanding of drug propagation and laid the foundation to  establish advanced therapeutic methods.

A typical example for laminar flow is Poiseuille flow, where a pressure drop exists between the inlet and outlet of a microfluidic channel. If the flow only moves along the $x$ direction, and if the channel cross-section is circular, then the velocity distribution $u_x (r)$ can be expressed as {\cite[eq.~(3.42a)]{Bruus2008}}
\begin{equation}
u_x (r)=\frac{{\Delta}P}{4\mu L}(R^2-r^2),
\label{eq4_r}
\end{equation}
where $\Delta P$ is the pressure drop, $L$ is the channel length, $R$ is the radius of the cross-section, and $r$ is the radial location. Eq.~\eqref{eq4_r} indicates that the velocity follows a parabolic distribution, such that the flow velocity increases from the boundary towards the center of the channel. The velocity distributions for other cross-section shapes can be found in \cite[Ch.~3]{Bruus2008}.

\subsubsection{P\'{e}clet number}
The P\'{e}clet number (Pe) compares the relative dominance of advection versus diffusion and is defined as {\cite[eq.~(5.53)]{Bruus2008}}
\begin{equation}
\text{Pe}=\frac{u_{\text{eff}}L}{D}.
\label{eq12}
\end{equation}
For $\text{Pe}=0$, the molecular movement is purely diffusive; for  $\text{Pe}\to\infty$, the movement becomes a pure bulk flow process.

\begin{figure}[!t]
  \captionsetup[subfigure]{labelformat=empty}%
	\centering
	\subfloat[(a)]{\includegraphics[scale=0.45]{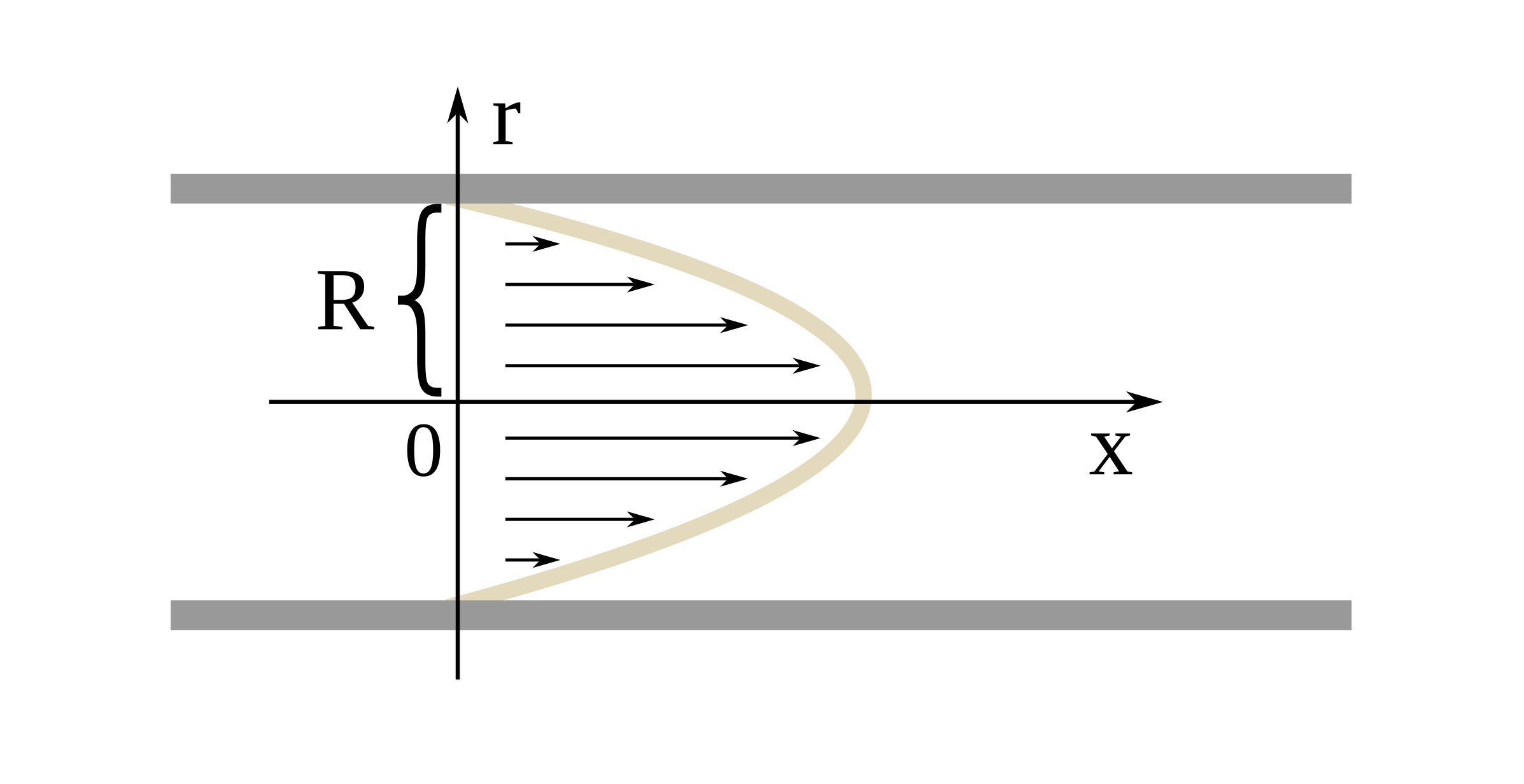}}\\
	\vfill
	
	\subfloat[(b)]{
	  \subfloat[(b1)]{\includegraphics[scale=0.63]{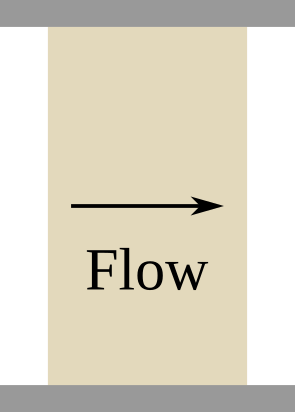}}
	  \hspace{2mm}%
	  \subfloat[(b2)]{\includegraphics[scale=0.63]{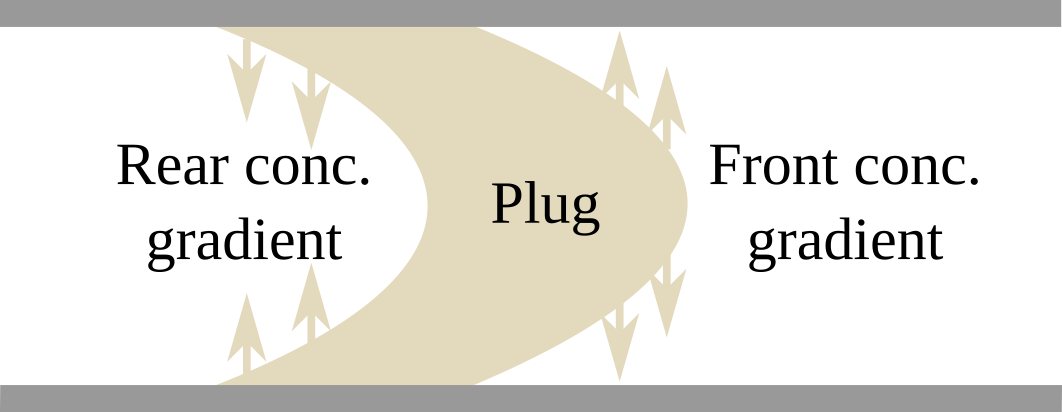}}
	  \hspace{2mm}%
	  \subfloat[(b3)]{\includegraphics[scale=0.63]{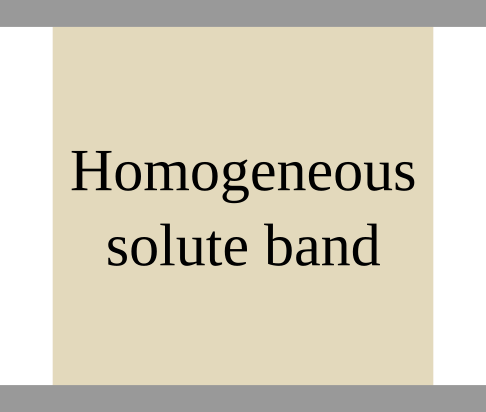}}
	}
	\caption{The schematic of Taylor dispersion in Poiseuille flow. (a) In a microfluidic channel, the velocity increases from the boundaries inwards, following the parabolic distribution in \eqref{eq4_r} ($R$: cross-section radius, $r$: radial location, $x$: direction of flow). (b) Taylor dispersion progression inside a microfluidic channel: (b1) A homogeneous solute band is injected into the channel. (b2) After injection, the solute band is stretched into a parabolic plug due to the parabolic velocity distribution. Then, the concentration gradients established at the front and back ends, cause the net motion of solute molecules to counteract the parabolic plug. (b3) Finally, the molecules are uniformly distributed over the cross-section.
	}
\label{f_taylor}
\end{figure}

The P\'{e}clet number is useful to predict the molecular distribution under Taylor dispersion, which describes how axial advection and radial diffusion jointly affect molecular transport in pressure-driven bulk flow \cite{Bruus2008}. Specifically, as shown in Fig. \ref{f_taylor}, a homogeneous band of solute is injected at $x=0$ to travel through a cylindrical microchannel with radius $R$.
A very short time after injection, the solute molecules are stretched into a parabolic plug by the flow having the velocity profile in (\ref{eq4_r}). Subsequently, two concentration gradients are established at the front and back ends of the solute plug. Due to these gradients, there is a net migration of solute molecules at the front end from the high concentration area (i.e., the channel center) to the low concentration area (i.e., the channel boundary). On the contrary, there is a net migration of molecules at the back end from the channel boundary to the area around the channel center. We use $R^2/(4D)$ to characterize the expected diffusion time along the radial direction, and use $L/u_\text{eff}$ to represent the time of molecule transport at average fluid velocity $u_\text{eff}$ over distance $L$. 
If $R^2/(4D) \gg L/u_\text{eff}$ (i.e., $\text{Pe} \gg 4L^2/{R^2}$), then the cross-sectional diffusion cannot be ignored and fully counteracts the parabolic plug, which leads to a uniform distribution of the solute over the cross-section \cite{Probstein1994,Wicke2018}. Thus, in this case, the 1D advection-diffusion equation with a modified diffusion coefficient can be used to approximate the 3D Poiseuille flow \cite{Bi2020}.

\subsection{Advection-Diffusion-Reaction-Based Propagation}
\label{Convection-Diffusion-Reaction}
In addition to the diffusion and advection processes, chemical reactions often occur simultaneously during molecular movement. Examples include the polymerase chain reaction (for synthetically copying DNA \cite{Yariv2005}) and surface capture \cite{Gervais,Squires}. To analyze molecular transport under chemical reactions, we consider the example of a second-order (bimolecular) reaction $S_i+S_j \stackrel{k_f}{\to} S_l$, where species $S_i$ reacts with species $S_j$ to generate product $S_l$ under the rate constant $k_f$. If molecular transport is subjected to diffusion and reaction, then the concentration changes of the reactant $S_i$ (analogously $S_j$) and the product $S_l$ can be expressed as
\begin{subequations}
\begin{alignat}{2}
\frac{\partial C_{S_i}(\textbf{d},t)}{{\partial}t}=D{\nabla}^2 C_{S_i}(\textbf{d},t)-{k_f} C_{S_i}(\textbf{d},t) C_{S_j}(\textbf{d},t), \\
\frac{\partial C_{S_l}(\textbf{d},t)}{{\partial}t}=D{\nabla}^2 C_{S_l}(\textbf{d},t)+{k_f} C_{S_i}(\textbf{d},t) C_{S_j}(\textbf{d},t).
\end{alignat}
\end{subequations}

A general diffusion-reaction equation is given by
\begin{equation} \label{dr}
    \frac{\partial C_{S_{i/l}}(\textbf{d},t)}{{\partial}t}=D{\nabla}^2 C_{S_{i/l}}+qf[{k_f},C_{S_i}(\textbf{d},t)],
\end{equation}
where $q=-1$ holds for reactants, $q=1$ holds for products, and $f[\cdot]$ is the reaction term which in general can account for the presence of multiple reactions. Furthermore, if molecular propagation is simultaneously governed by advection, diffusion, and chemical reaction, then the spatial-temporal concentration distribution can be expressed by the following advection-diffusion-reaction equation \cite{Jamali2019b}:
\begin{equation} \label{cdr}
    \frac{\partial C_{S_{i/l}}(\textbf{d},t)}{{\partial}t}=D{\nabla}^2 C_{S_{i/l}}(\textbf{d},t)-\textbf{u} \cdot \nabla C_{S_i}(\textbf{d},t)+qf[{k_f},C_{S_i}(\textbf{d},t)].
\end{equation}

With certain initial and boundary conditions, the expected time-varying concentration of each type of molecule can be derived. Some analytical solutions of the 1D form of \eqref{cdr} can be found in \cite{Genuchten1982a}.

\subsection{Cargo-Based Propagation}
\label{sec_cargo}

In contrast to the free transport of signaling molecules, there are also signal propagation mechanisms that rely on the molecules of interest being transported as cargo within some kind of biological container. These can be classified as \textit{active} transport, as energy is expended for loading, unloading, and moving the container. Such propagation mechanisms are used by nature to overcome the slowness and limited directionality of diffusion-based signaling, particularly over larger distances. In this subsection, we review cargo-based propagation using molecular motors and using bacteria participating in chemotaxis.

\subsubsection{Molecular Motors}

\begin{figure}[!t]
	\centering
	\includegraphics[width=2.5in]{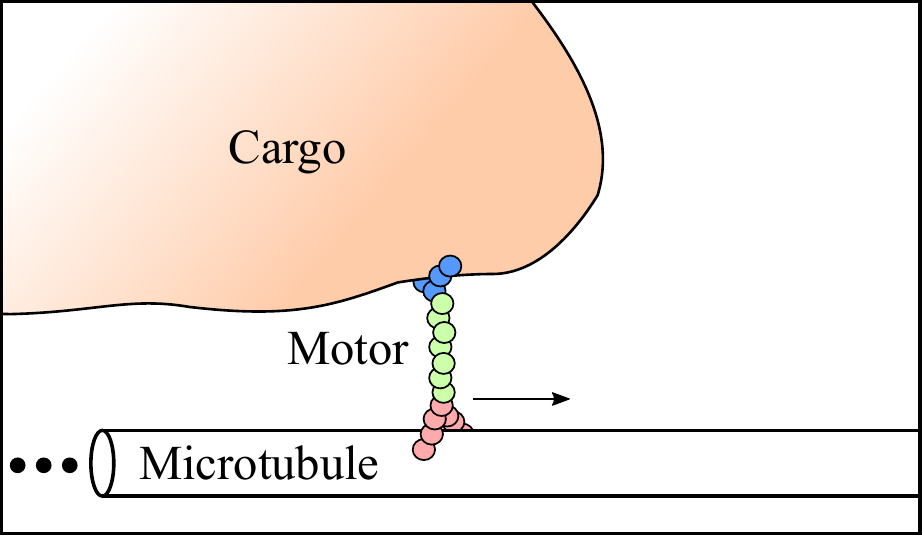}
	\caption{Representative example of a motor protein carrying cargo along a microtubule. Motor proteins are used for directed transport within a cell. Motor proteins typically have a head (shown in red) that ``walks'' along the cytoskeleton, a tail (shown in blue) that binds to its cargo, and a long stalk (shown in green) that joins the head and the tail. The cargo can be much larger than the motor, e.g., vesicles and mitochondria.}
	\label{fig_molecular_motor}
\end{figure}

A common example in intracellular signaling, particularly within eukaryotes (including plant and animal cells) as these cells are generally larger than bacteria and other prokaryotes, is transport via translational motors along the cytoskeleton. The cytoskeleton has different ``track''-like components that help to control the shape, movement, and division of the cell. The motors are proteins and a given type of motor protein can move in a single direction along a particular cytoskeleton element. Motor proteins can be classified into one of three families: myosin, kinesin, and dynein \cite[Ch.~16]{Phillips2013}. Myosin motors travel along actin filaments whereas kinesin and dynein motors move along microtubules (e.g., see Fig.~\ref{fig_molecular_motor}). Each type of motor binds to the cytoskeleton element at one end (i.e., the head) and binds to its cargo or cargo container at the other end (i.e., the tail). A sequence of energy-consuming reactions hydrolyzes ATP to induce changes in the motor binding conformation at the head so that the motor moves along the cytoskeleton.

The possibilities for cargo are quite diverse and can include mitochondria (which produce ATP), messenger RNA (intermediate molecules needed to produce proteins), and different types of vesicles (a more comprehensive discussion on transport with vesicles for Level 2 is given in Section~\ref{sec_interaction}). A common function is to transport the cargo from where it is synthesized in the cell to where it is used, e.g., a secretory vesicle could bind to a motor to carry external signaling molecules from the Golgi apparatus where they are produced to the cell surface where they are released. Even though the cargo can be much larger than the motor, relatively fast transport speeds are possible; propagation of secretory vesicles has been measured at speeds of up to $10\,\mu\textrm{m}/\textrm{s}$ \cite{Smith2008}.

Despite the infrastructure in place for molecular motors to move along cytoskeleton elements, the propagation still has statistical behavior over short time scales due to the precise timing of the ATP hydrolysis reactions \cite[Ch.~16]{Phillips2013}. It is even possible for a motor to detach, diffuse around, and rebind to the cytoskeleton \cite{Nieuwenhuizen2004}. Thus, as with diffusion-based transport, random walk models apply, and the expected behavior can be described analogously to an advection-diffusion channel (as discussed in Section~\ref{Convection-Diffusion}).

From a communications engineering perspective, molecular motor-based signaling was one of the first mechanisms proposed for synthetic nanoscale communication \cite{Hiyama2005a} and also considered in early experimental work \cite{Hiyama2008}. However, there has been limited work on characterizing communication in a system where devices are connected by cytoskeleton-like elements that are traversed by motors \cite{Chahibi2016,Suda2019}. Some attention has been given to the design of synthetic systems that reverse the roles of cytoskeleton and motor, such that cytoskeleton elements become the information molecules. In these systems, a surface is covered with motor proteins and these proteins push microtubules between transmitter and receiver devices \cite{Farsad2012a,Farsad2015a}.  Such designs are envisioned to be more suitable for implementation in lab-on-a-chip platforms.

\subsubsection{Chemotaxis}
\label{Chemotaxis}

A cellular mobility mechanism that has been proposed for adoption as a cargo-based transport method is chemotaxis. The most common example associated with chemotaxis is that of  bacteria moving along concentration gradients, but more generally chemotaxis refers to any organism movement in response to a chemical stimulus. In the case of bacteria, they engage in a series of \textit{runs} and \textit{tumbles} to perform local concentration sensing and bias their motion towards food sources or away from toxins; further details can be found in \cite[Ch.~19]{Phillips2013}. While moving, bacteria coming into contact with each other are able to exchange genetic information. As a result, there have been proposals to capitalize on this and use bacteria as cargo-carrying organisms between networking nodes that need to communicate \cite{Balasubramaniam2013a,Castorina2016}.   Under such a process, a receiver node releases molecules to attract cargo-carrying bacteria to \textit{run} toward it from the corresponding transmitter, and information (i.e., the cargo) is shared between the bacteria and nodes using plasmid conjugation (we discuss conjugation further in the context of contact-based communication in Section~\ref{sec_contact}). The more general concept of guiding nanomachine motion to target sites using the release of attractant or repellent molecules has been presented in \cite{Nakano2017,Nakano2019}, where the nanomachines are envisioned to be carrying a drug payload or perform some other therapeutic tasks.

Signal propagation under chemotaxis has multiple components, i.e., the diffusion of attractant and repellent molecules, the motion of cells (or more generally nanomachines) in response to molecule gradients, and the process of conjugation upon contact. The processes of bacteria runs and tumbles can be mathematically modeled as a biased random walk \cite{Schnitzer1993}, and such a model was also adopted to describe cargo-carrying bacteria in nanonetwork design \cite{Gregori2011,Petrov2015}.

\subsection{Contact-Based Communication}
\label{sec_contact}

While the diffusion-based and cargo-based transport mechanisms described thus far take place over the open space separating communicating devices, there are mechanisms that rely on direct or indirect contact between devices. Such mechanisms tend to have better reliability as a result. Here, we discuss direct contact-based signaling mechanisms including the use of gap junctions and plasmodesmata. We also discuss appendage-based processes, where cells have mechanisms to ``reach'' toward or into each other to make contact, including conjugation, tunneling nanotubes, and telocytes.

In biological systems, the distinction between propagation mechanisms might not always be clear. For example, a signal propagating through any number of gap junctions (GJs) or plasmodesmata (PD) (both described below) could be targeting a population of cells that are relatively far away from the source. One could then argue that a series of GJs \textit{is} in effect  the communication channel (in contrast to a series of local diffusion events). As the biophysical constraints of channel regulation across many sites along the signal path can greatly affect signal propagation and reception, we describe these biological systems and their regulation in moderate detail to support future work for channel modeling. This also leads to several open research problems discussed in Section~\ref{sec_open}.

\subsubsection{Gap Junctions}
Gap junctions are clusters of membrane channels that connect adjacent cells and they are present in every tissue of any multicellular organism, a hint on the importance of these communication channels. They are tightly regulated and permit the diffusion of small molecules directly from the cytoplasm of one cell to that of its neighbor, thereby avoiding the extracellular matrix. For general reviews on gap junctions, see \cite{Saez2003, Goodenough2009}.

\begin{figure}[!t]
	\centering
	\includegraphics[width=2.8in]{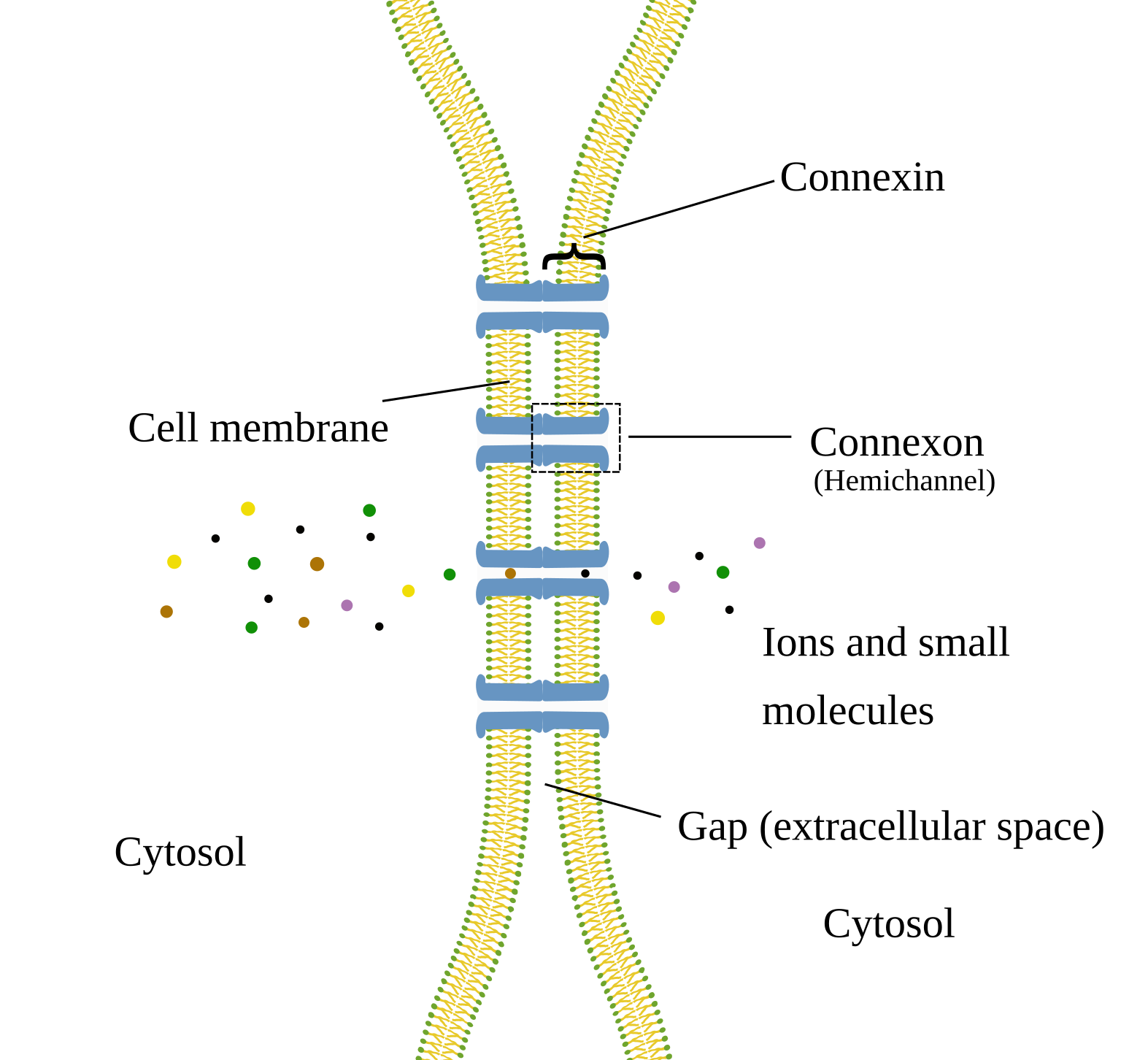}
	\caption{Gap junction structure. Individual proteins (connexins) form a hemichannel, or connexon. Two connexons from adjacent cells are joined to create one gap junction. A small gap remains between the cells, giving gap junctions their name. The channel opens selectively to allow ions (e.g., calcium ions) and other small molecules to pass through.}
	\label{fig_gap_structure}
\end{figure}

GJ structure was first visualized in the 1960s using electron microscopy \cite{Dewey1962}. Subsequent experiments revealed the three-dimensional structure of gap junctions in sufficient detail to enable the description of an individual GJ as the combination of two hemichannels constructed by specific proteins (\textit{connexins}), one from each cell, that are connected head-to-head \cite{Unwin1980,Maeda2009}; see Fig.~\ref{fig_gap_structure}. Clusters of gap junctions arranged in a hexagonal lattice connect the plasma membranes of adjacent cells. Each hemichannel (a \textit{connexon}) is comprised of six subunits that create a cylindrical pore connecting the cells.

By creating direct pathways between cells, it is known that gap junctions play an important role in intercellular communication. They permit exchange of ions, miRNA, and other small molecules such as metabolites and second messengers. Due to this important function, GJs have to be tightly regulated. Opening and closing of GJs can be achieved via chemical, electrical, and mechanical means. Two different mechanisms are known to be involved in the regulation of GJs' opening and closing \cite{Trexler1996, Oh2000}:
\begin{enumerate}
    \item A \textit{fast-gating} mechanism, where rectification of ionic currents passing through a fully-opened channel occurs due to selective permeability. Transitions are fast and at least three intermediate states between open and closed are known.
    \item A \textit{slow voltage-sensitive} mechanism, also termed \textit{loop-gating}. With this mechanism, transitions between states occur in many small steps, resulting in a slower response.
\end{enumerate}

 Regulation of GJ permeability is achieved by means of an electrostatic barrier created by Ca$^{2+}$ \cite{Bennett2016}. Calcium ions binding to specific side chains on each hemichannel create a positive gradient that inhibits any other positive ion such as K$^{+}$ from entering the pore. The mechanism, though not yet fully understood, involves the interaction between parts of the intracellular domains of GJ and  Ca$^{2+}$-bound calmodulin \cite{Peracchia2004}. Connexin subunits in gap junctions can bind Ca$^{2+}$ ions and create a strong positive surface potential. The result is an effective electrostatic barrier that can block the entrance of other positive ions. This allows a rapid response of the gating mechanism, much more so than if large conformational changes were needed. 

Protein phosphorylation (inducing structural changes by the addition of a phosphate group) is another major mechanism for GJ regulation. It acts at several levels affecting the trafficking of connexins from inside the cell to the plasma membrane, and also the clustering, localization, and recycling of GJs \cite{Moreno2007}. Individual GJ channels have a fast turnover rate, enabling the adjustment of the communication level between the two cells by modulation of the number of connexons produced by each cell and thus the surface area of the lattice \cite{Solan2018}.

Despite the crucial role of GJs in cell communication, communications engineering-based analysis of systems using GJs is still at an early stage. The authors of \cite{Nakano2008} considered the assembly of a synthetic communication network based on GJs. They demonstrated successful transmission via GJs expressed in genetically modified HeLa cells by propagating a calcium wave at about $5\,\mu$m/s. In \cite{Kilinc2013a}, an information-theoretic model was developed to derive a closed-form expression for the GJ channel capacity where the GJs create and propagate action potentials. The model was applied to correlate increases in the incidence of cardiac diseases with dysfunction in communication. GJs were also included in the channel modeling of calcium propagation in \cite{Bicen2016}.

\subsubsection{Plasmodesmata}
\begin{figure}[!t]
	\centering
	\includegraphics[width=3.0in]{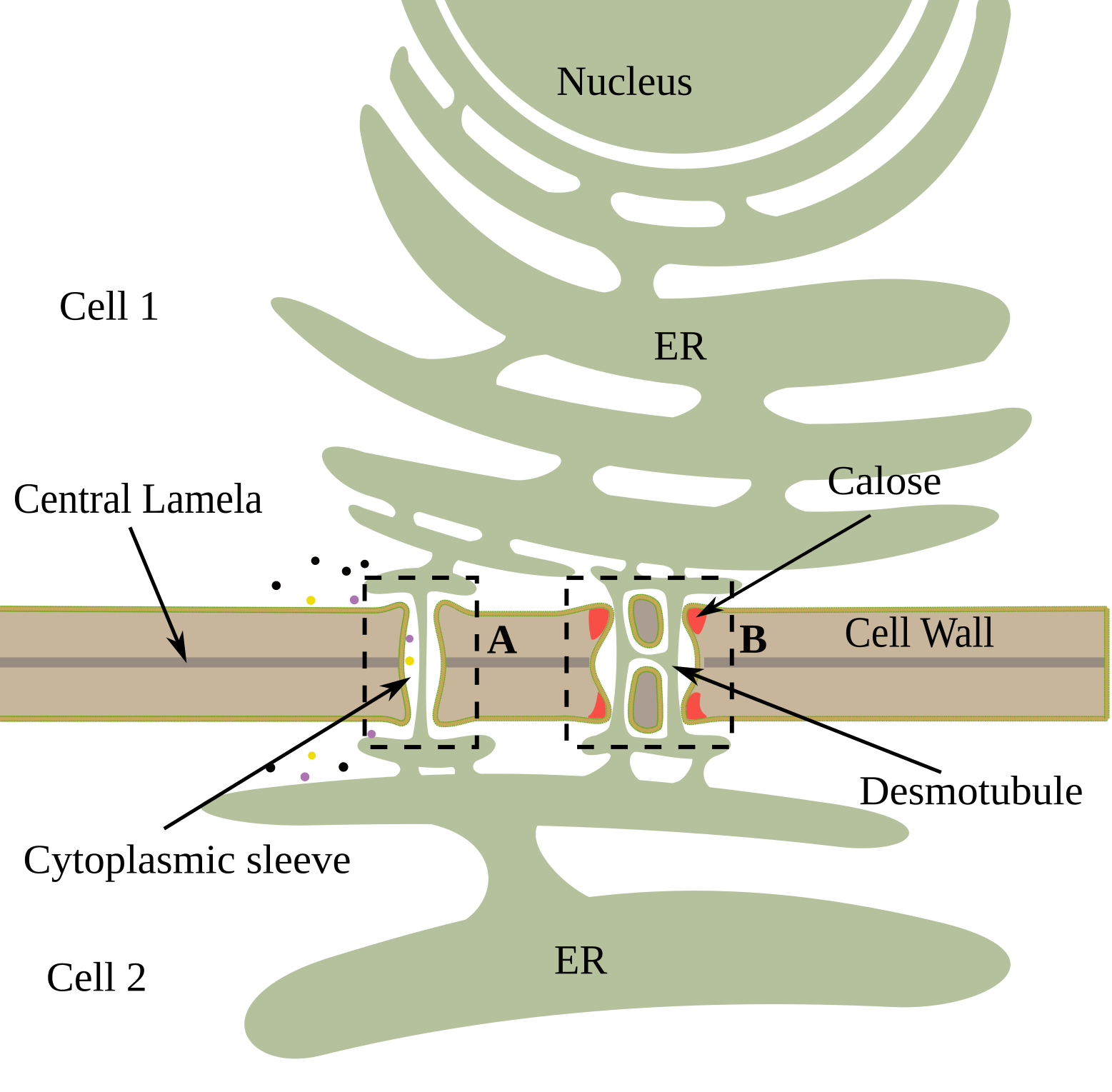}
	\caption{Plasmodesmata structure (not to scale). Both simple and complex forms are depicted in the diagram (dashed lines A and B, respectively). Adjacent cells have connected cytoplasms and ERs via desmotubules. Molecules can pass through the cytoplasmic sleeve. A buildup of calose (in red) closes the channel. The walls of each cell are separated by the central lamela.}
	\label{fig_plasmodesmata_structure}
\end{figure}

In plants, structures comparable to gap junctions are called \textit{plasmodesmata} (PDs, singular \textit{plasmodesma}). Plant cells are surrounded by a rigid cell wall that provides structural rigidity but at the same time constrains the passage of molecules and hence communication. To overcome this barrier, PDs serve as channels in the cell wall that connect adjacent cells; see Fig.~\ref{fig_plasmodesmata_structure}. By connecting virtually every cell within a plant, PDs create an avenue that permits the transfer of metabolites, nutrients, and signals to the remotest tissue \cite{Brunkard2015,Nicolas2017}.

Plasmodesmata are nanoscale structures, so they are not clearly visible using an optical microscope. However, with advances in electron microscopy it became possible to obtain detailed images of their structure \cite{Bell2011}. The channel is membrane-lined and connected tightly with the cell wall. This means that there is a continuum of plasma membrane between the different cells that enables the endoplasmic reticulum (ER) of each cell to connect with that of its neighbor. In the center and along the length of the channel there is a structure resembling a pole and called the \textit{desmotubule}. This is connected to the surrounding membrane by spikes. Molecules can travel through the gap between the desmotubule and the plasma membrane, i.e., the \textit{cytoplasmic sleeve}. Regulation of traffic through PDs is currently believed to happen by the adjustment of the width of the cytoplasmic sleeve, and in turn this is due to deposition or removal of the protein \textit{calose} around the mouth of the PD channel, thus restricting access as needed \cite{Amsbury2018}.

\subsubsection{Appendage-based communication}

We now briefly discuss several contact-based signaling mechanisms (namely conjugation, tunneling nanotubes, and telocytes) where cells connect with neighbors via appendages. Conjugation is a widespread mechanism of genetic material exchange between bacteria, and a key factor of microbial genomic plasticity \cite{Guglielmini2011,Guedon2017}, as it facilitates the transfer of DNA between cells in close range \cite{Cabezón2015}. Different types of mobile genetic elements (MGE) are responsible for initiating and establishing conjugation \cite{Frost2005}. Conjugative plasmids are the most widely known and studied MGE, as they are both easily identified and ubiquitous across bacterial species. Plasmids are also one of the most important factors of pathogenicity and of the development of antibiotic resistance in prokaryotes. MGE plasmids are small double-stranded DNA elements separate from the rest of the bacterial genome. They contain all the necessary sequences for coding their own replication and transfer to other cells \cite{Smillie2010}. Two distinct mechanisms are known to be used by plasmids for their transfer. One is particular to actinobacteria such as species of \textit{Streptomyces} and is mediated by a protein related to DNA re-positioning during cell division \cite{Thoma2015}. The second mechanism is more complex and involves a single-stranded DNA transfer apparatus that is widespread among diverse bacterial species \cite{Smillie2010}.

\begin{figure}[!t]
	\centering
	\subfloat[][]{\includegraphics[width=0.47\linewidth]{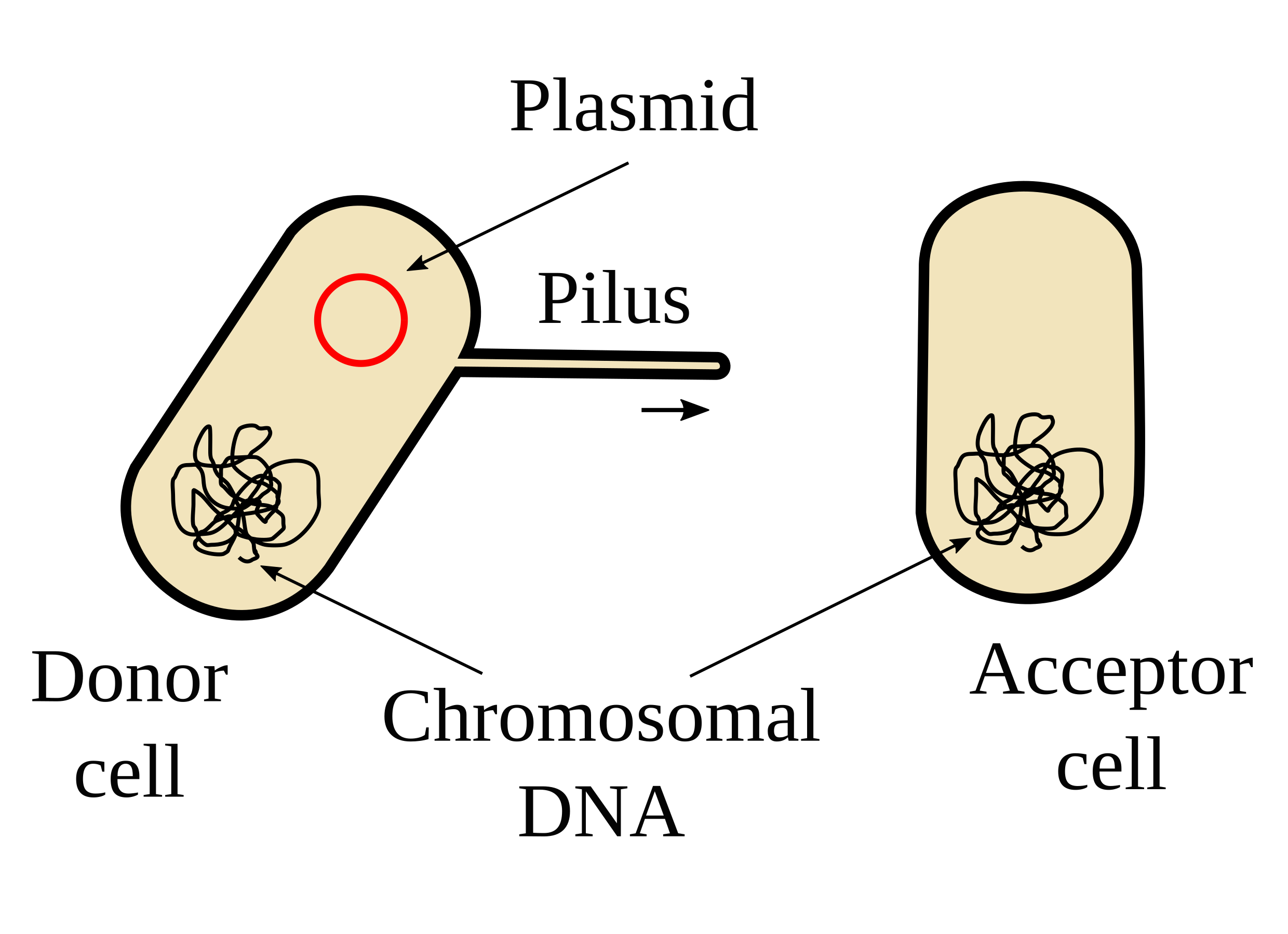}}
	\subfloat[][]{\includegraphics[width=0.43\linewidth]{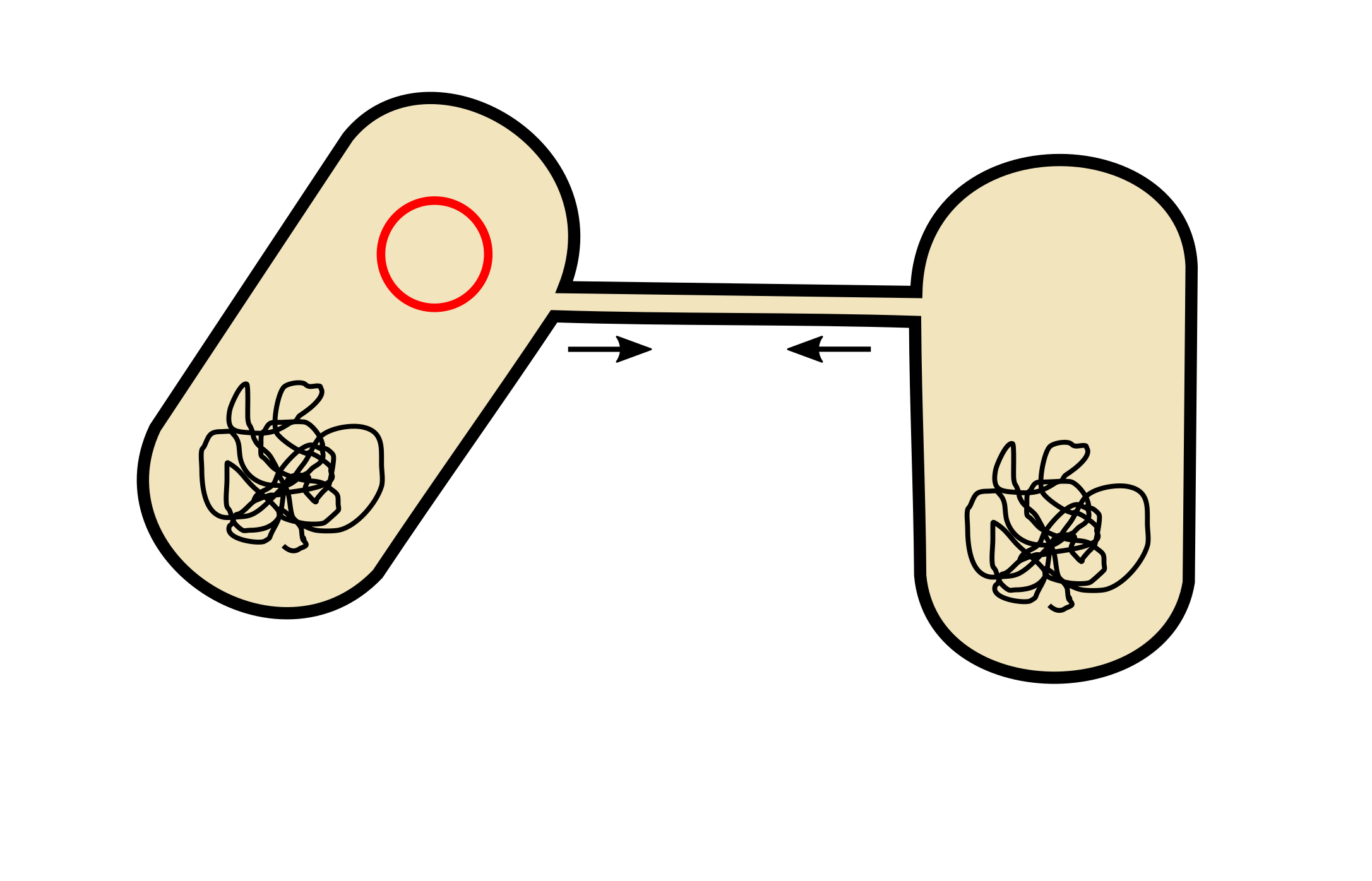}}\\
	\subfloat[][]{\includegraphics[width=0.45\linewidth]{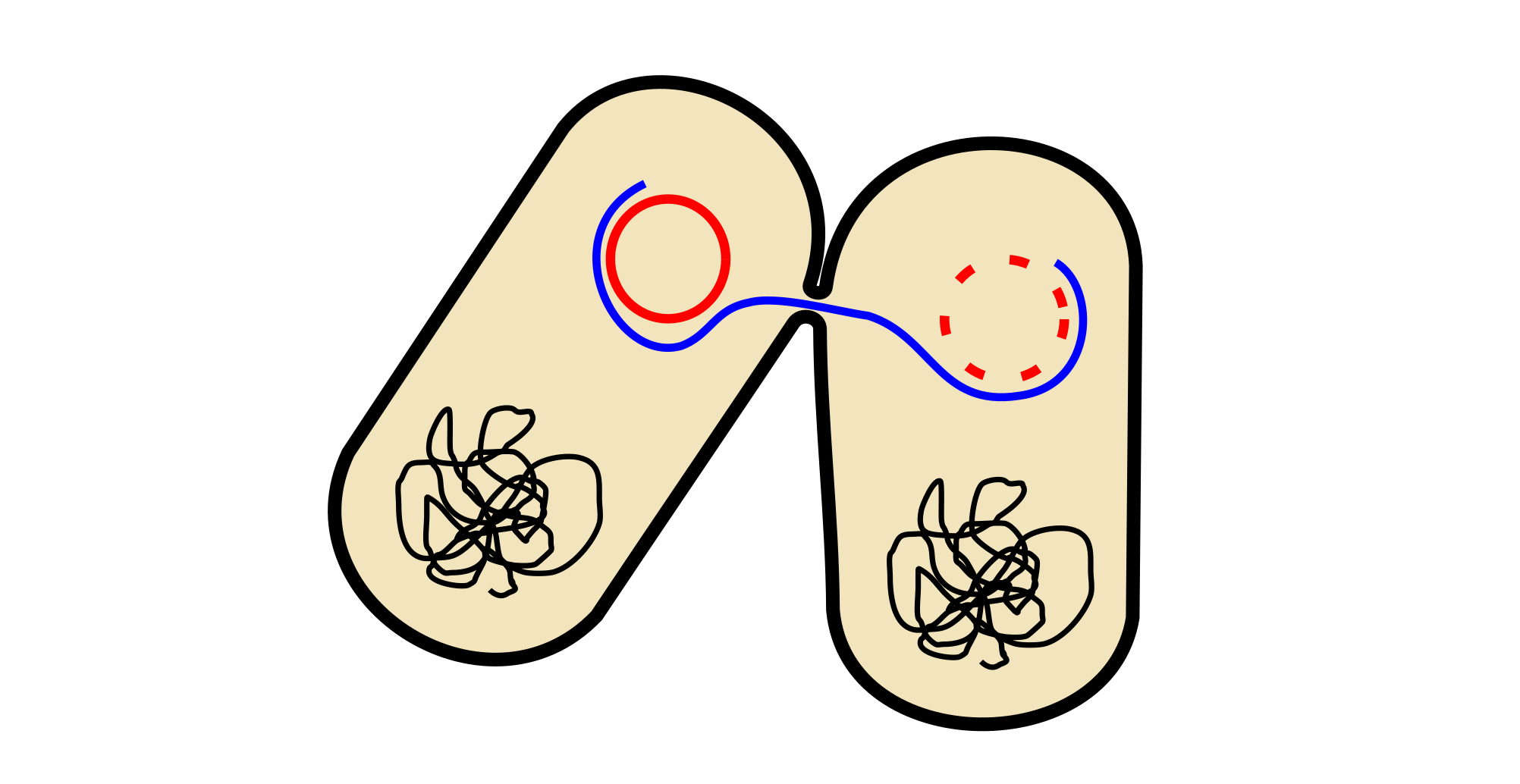}}
	\subfloat[][]{\includegraphics[width=0.45\linewidth]{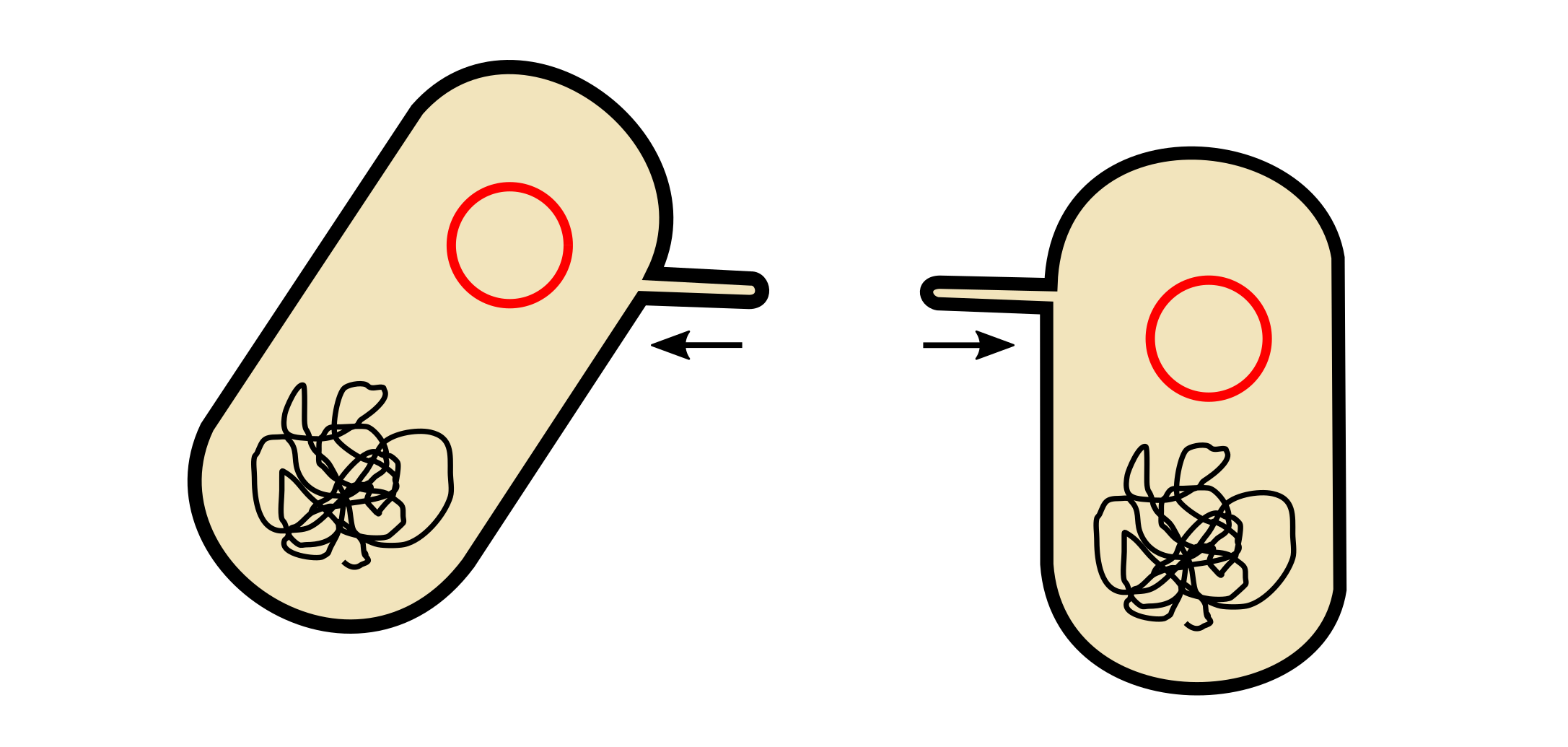}}
	\caption{Steps of bacterial conjugation. (a) The donor cell, carrying the conjugative plasmid (red circle), extends a pilus that finds its way to the acceptor (i.e., recipient) cell through chemotaxis. (b) After the connection, the pilus contracts to bring the two cells in close contact. (c) As the plasmid DNA is being replicated (blue) in the donor cell, it is being simultaneously passed through the open channel to the acceptor as a single strand, where it is also immediately converted back into double-stranded cyclic DNA (dotted red circle). (d) At the end of the process, the cells separate by extending the pilus which then breaks up and is recycled.}
	\label{fig_conjugation}
\end{figure}

Conjugation is typically initiated by the donor cell carrying the plasmid to be copied. A filamentous hollow structure known as a \textit{pilus} extends towards the acceptor cell; see Fig.~\ref{fig_conjugation}. Upon contact the pilus fuses into the recipient cell's membrane and is destroyed at both ends simultaneously to bring the two cells into close proximity. The conjugative plasmid is then separated into single strands of DNA, one of which is transported through the open channel into the acceptor cell. At the same time, in both donor and acceptor cells, the single strands are converted back into double-stranded DNA. With the conclusion of DNA transport the pilus extends from both ends to separate the two cells and is then recycled after the final separation \cite{Cabezón2015, Waksman2019}.

Recent evidence suggests that other types of MGE, such as the integrative and conjugative elements (ICE) also known as conjugative transposons, might in fact be more abundant and more important for bacterial communication than plasmids \cite{Guglielmini2011,Bellanger2014,Ambroset2016}. Similar to plasmids, ICE encode all the necessary machinery for their excision from host DNA, transfer to, and integration into the host DNA. Unlike plasmids, however, ICE are incorporated into the host DNA (or an existing plasmid) \cite{Bellanger2014}.

There have been several works within the MC engineering community that proposed conjugation as a propagation mechanism for synthetic networks, including \cite{Lio2012,Castorina2016,Unluturk2017}. These contributions have tended to focus on behavior that occurs higher in our proposed hierarchy, i.e., quantifying the transmission of information using bacterial conjugation, and not characterizing signal propagation. 

\begin{figure}[!t]
	\centering
	\subfloat[][]{\includegraphics[width=0.85\linewidth]{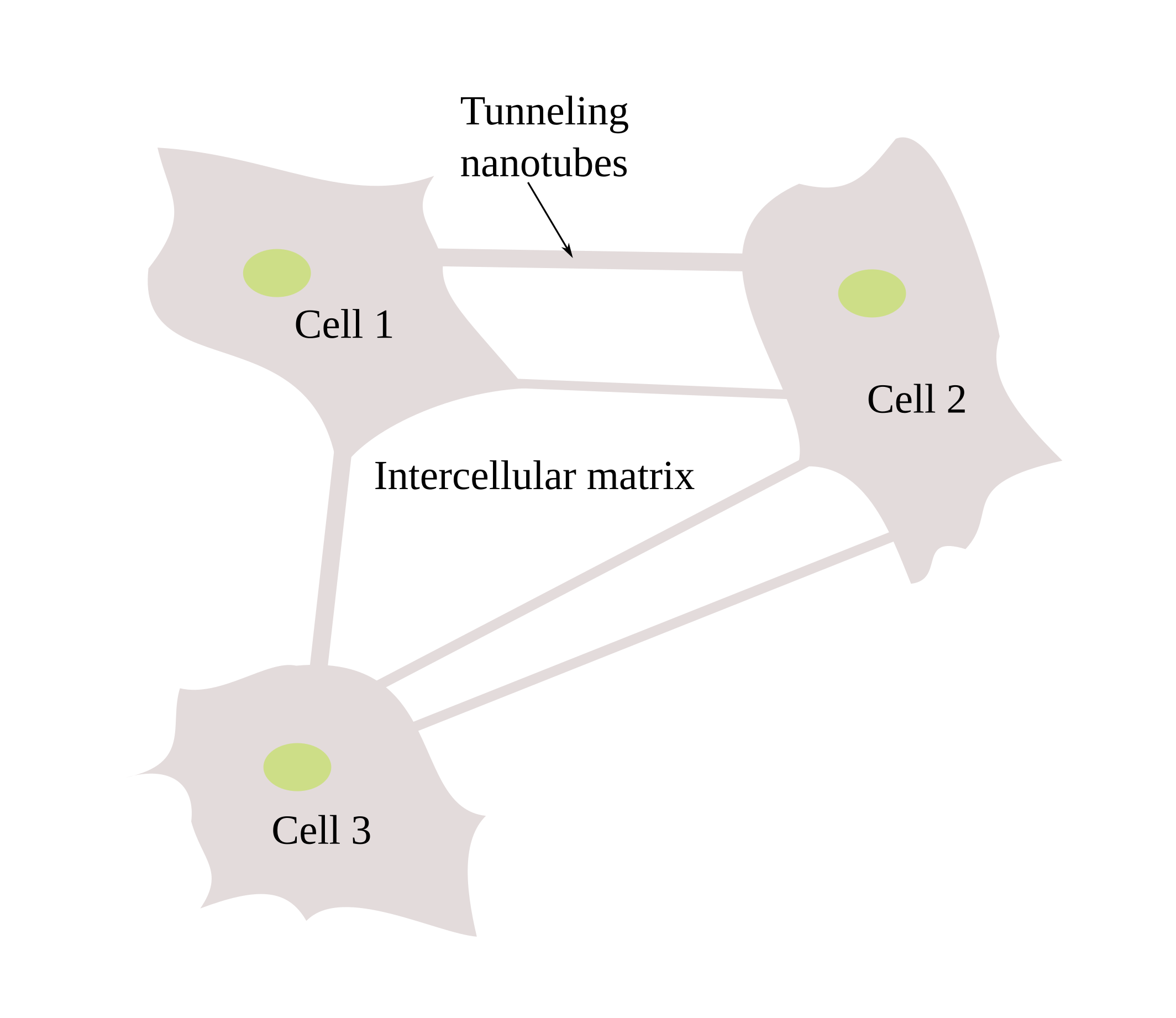}}\\
	\subfloat[][]{\includegraphics[width=1\linewidth]{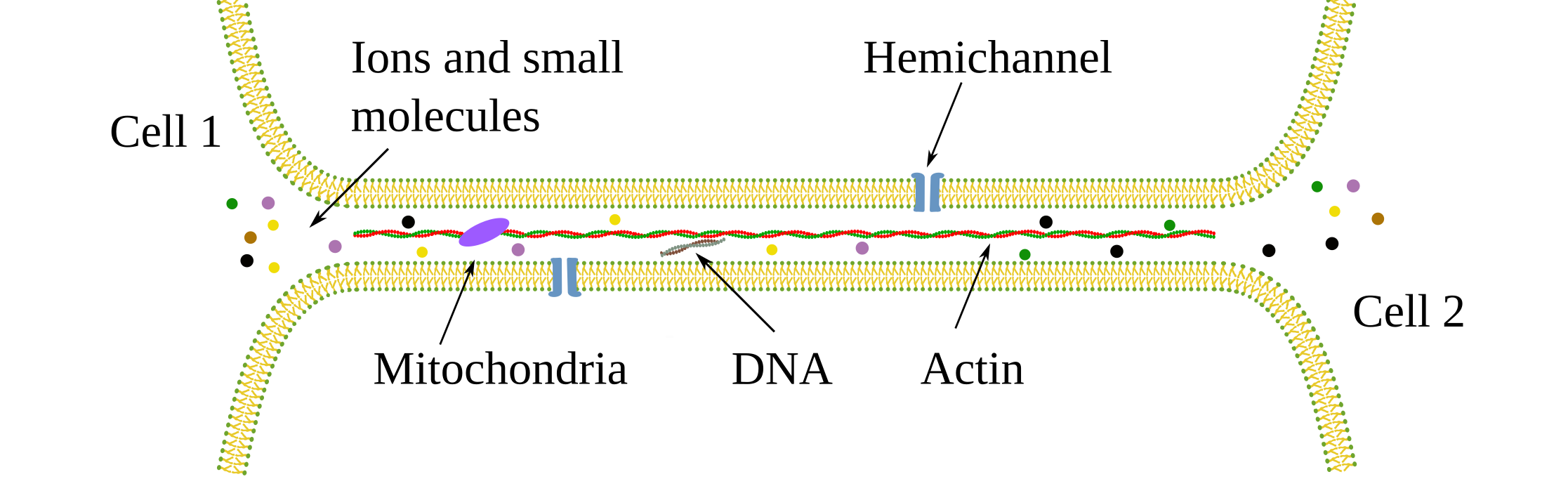}}
	\caption{Tunneling nanotube (TNT) structure (not to scale). (a) TNTs connect adjacent cells at relatively long distances. (b) Ions and small molecules are able to freely diffuse through the channel, while larger molecules and organelles are actively transported via the actin filament.}
	\label{fig_tunneling_nano}
\end{figure}

Tunneling nanotubes (TNTs) are long channels formed between cells that can be several micrometers apart. They are temporary structures that can dynamically form in a few minutes and directly connect the cytoplasms of the cells involved \cite{Rustom}; see Fig.~\ref{fig_tunneling_nano}. While ions and small molecules can diffuse freely along a TNT, TNTs also facilitate the active transport of a diverse range of molecules, organelles, and micro-vesicles, e.g., mitochondria and membrane components \cite{Sherer2008,Gurke2008}.

\begin{figure}[!t]
	\centering
	\includegraphics[width=0.95\linewidth]{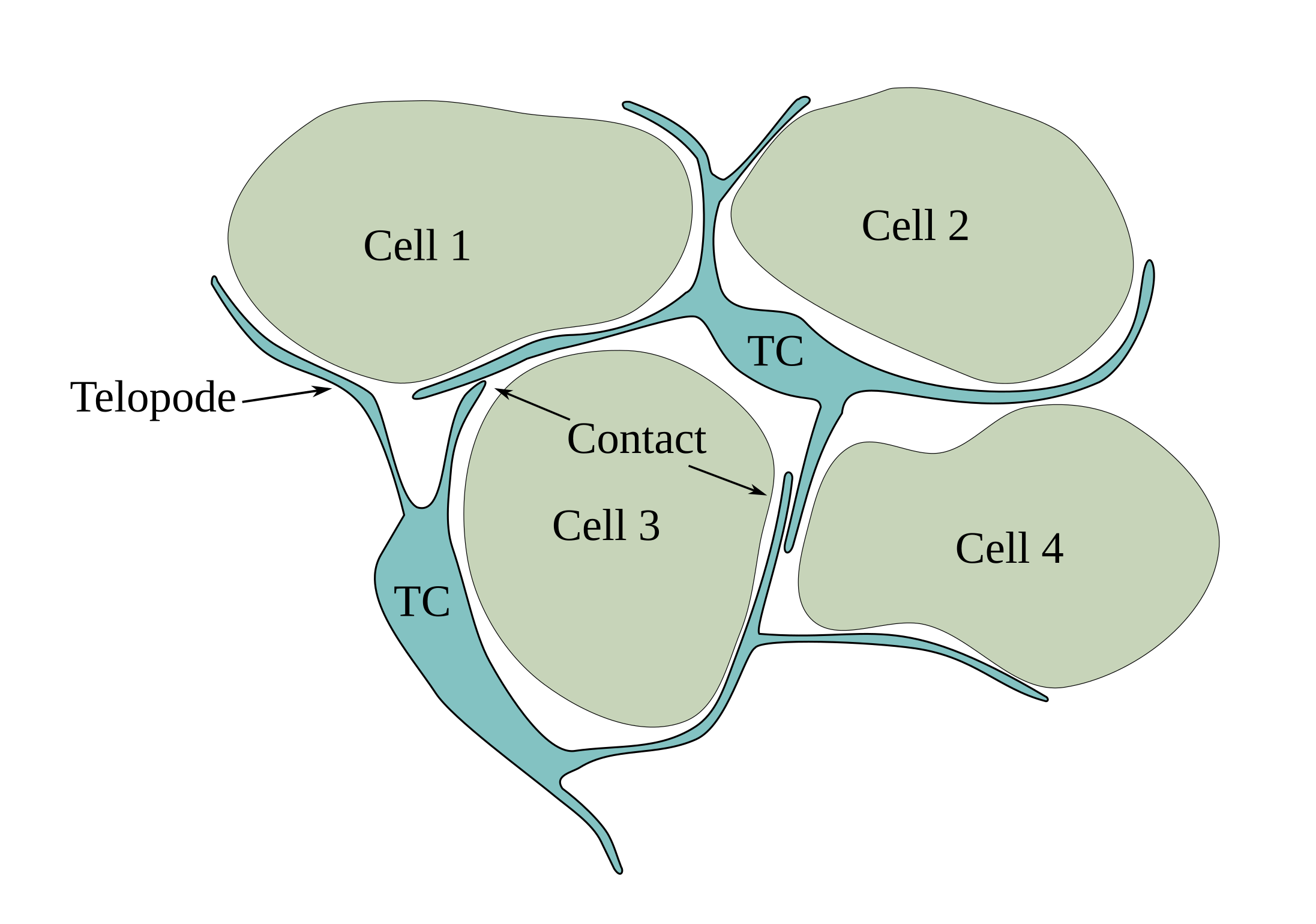}
	\caption{Telocytes (TC) connect various types of cells via long appendages (telopodia). Telopodia can connect with each other, enabling long distance communication between cells.}
	\label{fig_telocytes}
\end{figure}

Relatively recently, a specialized type of cell called a \textit{telocyte} was discovered \cite{Cretoiu2014}. Telocytes form a network on the extracellular matrix of all body tissues. This network is comprised of very long thin channels (\textit{telopodes}) between telocyte cells; see Fig.~\ref{fig_telocytes}. They form connections with the other cells in the surrounding environment and permit intercellular communication by diffusive, contact, electrical, and mechanical signaling. Their confirmed and speculated roles span a diverse array of processes in animal physiology, including cell signaling, extracellular vesicle release, mechanical support to surrounding tissues, muscle activity, guidance for migrating cells, tissue homeostasis, and even the transmission of neuronal signals in cooperation with other specialized cells \cite{Cretoiu2014, Cretoiu2017, Edelstein2016, Faussone-Pellegrini2016}. Thus, telocytes could play an essential role in several signal propagation mechanisms and are hence a very interesting target for future MC research.

\section{Level 2 - Physical and Chemical Signal Interaction}
\label{sec_interaction}

Communicating devices that use MC channels require interfaces to interact with the channels. A transmitter needs a mechanism to insert molecules into the channel, and a receiver needs a mechanism to observe (i.e., sample) the molecules that are in the channel. It is common in the MC literature to assume that these processes are perfect by making ideal assumptions about the generation and sampling of molecules; see \cite{Jamali2019b}. These typically include the instantaneous creation of a desired number of molecules at a fixed point, and then a perfect counting of the number of molecules that arrive at the receiver (whether or not they are removed from the physical propagation channel). These assumptions shift the focus of the analysis to the characterization of the physical signal propagation (as discussed for Level 1 in Section~\ref{sec_physical}), and can be accurate if the physical and temporal scales of molecule release and molecule sampling are sufficiently small relative to the physical channel. However, if these constraints are not satisfied, then the interface to the physical signal has to be an integral component of the end-to-end channel characterization.

The biochemical machinery for generating and receiving cellular signals can be rather complex and serve important roles in cellular function. For example, ions are commonly used for signaling and also directly regulate behavior via biochemical signaling pathways. Ca$^{2+}$ ion signals control, among others, muscle contraction, cell division, exocytosis, fertilization, metabolism, neuronal synaptic transmission, cell movement, and cell death \cite{Giorgi2018}. The propagation of ions, usually via diffusion, is only part of the role that they play. Ions actively interact with the mechanisms that release and receive them (e.g., ion-activated gap junctions and cell surface molecular pumps), so they are an essential part of the signal production and reception apparatuses.

In this section, we review mechanisms for generating and sampling the physical signal, including intermediate biochemical and biophysical processing and pathways where the physical signal is an input or output. These mechanisms correspond to Level 2 of our proposed communication hierarchy (see Figs.~\ref{fig_hierarchy} and \ref{fig_hierarchy_overlay}). We discuss the physical storage and release of molecules, in particular via vesicles (Section~\ref{sec_generation}). We present the common methods for signaling molecules to be detected and the diversity of biochemical responses (Section \ref{sec_reception}). We mathematically link Level 1 with Level 2 by discussing commonly-considered initial and boundary conditions for diffusion-based propagation, which describe how molecules are added to and how (if any) molecules are removed from the environment (Section \ref{sec:ICs_BCs}). The boundary conditions are needed to derive the channel response and thus directly constrain analytical solutions if they exist. We finish this section by introducing the biochemical influence of molecular signals on transcription networks, which control protein production, and discuss how we can study transcription networks by separating them into isolated network motifs (Section \ref{sec_transcription_network}).

Much of the functional communication complexity for computation and control in cell biology pertains to Level 2. The behavior in natural organisms that we associate with the higher levels of the hierarchy is generally achieved via mathematical abstraction. So, although our focus in this section is on natural mechanisms, we briefly note that the experimental (i.e., macroscale) addition and observation or capture of molecules is also at Level 2, but we elaborate further on experimental methods in the context of Level 3 in Section~\ref{sec_interface}. In synthetic devices, we are more likely to have direct (e.g., digital) implementations of higher-level behavior and require less complexity at the channel interface, even for nanoscale designs of electrical-transistor-based biosensors \cite{Kuscu2016a}. Thus, even though Level 2 is the lowest level within an individual device, we can already observe distinctions between natural and artificial behavior.

It is helpful in this section for the reader to have some understanding of cellular composition and the importance of lipid bilayers. Lipid bilayers are thin yet stable polar membranes that are hydrophilic on the outside (i.e., water soluble) and hydrophobic on the inside (i.e., they repel water) \cite[Ch.~10]{Alberts2015}. Lipid bilayers are the key basic component of biological membranes and they help to compartmentalize cells and maintain molecule gradients because many molecules cannot pass through them, particularly if they are charged or strongly polar. The outermost boundary of a cell, i.e., the plasma membrane, is comprised of a lipid bilayer and many other types of molecules whose functions can include maintaining the membrane's structure or to facilitate the transport of specific molecules across the membrane (e.g., through gap junctions or plasmodesmata as shown in Figs.~\ref{fig_gap_structure} and \ref{fig_plasmodesmata_structure}, respectively). Thus, cellular mechanisms for generating or sampling molecular signals need to account for the plasma membrane. For example, a typical cytosolic  Ca$^{2+}$ concentration is $0.1\,\mu$M, while in the extracellular fluid it is more than 10,000 times higher at about 1.2\,mM. This creates a very powerful ion gradient that results in a rapid influx of Ca$^{2+}$ towards the interior of the cell when there is a chance to do so. This difference is tightly controlled using pumps that actively transfer Ca$^{2+}$ ions out of the cytosol and Ca$^{2+}$ channels that are normally closed and impermeable to the ions \cite[Ch.~15]{Alberts2015}.

\subsection{Molecule Generation and Release Management}
\label{sec_generation}
The transmitter in an MC system needs to be able to generate and release a molecular signal. These molecules may be harvested from within the transmitter or its surrounding environment, or synthesized from its constituent components. If the molecules do not need to be released as soon as they are ready, then the transmitter also needs a mechanism for storing the molecules until they are needed. For example, Ca$^{2+}$ ions stored in the ER are released via Ca$^{2+}$ gates to restore the cytosolic ion concentration when it is depleted \cite{Putney2005}.

A common technique for storing molecules within eukaryotic cells, either for transportation or until the stored molecules are needed, is within vesicles. Vesicles are usually spherical or near-spherical shapes that are composed of a lipid bilayer. Thus, they can securely hold many types of molecules, e.g., cholesterol, proteins, neurotransmitters, or even invading bacteria. Vesicles can vary in size from about $50\,$nm (synaptic vesicles) to several microns in diameter \cite{Phillips2013}, and even smaller vesicles can contain many thousands of molecules. To empty their contents, vesicles merge with another bilayer (such as a cell's plasma membrane) and release their molecules onto the other side of the other bilayer (e.g., outside the cell as shown in Fig.~\ref{fig_hierarchy_overlay}) via exocytosis. Thus, molecules can be directly released from an intracellular vesicle into the extracellular space, which can occur very quickly; synaptic vesicles released by neurons can empty their contents within about a millisecond or less \cite{Martin2003}.

While many transport vesicles are produced at a cell's Golgi apparatus, processes that rely on rapid and precise vesicle release can fabricate them locally \cite[Ch.~13]{Alberts2015}. For example, synaptic vesicles are produced locally from budding at the plasma membrane to help ensure a steady supply. No matter where they are produced, vesicles are generally too large to efficiently move by diffusion alone. So, they are carried along cytoskeletal fibers by motor proteins (as introduced for cargo-based transport for Level 1 in Section~\ref{sec_cargo}). Proteins that ``coat'' the outside of a vesicle are used to identify its intended destination so that it can bind to a suitable molecular motor. For example, a vesicle could be intended for an endosome instead of the plasma membrane. Additional surface proteins are used to control both vesicle docking and fusion once it has reached its target.

A key advantage for using vesicles is the precise regulation that is provided for molecule release, since particular proteins need to be available and in the correct state for a vesicle to be transported, docked, and fused with the destination membrane. However, vesicles in the constitutive exocytosis pathway are used for immediate uncontrolled release of their contents when fusing with the plasma membrane \cite[Ch.~13]{Alberts2015}. These provide materials to grow a plasma membrane, but can also carry proteins for secretion to outside the cell. In this pathway, proteins can be secreted as fast as they are produced; the only delay is in transport. In other cases, released molecules can bypass vesicle pathways entirely if they are able to directly pass through the plasma membrane \cite[Ch.~11]{Alberts2015}. This is true for small uncharged or weakly polar molecules, e.g., nitric oxide, or molecules that have dedicated transmembrane channels, e.g., the common ions sodium, potassium, and calcium.

As noted, MC models typically treat molecule generation and release as instantaneous processes, or at least as steps that take negligible time relative to molecule propagation across the channel of interest \cite{Jamali2019b, Kuscu2019a}. Exceptions include \cite{Chou2015,Arjmandi2016}, which have modeled transmitter molecule release with chemical reaction kinetics. The authors of \cite{Ramezani2018} modeled the impact of vesicle preparation and release on the information capacity in a chemical neuronal synapse.

\subsection{Molecule Reception and Responses}
\label{sec_reception}

The receiver in an MC system needs to be able to detect and respond to a molecular signal. Depending on the type of received molecule and the receiver's sensitivity, some threshold signal quantity may need to be observed in order to stimulate a corresponding response.

\subsubsection{Molecule Reception}
\begin{table*}[!t]
	\renewcommand{\arraystretch}{1}
	\caption{Reception Mechanism Summary.} 
	\label{table_reception}
	\centering	
	{\renewcommand{\arraystretch}{1.2}
		\scalebox{1}
	{\begin{tabular}{l|l|l}
		\hline
		\multicolumn{2}{c|}{Reception Type}           & \multicolumn{1}{c}{\multirow{2}{*}{Example}} \\ \cline{1-2}
		\multicolumn{1}{c|}{Reception Site}           & \multicolumn{1}{c|}{Receptor Protein} & \multicolumn{1}{c}{}                         \\ \hline \hline
		\multirow{2}{*}{\begin{tabular}[c]{@{}l@{}}Intracellular reception: Molecules can \\cross cell membrane\end{tabular}} & Intracellular enzyme                  & Dissolved gases                               \\ \cline{2-3} 
		& Intracellular receptor                & Cortisol, estradiol, and
testosterone                              \\ \hline
		\multirow{3}{*}{\begin{tabular}[c]{@{}l@{}}Surface reception: Molecules cannot \\ cross cell membrane\end{tabular}}    & Ion-channel-coupled receptor          & Acetylcholine, glycine, $\gamma$-aminobutyric acid, ions                                \\ \cline{2-3} 
		& G-protein-coupled receptor            & Neurotransmitters, local mediators, hormones \\ \cline{2-3} 
		& Enzyme-coupled receptor               & Insulin, nerve growth factor                  \\ \hline
	\end{tabular}}}
\end{table*}

In cells, extracellular signal molecules generally fall into one of two families: 1) molecules that are small or hydrophobic enough to easily cross the receiver cell membrane, and 2) molecules that are too large or too hydrophilic to cross the receiver cell membrane, as summarized in Table \ref{table_reception}. The first family of molecules can directly pass the cell membrane to activate intracellular enzymes or bind to intracellular receptor proteins, while the second family of molecules relies on receptors at the surface of the target cell to relay their messages across the cell membrane \cite[Ch.~11]{Alberts2015}. In the MC literature, these two reception paradigms are usually referred to as passive and active reception, respectively \cite{Nakano2013c}.

Dissolved gases and steroid hormones are representatives of the first family \cite[Ch.~15]{Alberts2015}. Most dissolved gases can cross the plasma membrane and enter the cell interior to directly activate intracellular enzymes. For example, smooth muscle relaxation in a blood vessel wall can be  triggered by Nitric Oxide (NO). Unlike molecules that directly activate {intracellular enzymes}, the detection of steroid hormones (such as cortisol, estradiol, and thyroxine) relies on intracellular receptors. All of these molecules cross the plasma membrane of the target cell and bind to their protein receptors distributed either in the \textit{cytosol} (i.e., the liquid inside the cell) or the nucleus to regulate gene expression.

\begin{figure}[!t]
	\centering
	\subfloat[][]{\includegraphics[width=0.33\linewidth]{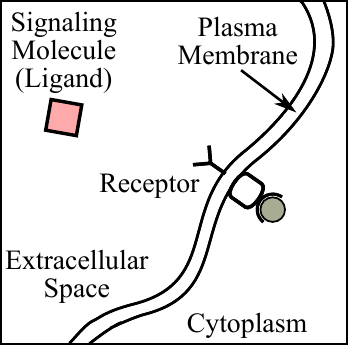}}
	\hfill
	\subfloat[][]{\includegraphics[width=0.33\linewidth]{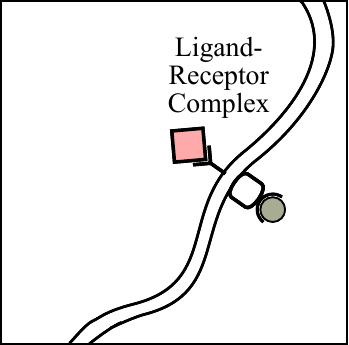}}
	\hfill
	\subfloat[][]{\includegraphics[width=0.33\linewidth]{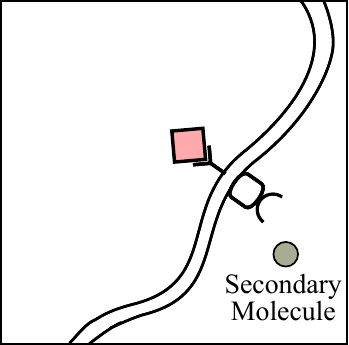}}
	\caption{Steps of a generic molecule reception process. (a) There is a receptor embedded in the plasma membrane that separates a cell's cytoplasm from the extracellular space. The receptor can bind to a ligand, which in this case is the signaling molecule of interest. (b) The ligand binds to the receptor to form a ligand-receptor complex. This instigates a conformational change in the receptor. (c) The conformational change leads to a response, e.g., the release of an internal secondary signaling molecule as shown.}
	\label{fig_receptor}
\end{figure}

The vast majority of extracellular signal molecules belong to the second family. They are either too large or hydrophilic to cross the plasma membrane, so their detection requires the use of surface receptor proteins; see Fig.~\ref{fig_receptor}. According to their biochemical signaling pathways, the surface-binding receptors can be further classified into three classes: ion-channel-coupled receptors, G-protein-coupled receptors, and enzyme-coupled receptors \cite[Ch.~15]{Alberts2015}. 
\begin{itemize}
    \item Ion-channel-coupled receptors are prevalent in the nervous system and other electrically excitable cells. This kind of receptor binds with ion molecules and can transduce changes in ion concentrations into changes in membrane potential.
    \item  G-protein-coupled receptors associate with a G protein in the cytosolic domain.  Once extracellular signal molecules are bound to G-protein-coupled receptors, these receptors are able to activate membrane-bound, GTP-binding proteins (G proteins), which then turn on or off an enzyme or ion channel on the same membrane and finally alter a cell's behavior \cite{Birnbaumer1990,Gilman1987}. Examples of this type of reception include the transduction of a heartbeat slowdown signal for heart muscle cells, a glycogen breakdown signal for liver, and a contraction signal for smooth muscle cells.  A recent review of G-proteins can be found in \cite{Syrovatkina2016}.
    \item  The cytoplasmic domain of enzyme-coupled receptors either acts on an enzyme itself or associates with another protein to form an enzyme once signaling molecules bind to the outer surface of the plasma membrane. Enzyme-coupled receptors play a significant role in the response to the growth factor molecules that regulate cell growth, proliferation, differentiation, and survival. 
\end{itemize}

\subsubsection{Reception Responses}
There is a broad diversity in how biochemical receptors respond to molecular signals, and even receptors sensitive to the same kind of signaling molecule can behave differently in different cells \cite[Ch.~15]{Alberts2015}. For example, responses to acetylcholine include decreasing the firing of action potentials, stimulating muscle contraction, and stimulating saliva production. Another example is calcium signaling. The same stimulus can trigger a Ca$^{2+}$ wave across one cell, local calcium oscillations in another cell, or cause only a localized increase in the concentration in yet another cell \cite{Clapham2007}. The different responses are due to the ability of Ca$^{2+}$ to bind to a large variety of different proteins. Thus, the same signal activates different signaling pathways depending on the cell type and the available proteins.

The diversity in biochemical responses means that a given type of receptor (or a collection of coupled receptors along a pathway) has several distinguishing properties \cite[Ch.~15]{Alberts2015}. The \textit{timing} of responses can vary by many orders of magnitude, from milliseconds for muscle control and other synaptic responses \cite{Lindner2016,Srivastava2017}, to seconds for bacteria using chemotaxis to respond to chemical gradient changes \cite{berg1993random}, to hours or even days for changes in the behavior or fate of a cell (e.g., gene regulation, differentiation, or cell death). Correspondingly, the \textit{persistence} of a response could be very brief (as is usually needed in synapses) or even permanent. \textit{Sensitivity} to a signal can be controlled by the number of receptors present or by the strength of a secondary signal created by an activated receptor. Similarly, a biochemical system's \textit{dynamic range} specifies its responsiveness over a range of molecular signal strengths. More complex responses can be achieved using biochemical \textit{signal processing}, e.g., applying feedback to implement switches and oscillators. Some responses are controlled by the \textit{integration} of multiple molecular signals, which we can study with a mathematical understanding of local data abstraction (i.e., Level 4 in Section~\ref{sec_data}). Conversely, a single molecular signal can \textit{coordinate multiple responses} simultaneously within the same cell, e.g., to stimulate both growth and cell division.


	%
\begin{table*}[!t]
	\caption{Comparison of Diffusion-based Propagation Mechanisms.}
	\label{table_BCs}
	\centering
	{\renewcommand{\arraystretch}{1.45}
	\scalebox{1}
{\begin{tabular}{l|l|l|l|l|l|l||l}
	\hline
	\multicolumn{2}{c|}{Release Strategy}                  & \multicolumn{3}{c|}{Propagation Environment}                                                            & \multicolumn{2}{c||}{Reception Mechanism}                              & \multicolumn{1}{c}{\multirow{2}{*}{Channel Response}} \\ \cline{1-7}
	TX Type                         & IC                    & Boundary                     & Equation                                & BC                   & RX Type                           & BC                                & \multicolumn{1}{c}{}                                  \\ \hline \hline
	\multirow{12}{*}{Point} & \multirow{12}{*}{IC$_1$} & \multirow{3}{*}{Unbounded}   & \multirow{6}{*}{Eq. (3)}            & \multirow{3}{*}{BC$_2$} &N/A            &N/A          	& \cite[Eq. (2.8)]{berg1993random}                                                   \\ \cline{6-8} 
	&                       &                              &                                                   &                      & Spherical fully absorbing         & BC$_4$                               & \cite[Eq. (22)]{yilmaz2014}                                                    \\ \cline{6-8} 
	&                       &                              &                                                   &                      & Reversible absorbing              & BC$_3$                               & \cite[Eq. (8)]{Deng2016}                                                    \\ \cline{3-3} \cline{5-8} 
	&                       & Spherical bounded            &                                                   & \multirow{2}{*}{BC$_1$} & Spherical fully absorbing         & BC$_4$                               & \cite[Eq. (13)]{dinc2019tmbmc}                                                    \\ \cline{3-3} \cline{6-8} 
	&                       & Rectangular bounded          &                                                   &                      & Fully absorbing walls             & BC$_4$                               & \cite[Eq. (19)]{ankit2020}                                                    \\ \cline{3-3} \cline{5-8} 
	&                       & Rectangular/Circular bounded &                                                   & BC$_1$, BC$_2$             & \multirow{3}{*}{N/A} & \multirow{3}{*}{N/A} & \cite[Eq. (14.4.4), (14.13.7)]{jaeger1959}                                                    \\ \cline{3-5} \cline{8-8} 
	&                       & Unbounded                    & \multirow{2}{*}{Eq. (6)} & \multirow{2}{*}{BC$_1$} &                                   &                                   & \cite[Eq. (18)]{adam2014icc}                                                    \\ \cline{3-3} \cline{8-8} 
	&                       & Cylindrical bounded          &                                                   &                      &                                   &                                   & \cite[Eq. (11)]{Wicke2018}                                                     \\ \cline{3-8} 
	&                       & \multirow{4}{*}{Unbounded}   & \multirow{3}{*}{Eq. (12)}  & \multirow{4}{*}{BC$_2$} & Reversible absorbing              & BC$_3$                               & \cite[Eq. (23)]{arman2016tnb}                                                    \\ \cline{6-8} 
	&                       &                              &                                                   &                      & Passive receiver                  &                   & \cite[Eq. (9)]{Noel2014}                                                    \\ \cline{6-8} 
	&                       &                              &                                                   &                      & Partially absorbing               & BC$_3$                               &  \cite[Eq. (16), (17), (29), (30)]{lanting2019}                                                    \\ \cline{4-4} \cline{6-8} 
	&                       &                              & Eq. (13)        &                      & {N/A} & {N/A} & \cite[Eq. (8)]{adam2014jasc}                                                    \\ \cline{1-8} 
	
	Volume & IC$_2$ & Unbounded & Eq. (3) & BC$_1$ &Passive \& active receiver &BC$_4$ & \cite[Eq. (12)]{Noel2016}\\ \hline
\end{tabular}}}
\end{table*}

\subsection{Mathematical Modeling of Emission, Propagation, and Reception}
\label{sec:ICs_BCs}
The release and reception processes can be mathematically modeled by defining initial conditions (ICs) and boundary conditions (BCs) for the propagation equations, such as those discussed for diffusion in Level 1 (Section~\ref{sec_physical}). In this way, the spatial-temporal concentration distribution can be obtained by solving the partial differential equations (PDEs) that describe propapgation channels with ICs and BCs. In other words, the release strategy, propagation channel, and reception mechanism jointly determine the channel response and the observed signal. The recent survey \cite{Jamali2019b} summarized channel impulse responses (CIR) under different models for the transmitter, physical channel, and receiver, where the CIR was formally defined as the \textit{probability of observation} of one output molecule at the receiver when one molecule is impulsively released at a transmitter. It is noted that although the CIR definition implies impulsive release of signaling molecules, the transmitter geometry and molecular generation method still affect the CIR. Unlike \cite{Jamali2019b}, here we focus on the mathematical formulation of specific (mostly ideal) conditions so that the ICs and BCs can be mapped to the discussions in Sections~\ref{sec_generation} and \ref{sec_reception}. With these conditions, we also provide a brief summary of some known channel responses in Table \ref{table_BCs}.

\subsubsection{ICs on Release Strategies}
As stated earlier, the simplest scenario is that $N$ molecules are released from a point in an impulsive manner at time $t_0$, so the IC can be expressed as
\begin{align}
  \text{IC}_1:  C(\textbf{d},t_0)=N\delta(\textbf{d}-{\textbf{d}_\text{TX}}), \label{IC1}
\end{align}
where $\delta(\cdot)$ is the Kronecker delta function and $\textbf{d}_\text{TX}$ is the location of the release point.

Although the point transmitter has been widely used in MC research, it is quite idealized. Another idealized transmitter is the volume transmitter, which occupies physical space and its surface does not impede molecular movement. Signaling molecules are released from a releasing space $\Tilde{\mathcal{V}}_\text{TX}$ or a releasing surface $\Tilde{\mathcal{S}}_\text{TX}$ of the volume transmitter. Therefore, a volume transmitter can be regarded as a superposition of many point transmitters that are located at different positions, and the corresponding IC can be expressed by extending \eqref{IC1} as follows:
\begin{align}
    \text{IC}_2:
    \int_{\textbf{d}_{\text{TX}} \in {\Tilde{\mathcal{V}}_\text{TX}}} N\delta(\textbf{d}-{\textbf{d}_{\text{TX}}})dV~ \text{or}~  \int_{\textbf{d}_{\text{TX}} \in {\Tilde{\mathcal{S}}_\text{TX}}} N\delta(\textbf{d}-{\textbf{d}_{\text{TX}}})dS, \label{IC3}
\end{align}
where $\textbf{d}_{\text{TX}}$ is a location within the releasing volume $\Tilde{\mathcal{V}}_\text{TX}$ or on the releasing surface $\Tilde{\mathcal{S}}_\text{TX}$. We note that \eqref{IC3} can also describe the molecule release from an ion-channel-based transmitter if it has many open ion channels \cite{Jamali2019b}.

\subsubsection{BCs on Propagation Channels}
An unbounded environment is a common assumption to simplify the derivation of the channel response. However, in practice, the molecular propagation medium is often much more complex. Molecular propagation can be constrained by various boundaries, such as the tunnel-like structure of a blood vessel, oval shape of liver cells, and the rectangular geometry of plant cells. A bounded medium can provide molecules with guided transmission, limits dispersion, and can have beneficial effects for long-range communication. The boundaries of a constrained medium are often assumed to be reflective, and the corresponding BC is given as
\begin{align}
  \text{BC}_1: \frac{\partial C(\textbf{d},t)}{\partial d_i}\Big|_{{d_i}={d_b}}=0, \label{BC1}
\end{align}
where $d_i\in[x,y,z]$ is an element of the position vector $\textbf{d}$ and $d_b$ is the position of the propagation boundary along direction $d_i$.

In addition, for both unbounded and bounded environments, the concentration at locations sufficiently far away from the releasing source is usually assumed to be zero, which can be mathematically described as\begin{align}
  \text{BC}_2: C( \left\| \textbf{d} \right\| \to \infty,t)=0. \label{BC2}
\end{align}

\begin{figure*}[!t]
	\centering
	\includegraphics[width=0.8\linewidth]{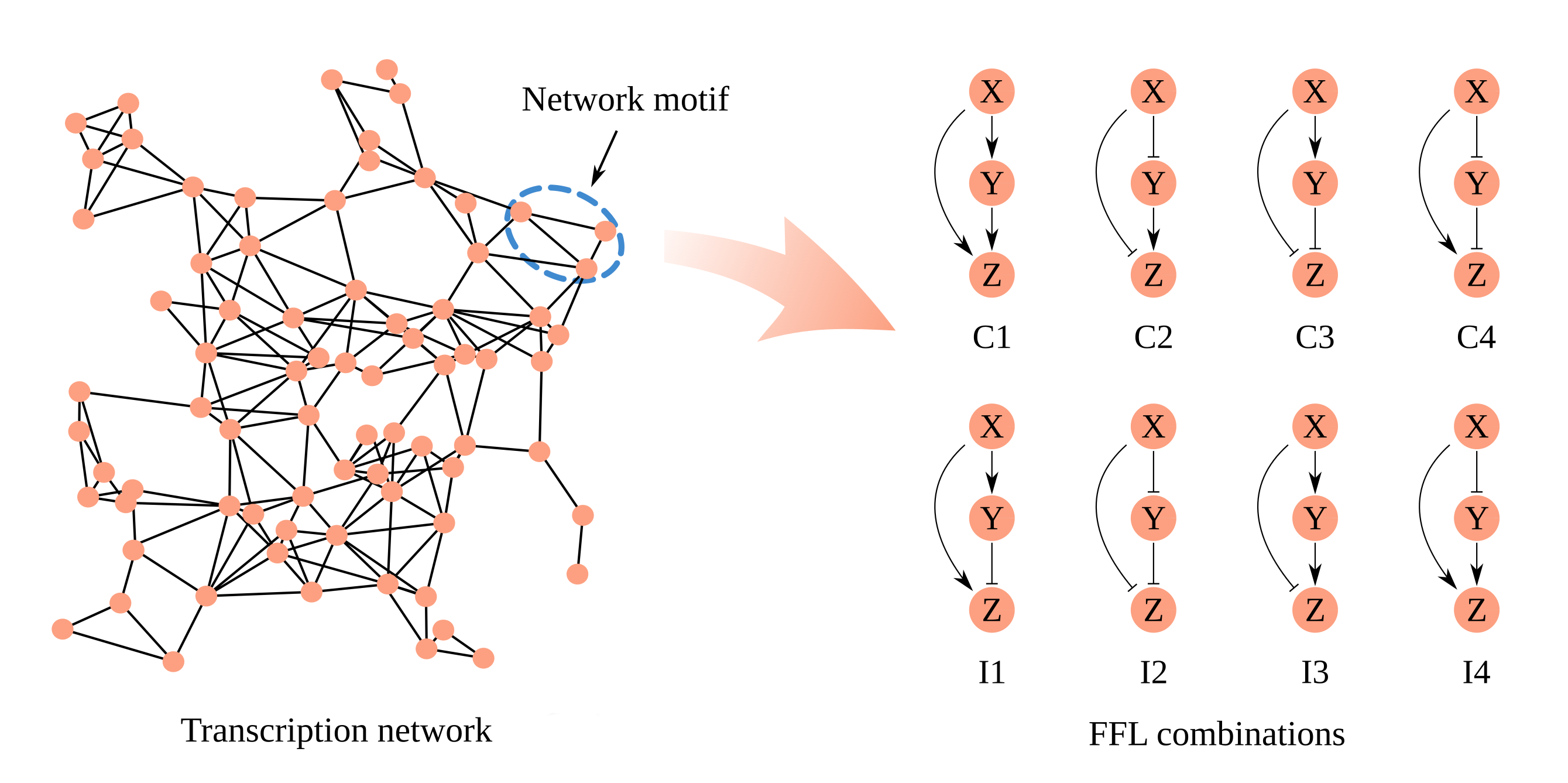}
	\caption{Overview of a transcription network, network motif, and feed-forward loops (FFL). In the transcription network, circles indicate genes and edges indicate gene interactions. Network motifs (dotted blue oval) are small sets of recurring interactions and are the building blocks of a transcription network. Feed-forward loops, one of the fundamental network motifs, are comprised of three nodes connected in one of eight possible configurations, i.e., Coherent FFL (C1-C4) and Incoherent FFL (I1-I4). Arrows denote activation and $\bot$ symbols denote repression of the corresponding node (gene).}
	\label{f_tn}
\end{figure*}

\begin{figure*}[!t]
	\centering
    \includegraphics[width=1\linewidth]{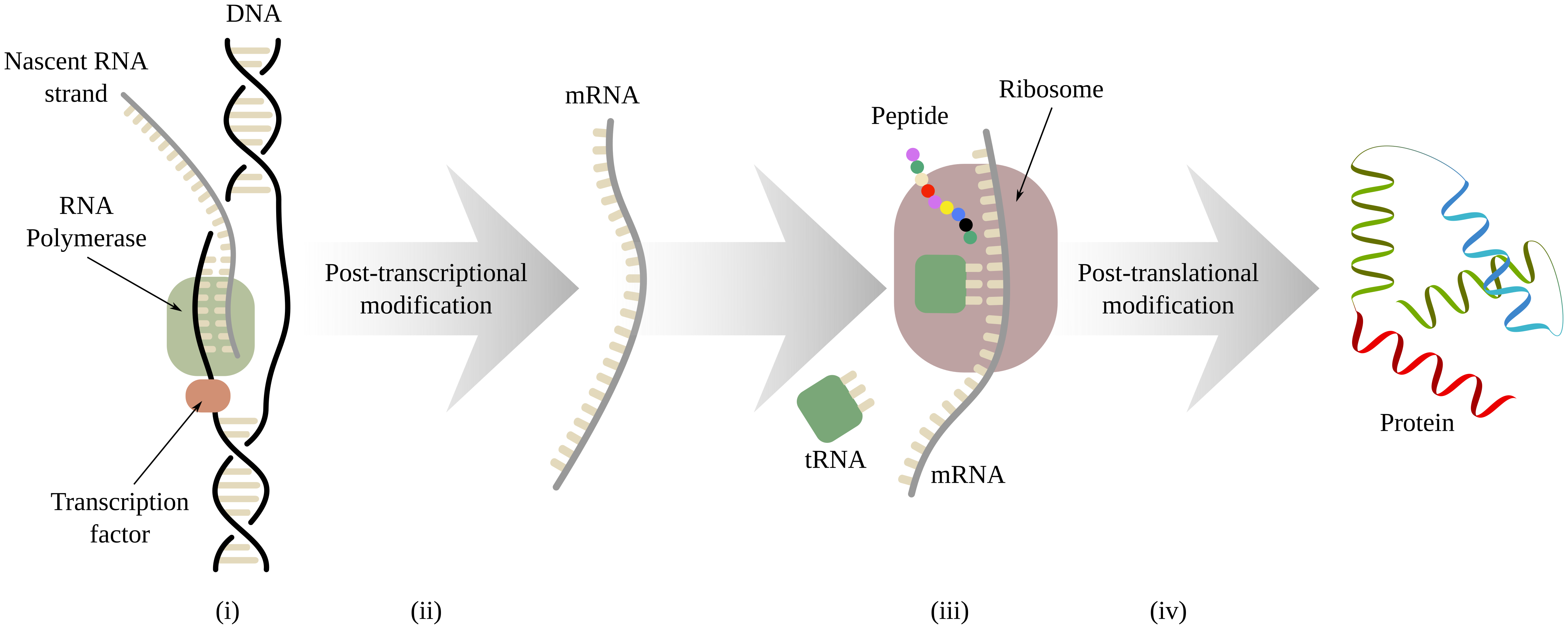}
	\caption{Overview of protein production from DNA. (i) RNA polymerase guided by a trasncription factor binds to the beginning of a gene and transcribes the DNA sequence into mRNA. (ii) The newly formed RNA molecule is being modified immediately after production to give rise to a mature mRNA which then, in eukaryotes, is transported outside the nucleus. (iii) In translation, a ribosome uses mRNA as a template for the assembly of peptides. Raw materials are brought in by tRNA. (iv) A number of peptides are combined into proteins. After further modifications and folding, a mature protein is produced.}
	\label{f_ge}
\end{figure*}

\subsubsection{BCs on Reception Mechanisms}
As stated earlier, the two conventional paradigms for molecule reception in the MC literature are active and passive, where molecules do and do not participate in chemical reactions at the receiver, respectively. If a receiver is passive, then molecules are transparently observed by the receiver without disturbing their propagation. If the receiver is active, then the molecules are usually detected by surface receptors via absorption. However, if molecules can be \textit{adsorbed} (i.e., ``stick'' to the surface) instead of just be \textit{absorbed} (i.e., removed from the surface), then it is also possible that the receiver is capable of desorbing the molecules that were previously adsorbed. This type of receiver can be called a reversible adsorption receiver and examples include the reception of hormones and neurotransmitters \cite{Kahn1976}. The corresponding BC is given as \cite{andrew2009}
\begin{align}
  \text{BC}_3: 
  D\frac{\partial C(\textbf{d},t)}{\partial \textbf{d}}\Big|_{\textbf{d} \in \Tilde{\mathcal{S}}_{\text{RX}}}=k_1 C( \textbf{d} \in \Tilde{\mathcal{S}}_{\text{RX}},t)-k_{-1}C_{a}(t). \label{BC3}
\end{align}
where $k_1$ is the adsorption rate, $k_{-1}$ is the desorption rate, $\Tilde{\mathcal{S}}_{\text{RX}}$ is the adsorbing surface of the receiver, and $C_{a}(t)$ is the average adsorbed concentration on the receiver surface at time $t$. We note that BC$_3$ in \eqref{BC3} is a general formulation and can be reduced to relevant special cases as follows. When $k_1 \to \infty$ and $k_{-1}=0$, i.e., every collision leads to \textit{absorption} and there can be no desorption, then the receiver becomes a fully absorbing receiver, and BC$_3$ in \eqref{BC3} reduces to \cite{yilmaz2014}
\begin{align}
  \text{BC}_4: C( \textbf{d} \in \Tilde{\mathcal{S}}_{\text{RX}},t)=0. \label{BC4}
\end{align}

When $k_1$ is a non-zero finite constant and $k_{-1}=0$, then the receiver becomes a partially absorbing receiver \cite{Deng2016}.

We note that the aforementioned ICs and BCs are very general, and one type of IC or BC can be represented in various forms. The reason for this is that the different models can be expressed in terms of different coordinate systems, e.g., Cartesian coordinates, cylindrical coordinates, and spherical coordinates, as appropriate for a given MC environment. For example, cylindrical coordinates are preferred in scenarios that have some rotational symmetry about the longitudinal axis, such as a circular duct channel.

\subsection{Biochemical  Signaling Pathway: Transcription Network}
\label{sec_transcription_network}
The molecule release and reception functions within a cell are carried out by proteins, such as the bacteriorhodopsin protein that functions as a light-activated proton pump and transports H$^+$ ions out of the cell, and the aforementioned surface receptors that control the passage of molecules into the cell \cite{Alberts2015}. Thus, the careful production and timely delivery of these proteins is of utmost importance for a cell's survival. Tight control of protein production is achieved through the interaction of a number of genes, forming what is known as a \textbf{\textit{transcription network}} \cite{Alon2006}. As shown in Fig. \ref{f_tn}, a transcription network can be represented by circles and edges, where circles represent genes and edges represent their interactions. The building blocks of a transcription network are a small set of recurring interactions between genes. These interactions are called \textit{network motifs}. In a network motif, the interaction between two genes is realized through gene expression and regulation, where the product of one gene acts as the transcription factor to regulate the expression of the other. In the following, we first provide mathematical descriptions of gene expression and regulation. Then, we describe the \textbf{\textit{feed forward loop}} (FFL), the typical network motif.

\subsubsection{Gene Expression and Regulation with Mathematical Descriptions}
\label{sec_gene_expression}
Gene expression initially starts with transcription, where DNA is used as a template to synthesize mRNA, and then mRNA will be converted to proteins through translation, as shown in Fig. \ref{f_ge}. DNA transcription begins when the enzyme RNA polymerase (RNAP) recognizes and binds to the promoter region. The promoter region is unidirectional and can be found at the beginning of a gene. In addition, it decides not only the starting point of mRNA synthesis, but also the synthesis direction. After RNAP binds to the promoter sequence, RNAP unwinds the DNA at the starting point and begins to synthesize a strand of mRNA. Once the mRNA is produced, it is translated by a ribosome into protein molecules with the help of transfer RNA (tRNA). The production of mRNA is controlled by transcription factors that bind to operator sites near promoter regions. The transcription factors act as activators (or repressors) to enhance (or obstruct) the binding ability of RNAP to promoter sites, thus controlling the targeted gene expression rate. 
It is important to note that both transcription and translation establish two major control points for protein regulation, as their products are being commonly modified or even degraded before reaching the next stage (\textbf{\textit{post-transcriptional}} and \textbf{\textit{post-translational modification}}).

The aforementioned gene expression and regulation can be mathematically modeled by the Hill function once the binding of a transcription factor to its site on the promoter reaches a steady state, i.e., equilibrium \cite{Alon2006}. Let $E_x$ denote the input signal that carries information from the external world, and $X$ denote a transcription factor with active form $X^*$. For \textit{activators}, the input-output relation is
\begin{equation}
	\label{hill1}
\frac{d[Out]}{dt}=\frac{\beta {X^*}^n}{K^n+{X^*}^n},
\end{equation}
where $[Out]$ is the concentration of the output protein, $K$ is the activation coefficient, $\beta$ is the maximal expression level of the promoter, and $n$ is the Hill coefficient. The activation coefficient $K$ has units of concentration and depends on the chemical affinity between the transcription factor and its operator region. With an increase in concentration of the transcription factor, it is more likely for the transcription factor to bind to the operator region. However, since the binding probability cannot be larger than $1$, the output protein level is unable to increase infinitely and approaches a saturated maximal expression level $\beta$. The Hill coefficient $n$ determines the steepness of the Hill function \cite[Fig. 2.4]{Alon2006}. For \textit{repressors}, a similar relation exists with the same parameters and can be expressed as
\begin{equation}
\label{hill2}
\frac{d[Out]}{dt}=\frac{\beta K^n}{K^n+{X^*}^n}.
\end{equation}

The values of $K$, $\beta$, and $n$ may change with cell evolution. For example, $K$ will change if a DNA sequence suffers from mutations that alter the transcription factor binding site. 

\subsubsection{Network Motif}
\label{Network_motif}
As the building blocks of transcription networks, network motifs have resisted mutations to persist over cell evolution, and the study of their dynamics can facilitate the understanding of complex transcription networks.
A typical network motif is the FFL \cite{Alon2006}, which has been studied in hundreds of gene systems in some organisms, such as  \textit{E. coli} \cite{Mangan,Shen-Orr} and yeast \cite{Lee,Milo}. In the structure of an FFL, transcription factor $X$ regulates proteins $Y$ and $Z$, and $Y$ is also a transcription factor for protein $Z$. Due to the possibility of three edges with each being either activation or repression, there are eight variations of this motif; see Fig.~\ref{f_tn}. The eight signaling pathways can be classified into two categories: coherent FFL and incoherent FFL, according to whether the regulation (i.e., activation or repression) of the direct path from $X$ to $Z$ is the same as the overall regulation 
going through $Y$ (i.e., the regulation from $X$ to $Y$ and the regulation from $Y$ to $Z$) \cite{Alon2006}. In the most well-studied transcriptional networks in \textit{E. coli} and yeast, the coherent type-1 FFL (C1-FFL) and incoherent type-1 FFL (I1-FFL) occur with a high frequency and so we discuss them in detail below.
\begin{itemize}
	\item \textbf{C1-FFL:} 
	The signaling pathway of C1-FFL is depicted in Fig. \ref{f_coherent}. A dynamic feature for C1-FFL is the ability to distinguish spurious input square signals. The accumulation time of $Y$ depends on the duration of the input signals. If a transient spike signal arrives, the accumulated concentration of $Y$ is too low to satisfy the threshold condition and $Z$ will not be produced, i.e., the system does not respond to this input signal. This feature prevents the C1-FFL motif from responding to spurious input signals.
	\begin{figure}[!t]
		\centering
		\includegraphics[width=3.35in]{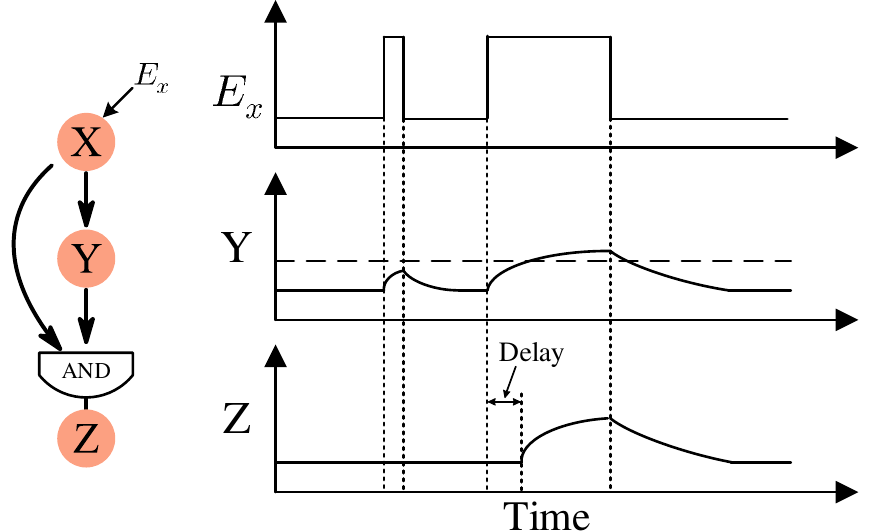}
		\caption{C1-FFL with an AND input function at the $Z$ promoter. $E_x$ is the input signal for $X$. Transcription factor $X$ is an activator ($\downarrow$) for $Y$ and $Z$, and $Y$ is also an activator for $Z$. The AND gate indicates that both $X$ and $Y$ are needed to regulate $Z$.}
		\label{f_coherent}
	\end{figure}
	\item \textbf{I1-FFL:}
    The signaling pathway of I1-FFL is depicted in Fig. \ref{f_noncoherent}. Compared with C1-FFL, $Y$ regulates $Z$ via repression instead of activation in I1-FFL, such that $Z$ shows a pulse-like profile in response to a sustained input signal.
	 	\begin{figure}[!t]
	 		\centering
	 		\includegraphics[width=3.35in]{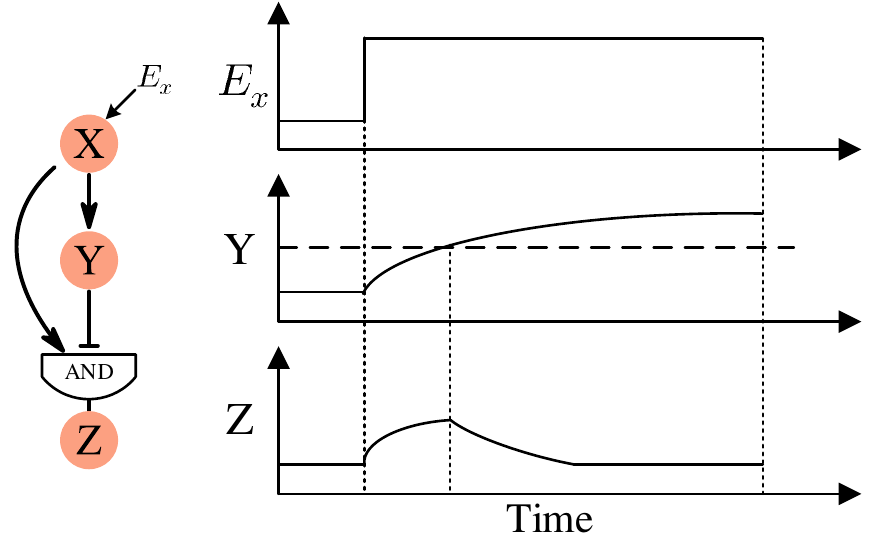}
	 		\caption{I1-FFL with an AND input function at the $Z$ promoter. $E_x$ is the input signal for $X$. Transcription factor $X$ is an activator ($\downarrow$) for $Y$ and $Z$, while $Y$ is a repressor ($\perp$) for $Z$. The AND gate indicates that both $X$ and $Y$ are needed to regulate $Z$.
	 		}
	 		\label{f_noncoherent}
	 	\end{figure}
	 	Once induced by an input signal $E_x$, the expressions of the genes encoding protein $Y$ and protein $Z$ are both activated, and $Z$ is instantly produced. Here, the delay that appears in C1-FFL is eliminated because protein $Y$ needs some time to reach the repression threshold for the $Z$ promoter, which gives a chance for protein $Z$ to accumulate. Once the concentration of $Y$ crosses the repression threshold, it starts to repress the protein production rate of $Z$. As a result, the concentration of $Z$ begins to decrease and either reaches a steady state or drops to zero depending on the repression strength of $Y$. 
\end{itemize}

\section{Level 3 - Physical/Data Interface}
\label{sec_interface}

\begin{figure*}[!t]
	\centering
	\includegraphics[width=\textwidth]{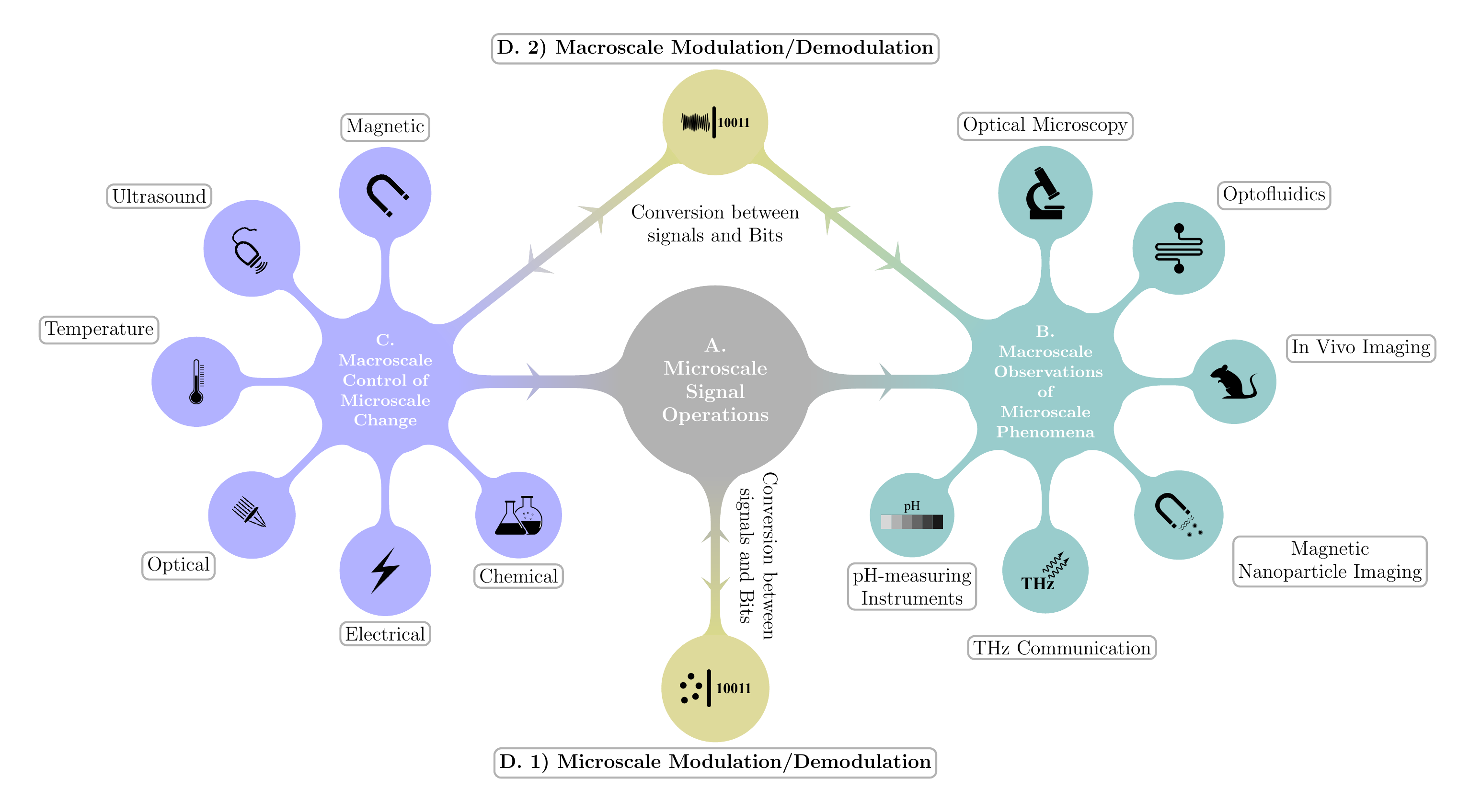}
	\caption{Schematic diagram showing typical workflows when dealing with microscale signal operations in MC. The label indexing in the figure corresponds to the subsections of Section~\ref{sec_interface}. Signal operations in MC systems (A) depend on microscale data (D.1) and can be detected using a number of available methods such as microscopy or optofluidics (B) and then be decoded (D.2). The signal operations (A) can also be controlled (C) based on target data (D.2) to modify the microscale data (D.1).}
	\label{fig:Level3}
\end{figure*}

Level 3 of the proposed hierarchy is the interface between physical signals at communicating devices and quantifying these signals mathematically.
From the perspective of communication systems, this includes: 1) how signals are modulated at a transmitter and demodulated at a receiver;  2) how modulated signals control the propagation (i.e., communication channel), and 3) how  the physical signals  (i.e., channel responses) are observed for demodulation. In other words, given that there is information to transmit, how does the transmitter translate this into a molecular (or some other physical) signal? Then, how does a receiver translate the observed signal back into information?

Depending on whether the data interface is at microscale or macroscale, with reference to Fig.~\ref{fig:Level3} we can categorize the physical/data interface of a microscale communication system into 1) microscale modulation and demodulation (Section~\ref{sec_microscale_mod}), and microscale signal operation (Section~\ref{microscale_signal_operations}); and 2) macroscale modulation and demodulation (Section~\ref{sec_macroscale_mod}), macroscale control of microscale change (Section~\ref{sec_macro_control}), and macroscale observation of microscale phenomena (Section~\ref{sec_macro_obs}).

In this section, we start with a general overview of quantifying microscale signals from the perspective of gene regulation and metabolic control. We proceed to review methods for observing and quantifying microscale phenomena from the macroscale, i.e., in a laboratory environment. This leads to a discussion of macroscale control of microscale change. We finish the section by discussing the quantification of cellular signals as information, with both modulation and demodulation processes.

\subsection{Microscale Signal Operations}
\label{microscale_signal_operations}

The interface between physical signals and their mathematical quantification can be perceived as being relatively simple for many cell signaling processes. It is often a matter of detecting whether the signal is stronger than some threshold, e.g., detecting a sufficiently high autoinducer concentration in quorum sensing (which we describe as a case study in Section~\ref{sec_case}). The creation of a signal and then the detection of its presence is a common communication methodology for cells, and is sufficient to link many processes at Level 2 (i.e., biochemical pathways detecting signals) with activity at Level 4 (i.e., the device-level state and the actions that the device takes). For example, there are biochemical signaling pathways that modify protein function directly (i.e., without requiring changes to gene expression) \cite[Ch.~15]{Alberts2015}. However, receiving a signal can also require more precision than simple detection, as can often be observed in the context of \textit{gene regulation}, i.e., the activation and deactivation of different genes\footnote{We emphasise that gene \textit{regulation} is distinct from the genetic information embedded within DNA or RNA itself; gene regulation \textit{controls} which DNA sequences are made accessible for transcription into RNA.} to control the proteins that are produced within a cell, as we introduced for Level 2 in Section~\ref{sec_interaction}. Due the impact of gene regulation on cell behavior, in the following, we discuss the quantification of signals from the perspective of gene regulation. We choose control of the metabolism as a specific example of gene regulation, given the metabolism's importance for cell growth and reproduction. We further discuss conversion between quantified microscale signals and bits in Section~\ref{sec_microscale_mod}.

\subsubsection{Gene Regulation}

Regulation is often described using genetic circuits, which show how a combination of inputs (i.e., signals) leads to activation of the gene in question (e.g., see the generic transcription network in Fig.~\ref{f_tn}). Depending on the sensitivity to the inputs and on the possible ranges of outputs, the quantification of these processes can be understood as being analog or digital. For example, if there is an appreciable difference in response according to input signal concentrations, such that the output varies continuously with the input, then the quantified response is analog. This can occur in the fine tuning of metabolic processes by some hormones \cite[Ch.~15]{Alberts2015}. If there is a discrete (i.e., readily countable) number of response levels, regardless of input concentrations, then the response is digital. A threshold-based response (i.e., most existing work from the MC engineering community \cite{Jamali2019b}) is digital, whether the response threshold is a single detected molecule or some larger quantity. There are often only two response levels (e.g., on and off), where the bistability of the circuit is achieved through positive feedback that pushes the response to one of the two levels \cite[Ch.~15]{Alberts2015}.

\subsubsection{Metabolic Control}

We highlight the gene regulation of cell metabolism as an example of microscale signal operations. The metabolism of a cell refers to all chemical reactions that take place inside the cell and that are necessary for reproduction and growth \cite[Ch.~2]{Alberts2015}. These chemical reactions are highly interdependent and chained into signaling pathways, where the product of one reaction is the on-demand substrate for the next reaction in the pathway. These reactions require specialized proteins (i.e., enzymes) in order to proceed, thus they offer an effective means of regulation. By varying
the amount of enzyme that controls each reaction, a cell is able to control its metabolism precisely. This control using enzymes, in turn, relies on a cascade of reactions triggered by extracellular cues that eventually stimulate the release of transcription factors (TFs) inside the cell to activate or repress enzyme production. 

It is important to note that the release of TF is determined by a combination of different chemical components (e.g., hormones) with particular concentrations, which can result in a digital ON/OFF enzyme activation mechanism. Such mechanisms can be controlled by the cell's environment, where  variations in the surrounding chemical composition can prompt the up- or down-regulation of the enzymes to control cellular growth or production of chemical compounds within the cell. Each TF can enter the nucleus and interfere with the expression of specific clusters of genes to alter the type and amount of proteins produced, which ultimately establishes the cell's metabolism. For example, detection of the glucocorticoid hormone by a liver cell triggers an increase in energy production via the enzyme tyrosine aminotransferase \cite[Ch.~7]{Alberts2015}.

\subsection{Macroscale Observations of Microscale Phenomena}
\label{sec_macro_obs}

Observing signaling phenomena in laboratory experimentation is important to understand their behavior and infer information about the system\footnote{Although macroscale tools provide a way to observe microscale phenomena, imperfections in experimental tools can lead to a noisy and non-ideal interface. One approach to model this uncertainty and try to enhance the accuracy of observations are learning-based models \cite{Koo2020}.}. However, we are generally constrained by the level of detail that we can readily observe (especially microscale behavior). For example, a living cell has a typical size of about $100\,\mu$m, interactions between cells can occur at a scale from a few $\mu$m to a few mm, and there is also communication between different organs, which might span up to a few meters. Signaling molecules can vary in size from around $30\,$pm for individual ions to $100\,$nm for extracellular vesicles \cite{Raposo2013}. Furthermore, temporal scales vary widely. Chemical reactions occur typically in milliseconds \cite{Ferruz2016}, ionic diffusion in biological tissues occurs at a rate of a few tens of $\mu$m$^2$/s \cite{Allbritton1992}, and physiological tissue responses to stimuli can occur in milliseconds or over many hours.

Currently, there is no single technique that enables the inspection of biological signaling processes across all spatial and temporal scales simultaneously. Thus, observation and verification often relies on a combination of established methods. In the following, we review state-of-the-art methods for observing microscale phenomena, including optical techniques, magnetism, THz waves, and pH sensing.

\subsubsection{Optical Microscopy}
\label{Optical Microscopy}

Perhaps the most commonly known observation method is optical microscopy. Since its conception in the 16th century, the optical microscope is one of the most valuable instruments in any laboratory that investigates the microscale world. The major limitation of optical microscopy is the diffraction barrier, namely the inability of the lens to distinguish between objects which are separated by a distance less that half the wavelength of the light used. Confocal laser microscopy greatly improved image resolution using visible light \cite{Cremer1978}, though it was only since the end of the last century that it was finally possible to overcome this limit and obtain instruments such as the near-field scanning microscope, the scanning tunneling microscope, and the atomic force microscope. Subsequent refinements led to an increase in resolution to the point that a single molecule can now be distinguished \cite{Hao2013-micro, Zhou2019-Quantum}. Together with fluorescent microscopy, these tools remain among the most accessible and valuable in cell biology imaging.

There is increasing demand in modern science to visualize dynamic spatial and temporal events at the micro- and nanoscale. It is now possible to obtain nanometer size images of cells while simultaneously measuring subjected mechanical forces \cite{Sridharan2019,Lee2019-Disp}. Protein motion has also been observed in great detail at the microsecond timescale using interferometric scattering microscopy \cite{Taylor2019}. Concerning the release of molecules by the cell into extracellular space, various methods such as fluorescence microscopy \cite{Axelrod2001,Keighron2012} and electrochemical techniques \cite{Kibble1996,Lindau2012}, either alone or combined, are particularly suited for capturing the trafficking of molecules \cite{Thorn2012,Liu2019-Recent}.

\subsubsection{Optofluidics}
When analyzing biological samples, it is often desirable to separate and sort out individual molecules. Conventional microscopy is cumbersome and already close to its limit in terms of spatial resolution, therefore is not usually suitable for this purpose. Optofluidics technology was developed with this in mind, combining advanced optical microscopy with microfluidics \cite{Manz1990,Psaltis2006,Minzioni2017}. It was developed as a way to miniaturize analytical instruments and it later lead to lab-on-a-chip technology \cite{Manz1990,Harrison1993}. Optofluidics technology is particularly suitable for analysis of very small working volumes, in the range of nanoliters or femtoliters. This is because it combines the analytical mechanism with sample preparation. By taking advantage of low energy consumption, nanoscale sample handling, and being free from requirements for very specialized electronics, optofluidics has been combined with other techniques such as flow cytometry \cite{Mao2009-Single}, interferometry \cite{Lapsley2011}, and Raman spectroscopy \cite{Mak2013} with good results in cell and molecular microscopy imaging. Its characteristics have also enabled integration into biosensors \cite{Fan2008} and on-chip technologies \cite{Huang2014}.

\subsubsection{\textit{In Vivo} Imaging}
\label{in vivo imaging}
Despite the popularity of optical-based imaging, there is strong scattering of light by biological tissue and so its application is limited beyond optically transparent systems or cultured cells. A long-standing challenge is to make non-invasive observations of \textit{in vivo} activity. Conventional methods of \textit{in vivo} imaging continue to be studied to improve their sensitivity and spatial resolution, e.g., there is significant research in the design of contrast nanoparticles for magnetic resonance imaging \cite{Estelrich2015}. Recent advances in ultrasound imaging with synthetic biology have overcome ultrasound's lack of specificity to make it a strong candidate for \textit{in vivo} observation of cellular functions \cite{Maresca2018}. Air-filled protein nanostructures called gas vesicles have already been engineered for introduction in mammalian cells and have helped produce high-resolution ultrasound imaging of gene expression in living mice \cite{Farhadi2019}.

\subsubsection{Magnetic Nanoparticle}
Driven by the wide applications of magnetic nanoparticles in drug delivery systems \cite{Pankhurst2003,Arruebo2007,Sun2008}, the  properties of magnetic nanoparticles have been used  to observe microscale processes \cite{Nakano2014,Kisseleff2017}. 
In \cite{Kisseleff2017}, the authors presented a magnetic-nanoparticle-based interface and proposed a wearable susceptometer design to detect magnetic nanoparticles. In \cite{harald2018}, an experimental platform that used a susceptometer to detect magnetic nanoparticles was proposed, where the susceptometer can generate an electric signal if magnetic nanoparticles pass through it.

\subsubsection{THz Communication}
The integration of nanosensors and terahertz (THz) communication modules can also support macroscale observations of microscale phenomena. This is realized by the fact that chemical nanosensors are capable of measuring the concentration of a given gas or the presence of a specific type of molecule, which can then be communicated from intrabody to outside the body via THz signaling. For nanosensors that are made of novel nanomaterials, such as Graphene Nanoribbons (GNRs) and Carbon Nanotubes (CNTs), the sensed and absorbed molecules can change the electronic properties of the nanomaterials by either increasing or decreasing the number of electrons moving through the carbon lattice. With nano antennas, the change in the number of electrons can enable the conversion of molecular information to THz waves \cite{Akyildiz}. One example of using THz communication to observe microscale phenomena is \cite{Leea}, where a nano antenna array operating in the THz band was designed to detect different carbohydrate molecules and their concentrations.

\subsubsection{pH-Measuring Instruments}
Hydrogen ions (i.e., protons) are a  popular signal molecule type with advantages of small size and easy production. More importantly, hydrogen ions have the physical property that their accumulation can lead to a reduction of the solution pH. Therefore, the concentration variations of protons can be observed at macroscale using a pH meter \cite{Grebenstein2019,Grebenstein2019a}. The same approach is considered for communication in \cite{nariman2017}, where pH meter values are used to determine whether acids or bases are being transmitted.

\subsection{Macroscale Control of Microscale Change}
\label{sec_macro_control}

Macroscale instruments do not only enable us to observe microscale phenomena, but also make it possible to control microscale systems, which establishes an interface from the macroworld to the microworld and would expand the capability of MC. An example application that benefits from macroscale control is drug delivery, where precise guidance to the diseased cells and controllable release of drugs could largely improve their therapeutic effect. There are many approaches that have been developed towards macroscale control. Here, we briefly review the controlled release of signal molecules via macroscale stimulation.

\subsubsection{Macroscale Chemical Control}
It has been a common choice to use genetically modified \textit{E. coli} bacteria in experimental MC testbeds \cite{Krishnaswamy2013,Grebenstein2019, Grebenstein2019a,Osmekhina2018}. In \cite{Krishnaswamy2013}, a microfluidic chamber was used to trap \textit{E. coli}. These bacteria were genetically modified by introducing a plasmid from \textit{V. fischeri} to produce fluorescence in response to the C6-HSL signaling molecule. 

In \cite{Osmekhina2018}, communication between two physically separated populations of E. coli was controlled and observed. The populations were grown on a microfluidic chip and separated by a filter composed of cellulose nanofibrils between rows of polydimethylsiloxane (PDMS) pillars. The filter prevented the populations from mixing but enabled the passage of signaling molecules such as the quorum sensing molecule acyl-homoserine lactones (AHL). Furthermore, the microfluidic chip was designed to flush excess bacteria and thus constrain the population sizes. The sender population could produce AHL and fluoresce cyan in response to the addition of arabinose, and the receiver population fluoresced green in response to AHL. Fluorescence patterns of the two populations were observed with negligible delay, suggesting that rapid signaling from sender to receiver enabled the populations to behave synchronously.

\subsubsection{Macroscale Electric Control} 
External electric stimulus is a method  to bridge the macroworld and the microworld.  The electrically controllable release of DNA molecules immobilized in layer-by-layer (LbL) thin film was investigated in \cite{Tao2020}. Upon an electric signal on the LbL film, DNA molecules are disassembled and released with an electrodissolution of the layers. The DNA molecule release process can be switched off when the electric stimuli is removed, and the released number of DNA molecules is proportional to the amplitude of the electric signal, which allows for  a tunable release of signal molecules. It is noted that external electric stimulus can also trigger  biological responses via redox reactions, such as the patterning of biological structure and the induction of gene expression \cite{Kim2019b}.

\subsubsection{Macroscale Optical Control} 
Light-sensitive cellular entities can be controlled by external light sources. 
One example is the release of biomolecules from photoremovable containers upon illumination \cite{ellis2007}, which achieves a conversion of optical signal to chemical signal. Similar signal transduction has also been realized in the MC community  \cite{Grebenstein2019} and \cite{Grebenstein2019a}, where \textit{E. coli} was modified with light-driven proton pumps (i.e., bacteriorhodopsin), which can be excited  by external light sources to induce proton release, with increased pH value measured via a pH sensor.

\subsubsection{Macroscale Temperature Control} 
External temperature can be another macroscale stimulus to control microscale processes. Some nanocapsules that are temperature-sensitive, such as the liposomes in \cite{needham2000}, the dendrimers in \cite{kono2007}, and the polymersomes in \cite{liu2015}, can undergo a conformation or permeability change and release encapsulated signal molecules as a response to a temperature increase. In this way, thermal signals from the exterior of devices are converted into chemical signals. It is noted that the morphological changes in \cite{liu2015} are reversible, meaning that sustainable temperature control can be achieved. One method to achieve temperature control is via focused ultrasound, which has been proposed to control cellular signaling and the expression of specific genes \cite{Maresca2018}. Candidate targets include temperature-sensitive ion channels and transcription repressors.

\subsubsection{Macroscale Mechanical Control}
\label{mechanical_control}
In addition to temperature control, focused ultrasound can provide momentum and energy to interact with molecules, cells, and tissues via mechanical mechanisms \cite{Maresca2018}. For example, ultrasound waves can be amplified by microbubbles and provide mechanical forces on a millisecond timescale, i.e., with much greater precision than temperature changes. This approach has been used \textit{in vitro} to open mechanosensitive ion channels expressed in mammalian cells. A current constraint for use \textit{in vivo} is the difficulty in delivering such microbubbles beyond the bloodstream.

\subsubsection{Macroscale Magnetic Control} 
\label{magnetic_control}
Magnetic nanocarriers are important carriers for drug delivery. The magnetic behavior not only allows magnetic nanocarriers to be manipulated in space towards targeted locations by external magnetic fields, but also facilitates their visualization by increasing their imaging contrast in magnetic resonance imaging (MRI), which in turn provides a way of monitoring their movement through the body. After nanocarriers arrive at desired sites, the magnetic energy can be converted into internal energy to induce local heating, thus triggering the release of loaded drugs \cite{ulbrich2016}.

\subsection{Conversion from Signals to Bits}

Throughout this section, we have been referring to the quantification of physical molecular signals, how these signals are observed, and how such signals can be controlled. We have mentioned that these signals contain information, but we have not yet directly linked the mathematical abstraction to the quantification of information. Level 3 of the proposed hierarchy includes not only the mathematical abstraction of physical signals, but also how a quantified signal contains information. We now elaborate on this idea and re-visit some of our examples from this perspective.

The MC community already has an understanding of information transmission that is directly inspired by conventional telecommunication systems \cite{Kuran2011}. A transmitter in a communication system has information to send. Information that is in a quantifiable form is typically represented as a sequence of digital bits, i.e., 1s and 0s, and the sequence is packaged into a series of symbols, each of 1 or more bits. The transmitter needs a scheme to represent each symbol  as a different physical signal. Generating a physical signal that corresponds to the current information symbol is called \textit{modulation}. \textit{Demodulation} at the receiver then uses the observed signal to attempt recovery of the intended symbols and hence the original bit sequence. Thus, the observed physical signal is somehow quantified and then translated back to information. There have been significant research efforts to effectively and efficiently demodulate diffusion-based signals to recover sequences of digital bits \cite{Kuscu2019a}.

The simplest modulation scheme and also the most popular one in the MC literature is binary concentration shift keying (BCSK). In BCSK, the transmitter releases a certain number of molecules to send a 0 (i.e., bit-0), and a higher number of molecules to send a 1 (i.e., bit-1). When zero molecules are released to send a 0, then BCSK is also known as ON/OFF keying (OOK). BCSK sends a single bit of information with each symbol. Other modulation schemes use more variations in the number of molecules to send more bits, or they vary features, such as the type of molecule used or the precise time instant when molecules are released. Given the prevalence of ON/OFF signals in cell signaling, we can readily understand signaling in many biological systems as BCSK.

\subsubsection{Microscale Modulation and Demodulation}
\label{sec_microscale_mod}
The telecommunications engineering approach to modulation and demodulation does not always precisely align very well with signaling in cell biology, in particular when it comes to the ability to represent information with a long sequence of bits. This is evident in some existing platforms developed by the MC community, including the tabletop MC system \cite{Farsad2013a} and its iterations, which are actually macroscale systems, as well as droplet microfluidic channels \cite{Biral2013}; these testbeds focus on the physical or chemical properties of signal propagation and detection and not on integration with a biological system. While there are specific instances where biological data can readily map to sequences of bits, such as strands of DNA or RNA (where each base pair is a 2-bit symbol), many MC schemes are not structured in this manner and often 1 or a few bits is sufficient to represent all the information being modulated, e.g., whether a target threshold concentration has been reached to stimulate an action. Nevertheless, a digital representation can still be useful. For example, genes are often represented as switches that are turned on or off by transcription factors. Thus, there can be one bit of information for each switch, and this bit can change with the demodulation of the corresponding \textbf{\textit{gene regulation}} signal. This signal could come from within the cell, e.g., via a coupled internal signaling pathway, or from outside the cell, e.g., \textit{E. coli} demodulating a chemotactic signal from its surrounding environment to decide whether to proceed along its trajectory (i.e., \textit{run}) or change direction (i.e., \textit{tumble}) \cite{berg1993random}.

\subsubsection{Macroscale Modulation and Demodulation}
\label{sec_macroscale_mod}
Broadly speaking, making macroscale observations (as reviewed in Section~\ref{sec_macro_obs}) and trying to recover information about a cellular system corresponds to demodulation  at a receiver, whereas using macroscale methods to control such a system  (as reviewed in Section~\ref{sec_macro_control}) corresponds to modulation by a transmitter. At macroscale, we have the benefit of easy access to modern computing devices. Macroscale MC systems often include a connection with a microcontroller board (e.g., Arduino) or a computer to perform modulation to convert sequences of symbols into physical signals or to demodulate signals into received symbols. Thus, any of the macroscale methods discussed in this section could be abstracted and interpreted as a transmitter or receiver of quantified information. In the following, we highlight works that did so explicitly to quantify the transmission of bits.

At a macroscale transmitter, \textbf{\textit{electrical signals}} representing bit sequences can be directly modulated as chemical signals \cite{Tao2020} or through an intermediate signal form, such as an \textbf{\textit{optical signal}} in \cite{Grebenstein2019,Grebenstein2019a} and a \textbf{\textit{thermal signal}} in \cite{needham2000,kono2007,liu2015}. In \cite{Tao2020}, the authors realized OOK modulation by translating \textbf{\textit{electrical signals}} into biological DNA signals. For transmission of bit-1, a rectangular electrical signal with an amplitude of 5\,V and a duration of 10\,s was applied to stimulate the release of DNA molecules from a multilayer film, while for transmission of bit-0, the electrical signal was switched off. This setup could achieve a bit rate of 1\,bit/minute. In addition, the authors also found that the number of released DNA molecules was dependent on the amplitude of the electric stimulus, which could enable a higher order concentration shift keying modulation by modulating different symbols with different amounts of DNA.

In \cite{Grebenstein2019,Grebenstein2019a}, OOK modulation is achieved by an \textbf{\textit{optical-to-chemical}} conversion. The intended symbol sequence does not directly induce the light-driven proton pump to emit protons, but it is first modulated as an optical signal to switch an LED on or off. The LED is switched off during the entire symbol interval to represent bit-0 while it is turned on to transmit bit-1, thus controlling the release of protons by modified \textit{E. coli}. The proton releases were measured with a \textbf{\textit{pH sensor}}, and channel estimation techniques and adaptive transceiver methods were implemented at macroscale to demodulate the signal and recover the symbol sequences. A reliable throughput rate of about 1\,bit/minute was achieved, which was much faster than the 6-7 hours to recover ON/OFF \textbf{\textit{fluorescent patterns}} made by modified \textit{E. coli} bacteria in response to C6-HSL signal molecules in \cite{Krishnaswamy2013}.


\section{Level 4 - Local Data Abstraction}
\label{sec_data}

Level 4 of the proposed hierarchy is the interface between the mathematical quantification of physical signals (i.e., the output of Level 3) and how the information in these signals is manifested and manipulated within an individual communicating device. In other words, Level 4 is concerned with the context for information in cell biology signaling. By definition, this level is more mathematically abstract than the lower levels, but is also manifested as individual behavior. We expect that synthetic devices, whether they are at a microscopic or macroscopic scale, will generally have more functional complexity than natural microorganisms have at this level. For example, digital computing and memory devices can enable significant data processing capabilities. While nature does have means for storing and manipulating large quantities of information, e.g., DNA and memory in the brain, the functional complexity of communication is primarily in the biochemical processes that physically manipulate the signal, i.e., at Level 2 of our proposed hierarchy. Nevertheless, Level 4 describes data, where the data comes from, and how individual devices use it.

From a communication perspective, the transmitter is responsible for \textit{encoding} its information into a quantifiable form such as a bit sequence (that is then modulated, i.e., in Level 3). Once the receiver has demodulated the received molecular signal, it is then \textit{decoded} to recover the embedded information. The encoding and decoding processes are usually ignored in contributions by the MC community, because it is often assumed that a bit sequence of interest already exists (or one is randomly generated if needed). The fundamental communication problem is for the receiver to recover the bit sequence, typically without consideration of how this information is subsequently utilized (as this is beyond the scope of a conventional communication engineering problem). However, since behavior in cell biology is tightly coupled with the information that cells receive, it is particularly relevant for our holistic approach to consider the significance of the data.

The remainder of this section is organized as follows. We describe the meaning of information in cellular signals, including limits on how much information these signals can carry (Section \ref{information_cellular_signals}). Contexts for cellular information include genetic information in DNA and RNA, collecting information about the external cellular environment, and controlling actions such as cell division, cell differentiation, and cooperation. We then transition to a discussion of the design of analog and digital circuits based on chemical reactions and synthetic biology (Section \ref{sec:synthetic_circuits}), and how these can be used to realize communication functionalities in an engineered cell biology system (Section \ref{function_realization}). Finally, we elaborate on the physical structure of DNA and its potential for synthetic storage (Section \ref{sec_MicroStorage}).

\subsection{Information in Cellular Signals}
\label{information_cellular_signals}
While we consider DNA holistically as a case study in Section~\ref{sec_applications}, we summarized the translation and transcription processes in Section~\ref{sec_interaction}, and we will elaborate on microscale storage using DNA in Section~\ref{sec_MicroStorage}, it is worthwhile to briefly discuss it here in the context of local cellular data. Both DNA and RNA are linear polymers composed of nucleotide subunits with 4 distinct bases (DNA and RNA both use adenine, guanine, and cytosine; DNA has thymine while RNA has uracil) \cite[Ch.~6]{Alberts2015}. Thus, each subunit carries 2 bits of information, which get copied when DNA is transcribed to produce RNA. While some RNAs have specific standalone roles including reaction catalysis and regulation of other genes, mRNAs are RNAs that are created for translation to protein. In this latter case, triplets of bases called \textit{codons} are used to encode each of the 20 amino acids that are commonly found in proteins. Since 3 nucleotides, each having one of 4 bases, can be combined to make $4^3=64$ distinct codons, many amino acids are specified by multiple codons, and there are also codons that indicate the end of a sequence. While there are many biochemical steps to go from DNA to protein (some of which were described in Section~\ref{sec_transcription_network}), there is a clear mapping from nucleotide bases to amino acids.

Besides genetic information, many cellular signaling processes are driven by a single bit of information \cite[Ch.~21]{Alberts2015}, e.g., the presence or absence of an event or a change of state. From a communications perspective this can seem incredibly simplistic, but this is consistent with existing bounds on mutual information and capacity, including the estimation of environmental signals using biochemical reaction networks \cite{Hathcock2016}, intracellular signaling in an individual cell \cite{Suderman2017}, and an individual signal transduction channel \cite{Thomas2016}. However, natural options do exist to transmit information beyond such constraints. Generally, individual signaling pathways could be chained together to drive more complex functions and behavior. It has been shown that noise filtering in \textit{E. coli} enables it to detect antibiotic concentrations with up to 2 bits of resolution, thereby distinguishing sublethal levels \cite{Ruiz2018}. The authors of \cite{Selimkhanov2014} showed that temporal signal modulation can reduce the information loss induced by noise and increase the accuracy of biochemical signaling networks.

Communication between cells is also used to augment the available information \cite{Ellison2016}. Noise at a single-cell level can be exploited to increase information at a population level up to several bits by smoothing out individual cell responses that would otherwise lead to abrupt ON-OFF changes in population behavior \cite{Suderman2017}. There are limits to the gains available and this is in part due to the constraints imposed by communication reliability \cite{Mugler2016}. While it may be intuitive to think that communicating cells should be as close together as possible to maximize the precision in concentration estimation, it has actually been shown that sparse packing of a large population is optimal for concentration sensing \cite{Fancher2017}.

As we have noted, cellular information is tightly coupled with behavior. Even DNA, which is stored analogously to digital information, leads to the RNA and proteins that drive many cellular tasks. In the case of \textit{E. coli} measuring antibiotic concentrations, detection of sublethal levels can signal when to produce costly resistance mechanisms to improve population fitness \cite{Ruiz2018}. Information shared across a cellular population can include the fraction of the population that is preparing for a major event such as cell division or cell death \cite{Suderman2017}.

Other examples of the significance of local cellular data can be readily identified. For example, quorum sensing is used in many communities of bacteria to coordinate decisions by releasing signaling molecules \cite{Mukherjee2019b}. A simplified understanding of quorum sensing is that the accumulation of signaling molecules is treated as a proxy for the local estimation of current population density. While the density is not estimated precisely, bacteria can distinguish between ``high'' and ``low'' states and gene expression is switched to favor cooperative behavior when the estimate becomes sufficiently high. We discuss quorum sensing in greater detail as a case study in Section~\ref{sec_applications}.

Cell differentiation is the specialization of cells into particular roles and is a fundamental process for multi-cellular organisms \cite[Ch.~21]{Alberts2015}. One way in which cell differentiation is controlled is via the reception of signals from neighboring cells. Diffusion creates concentration gradients based on proximity to the source signal, enabling cells to specialize according to their location. Additional diversity can be provided by controlling differentiation with multiple types of signals, such that each molecule type corresponds to one bit of information.

Theoretical and experimental studies in \cite{Pierobon2016, Sakkaff2017, Sakkaff2018, Sakkaff2019} established methods to characterize the limits and information flow rates for cell metabolism, and quantify the amount of control that the external environment can exert on a cell in terms of metabolic fluxes. By using different combinations of chemical compounds with varying  concentrations,  temperatures, and acidity, the chemical composition of a cell's environment can be manipulated in order to trigger a specific response, such as the secretion of a useful metabolite.

Additional works that have sought to describe the information in natural cellular signals include calcium signaling specificity in \cite{Martins2019}, and the insulin-glucose system in \cite{Abbasi2017}. The authors of \cite{Awan2018,Awan2019a} modeled the information carried in the action potentials between plant cells.

\subsection{Digital and Analog Circuits }
\label{sec:synthetic_circuits}
An MC transmitter encodes information into a quantifiable form and then modulates it into a physical signal. An MC receiver demodulates a received chemical signal to recover the transmitted information. To  guarantee successful information delivery, signal processing units that process information flow over molecular concentrations are envisioned to be indispensable components for synthetic MC transmitters and receivers with complex communication functionalities, including
modulation-demodulation and encoding-decoding.

In general, biochemical signal processing functions can be realized in two fashions: 1) chemical circuits \cite{cook2009} based on ``non-living'' chemical reactions, and 2) genetic circuits \cite{Ron2003} in engineered living cells. In chemical circuits, a set of chemical reactions is designed for a target desired chemical response, whereas in genetic circuits, a gene regulatory network based on synthetic biology is designed to achieve desired function. {Considering the scalability of digital design and the discreteness of molecules, it is logical to start by designing circuits to process digital signals that switch rapidly from a distinct low state representing bit-0 to a high state representing bit-1. However, biological systems do not always operate with \textit{reliable} `1' and `0' signals; instead, many signals are processed probabilistically and show a graded analog response from low to high level \cite{Bradley2015}.} In addition, motivated by the fact that biological systems based on analog computation can be  more efficient compared with those based on digital computation \cite{Bradley2015,Sarpeshkar2014}, analog circuit design also receives attention from biologists and engineers. In the following, we review some synthetic digital and analog circuits which are designed based on chemical reactions and synthetic biology. These circuits can not only achieve certain computational operations by themselves, but can also be integrated to realize some communication functionalities. 

\subsubsection{Digital \& Analog Circuits via Chemical Reactions}
Many types of digital circuits have already been designed and realized via chemical reactions, demonstrating their capabilities to process molecular concentrations. Designing digital logic functions has also attracted increasing research attention. Combinational gates, including the AND, OR, NOR, and XOR gate, were designed in \cite{jiang2013} based on a bistable mechanism. 
For a single bit, the HIGH and LOW states are indicated by the presence of two different molecular species. 
The designed gates were mapped into DNA strand-displacement reactions and validated by generating their chemical kinetics. The authors of \cite{Ge2017} also used the bistable mechanism, where five general and circuit-free methods were proposed to synthesize arbitrary combinational logic gates. The AND, OR, NOR, and XOR gates were also realized via joint chemical reactions and microfluidic design in \cite{Bi2020b} with a different bit interpretation, where bit-1 is represented by a non-zero concentration value and bit-0 is represented by zero concentration. A mathematical framework was proposed in \cite{Bi2020a} to theoretically characterize the designed gates, and insights were also provided into design parameter selection (e.g., species concentrations) to ensure an exhibition of desirable behavior. 

An architecture of analog circuits to compute polynomial functions of inputs was proposed in \cite{song2016}, where the circuits were built on the basis of analog addition, subtraction, and multiplication gates via DNA strand displacement reactions. Relying on the help of Taylor Series and Newton Iteration approximations, these analog circuits can also compute non-polynomial functions, such as the logarithm. However, an accurate logarithm computation over a wide range of inputs requires a large number of reactions, due to the high-order power series approximation. In \cite{chou2017log}, the authors presented a method to accurately compute the logarithm with tunable parameters while maintaining low circuit complexity.
In \cite{luca2020}, a systematic approach to convert linear electric circuits into chemical reactions with the same functions was presented. The principle of the approach is that both electric circuits and chemical circuits can be described by ordinary differential equations (ODEs), no matter what quantities the ODEs represent (e.g., voltages or concentrations). Based on this, an electric high pass filter circuit was realized by a set of chemical reactions.

\subsubsection{Digital \& Analog Circuits via Synthetic Biology}
\label{circuits_syn}
A fundamental objective of synthetic biology is to control and engineer biochemical signaling pathways  to build biological entities that are capable of carrying out desired computing tasks. 
The single input logic gates were synthesized to carry out simple computations, and these include the BUFFER\footnote{A buffer gate can maintain the input and output logic relationship, and can be regarded as a delay gate.} gate \cite{Tamsir2011} and the NOT gate \cite{Wang2011}, which are directly inspired by mechanisms of gene expression induced by activators and repressors, respectively. To expand the information processing ability, multi-input logic gates, including a $2/6$-input AND gate \cite{Wang2011,Weinberg2017}, $2/3/4$-input NAND gate \cite{Win2008,Kim2019c}, and $4/5$-input OR gate \cite{Kim2019c}, were also designed. 
The authors in  \cite{Wang2011} further optimized their designed multi-input logic gates  with modularity (i.e., having exchangeable inputs and outputs to increase the reusability) and orthogonality (i.e., no crosstalk 
within the host cell to increase robustness and stability). For instance, the proposed $2$-input AND gate in \cite{Wang2011} can not only be rewired to different input sensors to drive various cellular responses, but can also show the same functionality in different types of cells. It is noted that multiple logic gates can be combined to realize much more complicated cellular tasks, such as multicellular biocomputing \cite{Tamsir2011} 
and the edge detection algorithm \cite{Tabor2009}.	

Many synthetic analog circuits have also been proposed. One example is the wide-dynamic-range, positive-logarithm circuit \cite{Daniel2013}, which consists of a positive-feedback component and a `shunt' component, demonstrating an ln($1+m$) input-output transfer characteristic for a scaled input concentration $m$. A comprehensive review of $17$ different analog circuits is provided in \cite{Teo2015}. An intuitive way to understand the design of analog circuits is to interpret the synthetic process as tuning the behavior or response curve of a biological component. In particular, the Hill function (introduced for Level 2 in Section~\ref{sec_gene_expression}) provides a semi-empirical approach in capturing the desired response curves \cite{Ang2013}. For example, in \cite{Marcone2018}, the parameters of the Hill function were optimized to tune the relationship between the temporal change of the output protein and the input transcription factor as close as possible to a hyperbolic tangent and a logarithmic function.

Integrating analog circuits with digital circuits is a strategy to achieve more complicated computations. A digitally controlled logarithm circuit was designed in \cite{Daniel2013}, where a positive or negative logarithm circuit is connected to a digital switch. This combined circuit achieves a positive or negative logarithm function in the presence of the input inducer IPTG/AraC, whereas it shuts OFF in the absence of the inducer. 

\subsection{Realizing Communication Functionalities}
\label{function_realization}
The digital and analog circuits realized either by chemical reactions or synthetic biology provide the communication community with novel tools for processing chemical signals. 
In the following, we review some theoretical circuit designs that enable modulation-demodulation and coding-decoding functionalities.

\subsubsection{Modulation \& Demodulation Functionalities} 
For concentration shift keying (CSK) modulation and demodulation, binary CSK (BCSK) and quadruple CSK (QCSK)\footnote{In modulation schemes for wireless communication, ``Q'' usually stands for ``quadrature'' and refers to phases, e.g., quadrature phase shift keying (QPSK) modulation. However, in MC, ``Q'' usually stands for ``quadruple'' and refers to four concentration levels.} realizations were presented in \cite{Bi2020} and \cite{Bi2020a}, respectively.
The BCSK transmitter designed in  \cite{Bi2020} was capable of modulating a rectangular input signal representing bit-1 as a pulse-shaped output, where the involved chemical reactions were directly inspired by the I1-FFL discussed for Level 2 in Section \ref{sec_interaction}. The corresponding receiver used an amplifying reaction to output a rectangular signal if the received signal exceeded a threshold. For the QCSK modulation and demodulation in \cite{Bi2020a}, the transmitter design was inspired by the electric 2:4 decoder that activates exactly one of four outputs according to a combination of two inputs. As an electric 2:4 decoder can be easily implemented using logic gates, the QCSK transmitter used the chemical reactions-based AND and NOT gates to modulate two inputs to four different concentration levels. At the receiver side, three detection modules proposed in \cite{Bi2020} with different thresholds were connected with two AND gates and an XNOR gate to achieve QCSK demodulation. In addition, the demodulation of rectangular signals having identical durations but different concentrations was analyzed in \cite{Chou2019a}, where the demodulator was based on the maximum a-posteriori probability (MAP) framework and can be implemented by several chemical species and reactions found in yeast.

In addition to the CSK modulation scheme, chemical circuits have also been applied to implement other modulation schemes, such as frequency shift keying (FSK), molecular shift keying (MoSK), and reaction shift keying (RSK). The realization of binary FSK (BFSK) demodulation was investigated in \cite{chou2012}. With two symbols encoded with different frequencies, the BFSK receiver consisted of two branches of enzymatic reaction circuits, which is analogous to the design of an electric BFSK decoder. The parameters of the two branches were carefully selected according to the transmitted symbols so that each symbol could only trigger one branch. For MoSK, the receiver architecture was presented in \cite{Egan2019}, where chemical reactions were exploited to determine if the sampled number of bounded signaling molecules exceeded a predefined level. For RSK, different chemical reactions were exploited for modulating transmission information into different signaling molecule emission patterns \cite{awan2015nanocom}. To demodulate RSK signals, the authors in \cite{awan2015nanocom} investigated two types of ligand-receptor based chemical circuits, and demonstrated the positive impact of feedback regulation on symbol error rate reduction. 
The amount of information transferred by chemical reactions-based transceivers was quantified in \cite{Awan2019}, where optimal transmitter circuits that maximize the mutual information of the whole communication link were derived for four types of receiver circuits (i.e., ligand binding, degradation, catalytic, and regulation reactions).

An engineered bacteria-based biotransceiver architecture with modulation and demodulation functionalities was proposed in  \cite{Unluturk2015}. In this architecture,  the transmitter employed a modulator to realize M-ary amplitude modulation, and was capable of generating a transmitted signal via a transmission filter; the receiver first processed a received signal via the receiver filter with low-pass filtering characteristic to reduce noise and then used the demodulator to decode transmitted bit sequences. 

\subsubsection{Coding \& Decoding Functionalities}
Classic coding schemes have been studied for MC to improve the reliability between communication links. A uniform molecular low-density parity check (LDPC) decoder to retrieve transmitted information from received signals was designed in \cite{Zhang2019} with chemical reactions. To execute the belief-propagation algorithm, a chemical oscillator was introduced to schedule the iterative message passing and trigger corresponding computations in each phase. The proposed LDPC decoder design is flexible and can deal with arbitrary code lengths, code rates, and node degrees.

A transceiver design with single parity-check (SPC) encoding and decoding functionalities was developed in  \cite{Marcone2018} using both chemical circuits and genetic circuits. The proposed transmitter is able to generate a parity check bit and modulate the corresponding codeword with CSK, and the proposed receiver acts as a soft analog decoder that calculates the a-posteriori log-likelihood ratio of received noisy signals to retrieve transmitted bits. During the aforementioned processes, chemical reactions are used to realize degradation, subtraction, and storage, while engineered gene expression processes are employed to implement some complicated operations, such as amplification, the hyperbolic tangent function, and the logarithm function. 

\subsection{Microscale Storage}
\label{sec_MicroStorage}
\begin{figure}[!t]
	\centering
	\includegraphics[width=3.2in]{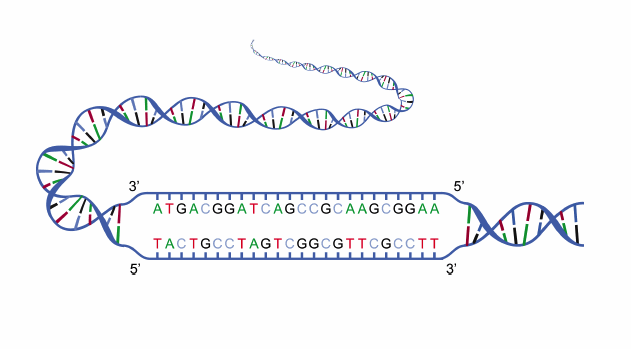}
	\caption{Antiparallel DNA strands consisting of combinations of the four bases: adenine (A, green), cytosine (C, light blue), guanine (G, black), and thymine (T, red). The pairing up of the bases along opposite strands is very specific (A with T and C with G) and this facilitates reliable replication. The primed numbers indicate the direction of each strand, as transcription always proceeds from the 5' to 3' end. Source: \cite{Human_Genome}.}
	\label{DNA}
\end{figure}

To end this section, we elaborate on the physical structure of DNA and the potential for DNA as a storage mechanism for synthetic systems. A DNA molecule is comprised of two antiparallel chains (DNA chains or strands), each composed of nucleotide subunits \cite[Ch.~4]{Alberts2015}; see Fig.~\ref{DNA}. Each subunit contains one of 4 bases: adenine (A), cytosine (C), guanine (G), and thymine (T), and it is common to use the name of the base to label an entire nucleotide subunit. Knowing the bases along one chain is sufficient to know the sequence along both chains, because an adenine subunit is always paired with a thymine subunit in the other chain, and cytosine is always paired with guanine. The chemical properties of the chains mean that they arrange in a ``double helix'' shape, which performs one complete turn for every 10 base pairs. The main skeleton of the chemical structure of each base consists of one or two carbon rings, with their carbon atoms denoted with a primed number from 1' to 5'. Depending on where the connection with the next base occurs, it is typical to describe the 5' and 3' ends of a DNA strand.

RNA is synthesized from DNA (in a process called transcription, proceeding from the 5' end to the 3' end; see Section~\ref{sec_transcription_network}) and produces a \textit{single} strand \cite[Ch.~6]{Alberts2015}. RNA is also composed of nucleotide subunits, but its bases are adenine (A), cytosine (C), guanine (G), and uracil (U) instead of thymine.

Utilization of DNA as an information storage medium is perhaps the most mature and promising of the applications of DNA in MC (for recent reviews see \cite{Rutten2018,Ceze2019}; we also discuss DNA storage as a case study in Section~\ref{sec_DNA}). Current magnetic drives are close to the limit of that technology with a storage capacity of 1 TB per square inch \cite{Milenkovic2018}. On the other hand, DNA has a maximum storage density of 2 bits per nucleotide. This means that each gram of single-stranded DNA has a theoretical maximum storage capacity of 455 exabytes. With such a technology, only 4 grams of DNA would be required to store the man-made digital information produced globally in 2016, including newspapers, books, and internet sites \cite{Shrivastava2014,DeSilva2016,Adleman1994}. However, in practice, the actual storage capacity of DNA is lower than the theoretical maximum. GC base pairs form one more hydrogen bond than AT. This results in a different melting temperature because GC bonds take more energy to break. Thus, the replication efficiency of DNA by polymerase chain reaction (PCR; discussed further in Section~\ref{sec_DNA}) varies depending on the ratio of GC to AT in that sequence which affects DNA synthesis and information retrieval \cite{Chen2019}. Similarly, repetitive occurrences of the same base (e.g., sequence AAAAAAAA) introduce errors during sequencing. These factors, together with natural DNA decay, mean that many copies of the same sequence are required for effective storage of information and this constrains storage capacity.

It has been shown that systems of DNA data storage have an error rate close to 1$\%$ \cite{Erlich2019}, a number comparable to current magnetic media. Errors are mainly produced during the writing and reading processes, so some redundancy with duplicated data is needed for reliable information retrieval. In natural systems, DNA polymerase II proceeds at a speed of about 70 bases per second \cite{Organick2018}, although actual DNA transcription happens at less than half of that rate, around 30 bases per second \cite{Tennyson1995}, mainly due to polymerase activity pauses. State-of-the-art engineered systems can achieve a writing-to-retrieval period of approximately 21 hours as demonstrated recently in a proposed end-to-end system \cite{Takahashi2019}. However, one of the key advantages of DNA storage lies with the large capacity for parallelization, e.g., any DNA sequence can be easily replicated into millions of copies in a short run of PCR.

\section{Level 5 - Applications}
\label{sec_applications}

When communication is used to send information, it is a means to an end. Organisms would not have evolved to engage in the costly activities to send and receive signals if there were no clear benefits of doing so. Thus, the top level of our proposed hierarchy (see Figs.~\ref{fig_hierarchy} and \ref{fig_hierarchy_overlay}) is the application level, which defines and describes the behavioral interactions between communicating devices. These interactions could be competitive or collaborative (e.g., a predator-prey dynamic versus coordination within a cooperative population). They can also apply over very different physical scales, e.g., within an individual cell, between a pair of cells, across a population of cells, between different species or kingdoms, or over a macroscopic-microscopic interface.

Interactions between communicating devices can either be natural or artificial. Since we provide detailed examples of applications in natural systems in the subsequent section on case studies (i.e., Section~\ref{sec_case}), in this section we focus on synthetic cell biology applications. Specifically, we select two promising applications of MC to demonstrate Level 5: \textit{biosensing} and \textit{therapeutics}. To reveal how these two applications rely on all of the lower levels of the proposed hierarchy, we use boldface font to refer to previously-covered topics in this survey and summarize these mappings in Table \ref{table_applications}. However, as indicated in this table, the potential for local data manipulation at Level 4 has not been fully explored in the current literature on these synthetic applications. Thus, at the end of this section, we present an envisioned automatic drug delivery system to demonstrate how MC system design might be applied to improve the state of the art of these cell biology applications. 
In this way, this section not only demonstrates how the components described at each level have already been utilized in biosensing and therapeutics, but also reveals promising collaboration opportunities for researchers from different communities, especially for those in the communication and synthetic biology fields. In particular, we describe how these two representative applications can be realized or facilitated with the tools from the microfluidics community discussed for Level 1 (in Section \ref{sec_physical}) and the synthetic biology community discussed for Level 4 (in Section \ref{sec_data}). 
In addition, this section includes applications of therapeutics based on magnetic fields as this macroscale control technique can reshape drug delivery systems and other \textit{in vivo} applications.

\begin{table*}[!t]
	\centering
	\caption{Application Examples.}
	\resizebox{\textwidth}{!}{%
		{\renewcommand{\arraystretch}{1.5}
		\scalebox{0.85}
	{\begin{tabular}{p{0.14\textwidth}||p{0.18\textwidth}|p{0.18\textwidth}|p{0.2\textwidth}|p{0.18\textwidth}}
				\hline		\thead{Application\\(Reference)} & \thead{Section \ref{sec_physical}: Level 1\\ Signal Propagation} & \thead{Section \ref{sec_interaction}: Level 2\\ Device Interface} & \thead{Section \ref{sec_interface}: Level 3\\ Physical/Data Interface} & \thead{Section \ref{sec_data}:
				Level 4\\ Local Data}  \\ \hline \hline
				\thead{Biosensing via \\Microfluidics\\ (\hspace{-0.004cm}\cite{Roggo2018})}
				& \thead{Diffusion-Based \\Propagation (\ref{Diffusion}); \\
				Advection-Diffusion-Based \\Propagation (\ref{Convection-Diffusion}); \\
				Chemotaxis (\ref{Chemotaxis})} & \thead{Gene Expression (\ref{sec_gene_expression})}
				& \thead{Optical Microscopy (\ref{Optical Microscopy})}
				& \thead{
				MC-Assisted\\Applications (\ref{MC_assisted_application})
				} \\ \hline
				\thead{Biosensing via \\Synthetic Biology\\ (\hspace{-0.004cm}\cite{Wang2013})}
				& \thead{Diffusion-Based \\Propagation (\ref{Diffusion})}
				& \thead{Gene Expression (\ref{sec_gene_expression})}
				& \thead{Optical Microscopy (\ref{Optical Microscopy})}
				& \thead{Digital/Analog Circuits via \\Synthetic Biology (\ref{circuits_syn})} \\ \hline
				\thead{Therapeutics via \\Microfluidics\\(\hspace{-0.004cm}\cite{Lo2009})}
				& \thead{Advection-Diffusion-Based \\Propagation (\ref{Convection-Diffusion})}
				& \thead{Molecule Reception and \\Responses (\ref{sec_reception})}
				& \thead{\textit{In Vivo} Imaging (\ref{in vivo imaging});\\
				Macroscale Mechanical \\Control (\ref{mechanical_control})
		        }
				& \thead{
				MC-Assisted\\Applications (\ref{MC_assisted_application})
				} \\ \hline
				\thead{Therapeutics via\\ Synthetic Biology\\(\hspace{-0.004cm}\cite{Anderson2006})}
				& \thead{Diffusion-Based \\Propagation (\ref{Diffusion})}
				& \thead{Molecule Reception and \\Responses (\ref{sec_reception});\\
				Gene Expression (\ref{sec_gene_expression})}
				& \thead{Macroscale Observations of\\ Microscale Phenomena (\ref{sec_macro_obs})} 
				& \thead{Digital/Analog Circuits via \\Synthetic Biology (\ref{circuits_syn})} \\ \hline
				\thead{Therapeutics via\\ Magnetic Field\\(\hspace{-0.004cm}\cite{Tietze2013})}
				& \thead{Advection-Diffusion-Based \\Propagation (\ref{Convection-Diffusion})}
				& \thead{Molecule Reception and \\Responses (\ref{sec_reception})}
				& \thead{\textit{In Vivo} Imaging (\ref{in vivo imaging});\\
				Macroscale Magnetic\\ Control (\ref{magnetic_control})}
				& \thead{
				MC-Assisted\\Applications (\ref{MC_assisted_application})
				}
			 \\ \hline
			\end{tabular}
	}}}
	\label{table_applications}
\end{table*}

A microfluidic device processes and manipulates small amounts of fluids using channels in dimensions of tens to hundreds of micrometers (i.e., $10^{-9}$$\sim$$10^{-18}$ litres). Advantages of microfluidic systems include rapid analysis, high performance, design flexibility, and reagent economy \cite{Nature}. Synthetic biology lies at the intersection of engineering, biological sciences, and computational modeling. It borrows tools and concepts from these disciplines to engineer non-existing biological systems or to redesign existing systems to achieve user-defined properties \cite{Polizzi2013}. Over the past few decades, microfluidics and synthetic biology have proven their potential as tools that offer unprecedented solutions for biosensing and therapeutics.

\subsection{Biosensing}
\label{sec_biosensing}
Biosensors are devices used to detect the presence of chemical substances. A biosensor is normally composed of a bioelement and a transducer \cite{Kumar2010}. The bioelement enables microscale detection by binding an analyte of interest, and the transducer modulates the variation of the analyte to an electrical or optical signal that can be observed at the macroscale. This application maps to Level 5 of our proposed hierarchy because communication occurs when we observe information encoded by the transducer, when the transducer receives information from the bioelement, and possibly also when biosensors communicate with each other. Biosensing plays a significant role in our daily life, and has been applied to many fields from  disease monitoring to  pollutant detection.

\subsubsection{Microfluidics}
Conventional biosensing methods are usually time-consuming and the corresponding equipment is big and expensive. There is a need for biosensors with faster analysis, higher cost-effectiveness, and smaller size \cite{Kwakye2003,Mairhofer2009}. Microfluidic platforms have become realizable to meet the above requirements.  
\begin{figure}[!t]
	\centering
	\subfloat[Schematic side view of the microfluidic sensing chip.]{\includegraphics[width=3.5in]{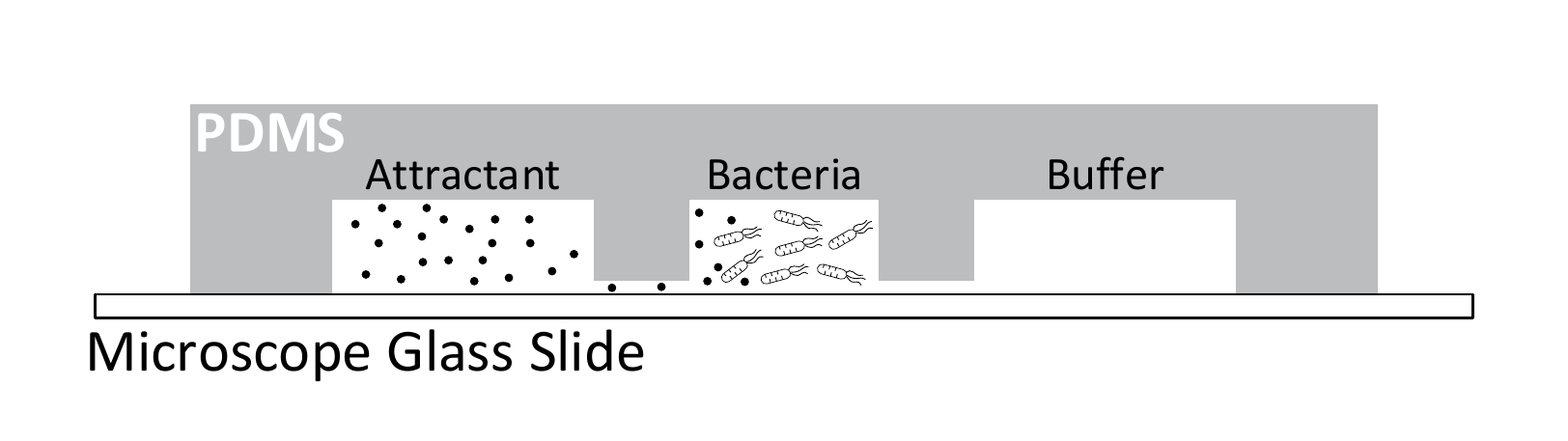}}\label{f_cs1}
	\hfil
	\subfloat[Schematic top view of the microfluidic sensing chip.]{\includegraphics[width=3.3in]{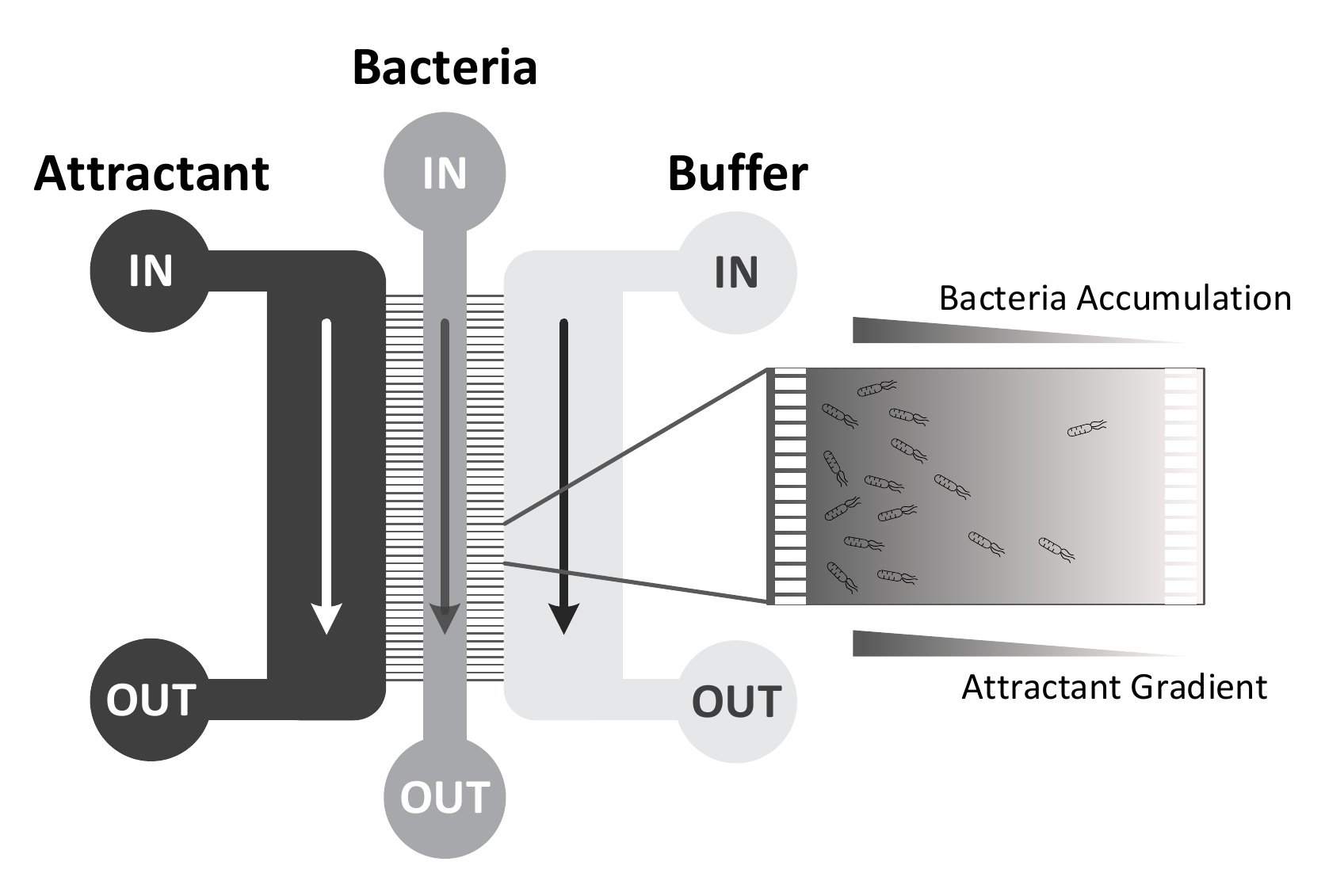}}\label{f_cs2}
	\caption{Illustration of the microfluidic bacterial chemotaxis biosensing system \cite{Roggo2018}.}
	\label{f_cs}
\end{figure}
The authors of \cite{Roggo2018} proposed a chemical biosensing microfluidic chip based on bacterial chemotaxis. As shown in Fig. \ref{f_cs}, the microfluidic chip\footnote{It is noted that the microfluidic architecture in Fig. \ref{f_cs} can also be used for biomedical research by including chemical reactions. One example is to emulate the scenario of oxygen-glucose deprivation to study stroke \cite{Chen2011}.} consists of one middle channel and two side channels. 
The bacteria are introduced in the middle channel, and the \textit{\textbf{flowing}} (Level 1) buffer with and without attractant (`source' and `sink', respectively) are injected into the side channels. The connection between the middle channel and the side channels only enables the \textit{\textbf{diffusion}} (Level 1) of attractant molecules, and this can create a microscale concentration gradient. As a response, the bacteria bias their motion towards the attractant using \textit{\textbf{chemotaxis}} (i.e., cargo-based transport guided by molecule gradients; presented for Level 1). The 
signaling attractant can also activate the \textit{\textbf{expression of a fluorescent gene}} (Level 2) embedded in bacterial cells so that the chemotactic intensity (i.e., the spatial distribution of bacterial cells) can be visualized using \textbf{\textit{fluorescent optical microscopy}} (Level 3). If a liquid sample that is taken from a natural environment (e.g., a river) is injected into the attractant side channel, then the fluorescent intensity provides a tool for estimating the cells' living conditions (i.e., \textbf{\textit{information in cellular signals}}, presented for Level 4). A higher attractant concentration leads to a stronger fluorescent intensity. The integration of chemotactic sensing in microfluidic chips enables rapid and quantitative sensing readouts, and the miniaturization of sensing devices also significantly reduces the power and reagent consumption. 

Microfluidic devices can also lower the cost and time of DNA detection. The authors of \cite{Reboud2019} demonstrated a paper-based microfluidic device that combined DNA extraction, amplification, and antibody-based detection of highly specific DNA sequences associated with malaria infection. The device was able to produce results in the field with high sensitivity ($>$98$\%$) in less than one hour. In addition, the manufacturing was simple and cost effective, enabling production of the device in great numbers. 

\subsubsection{Synthetic Biology}
\label{biosensing_synthetic_biology}
The selectivity of a biosensor describes its ability to distinguish targeted molecules among other similar chemicals. A biosensor with low selectivity can be activated by targets with similar chemical properties. For example, this is a concern when detecting toxic heavy metals for water pollutant monitoring \cite{Wang2013}. Examples of nonspecific metal biosensors include the triggering of the regulator CadC in \textit{S. aureus} by cadmium, lead, and zinc, the regulator CmtR in \textit{Mycobacterium tuberculosis} by cadmium and lead, and the regulator ArsR in \textit{E. coli} by arsenic, antimony, and bismuth \cite{Saha2017}.

From a communication engineering perspective, a general solution to increasing the selectivity is to endow biosensors with more signal processing capabilities, and the  engineered \textit{\textbf{digital synthetic biological circuits}} reviewed for Level 4 can be  applied  to build biosensors with increased selectivity \cite{Wang2013}. 
\begin{figure}[!t]
	\centering
	\includegraphics[width=3.55in]{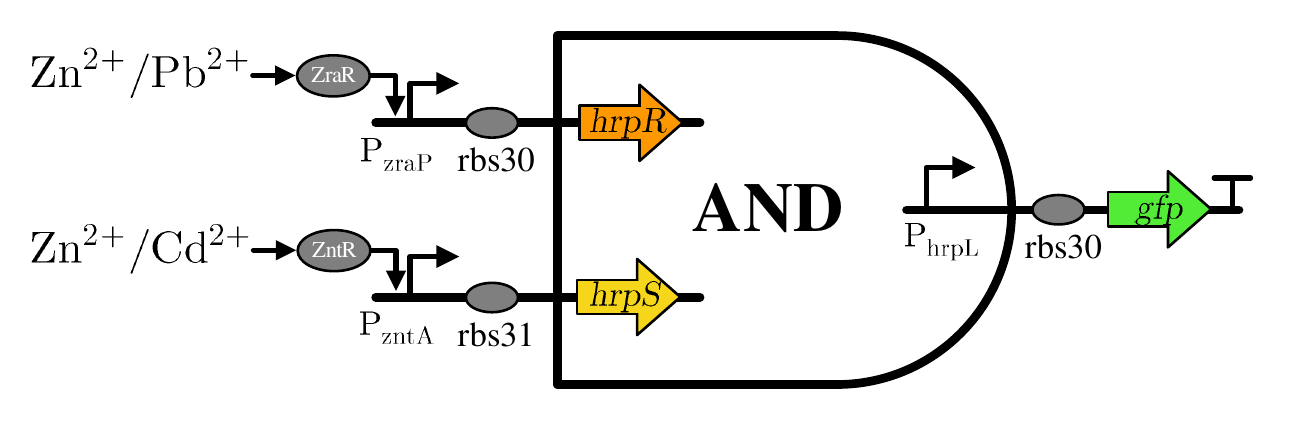}
	\caption{Zn$^{2+}$ specific biosensor using an engineered AND logic gate, where P$_\text{zraP}$ and P$_\text{zntA}$ are cognate promoters for ZraR and ZntR, rbs$30$ and rbs$31$ are both ribosome binding sites, P$_\text{hrpL}$ is another promoter that is activated 
	when the genes \textit{hrpR} and \textit{hrpS} are both expressed, and \textit{gfp} is the gene encoding a green fluorescent protein that works as a biosensor readout \cite{Wang2013}.}
	\label{AND}
\end{figure}
Fig.~\ref{AND} shows the schematic design of a biosensor that is only sensitive to zinc (Zn$^{2+}$) but not to palladium (Pd$^{2+}$) or cadmium (Cd$^{2+}$).
The sensor senses the targeted metal ions that \textbf{\textit{diffuse}} (Level 1) in the extracellular environment and can phosphorylate their respective regulators (i.e., ZraR for Zn$^{2+}$ and Pd$^{2+}$, and ZntR for Zn$^{2+}$ and Cd$^{2+}$). The transcription factors ZraR and ZntR regulate the \textbf{\textit{gene expression}} (Level 2) of \textit{hrpR} and \textit{hrpS}, respectively, and the protein products become the inputs of the engineered AND logic gate. At the AND gate, the expression of gene \textit{gfp} is activated only when both \textit{hrpR} and \textit{hrpS} are expressed. In this way, the readout green fluorescent protein is driven by a single bit of information, i.e., the presence or absence of Zn$^{2+}$ (\textbf{\textit{information in cellular signals}}, presented for Level 4), and can be observed via \textit{\textbf{optical microscopy}} (Level 3).

\subsection{Therapeutics}
\label{sec_therapeutics}
Therapeutics is a discipline developed to treat and care for a patient  with the purpose of preventing and combating diseases or alleviating pain. Drug delivery systems play an important role in therapeutics by controlling  the release and adsorption of pharmaceutical compounds to achieve desired therapeutic effects. They map to Level 5 of our proposed hierarchy because communication occurs when we observe changes at the disease site (e.g., reduction in tumor size), when therapeutic agents bind with their receptors at diseased cells, and when drug delivery devices receive information from the extracellular environment or body-area stimuli.
In the following, we present some therapeutic drug delivery methods powered by microfluidic platforms, synthetic biology, and magnetic fields, which show improved therapeutic efficiency compared with conventional systems, such as oral ingestion and intravascular injection.

\subsubsection{Microfluidics}
Microfluidic systems are beneficial for novel drug delivery applications by  improving drug delivery accuracy and reliability at reduced size \cite{Nguyen2013,Park2010,Gensler2012}. A manually-actuated drug delivery device for the treatment of chronic eye diseases was developed in \cite{Lo2009}.

\begin{figure}[!t]
	\centering
	\includegraphics[width=3.4in]{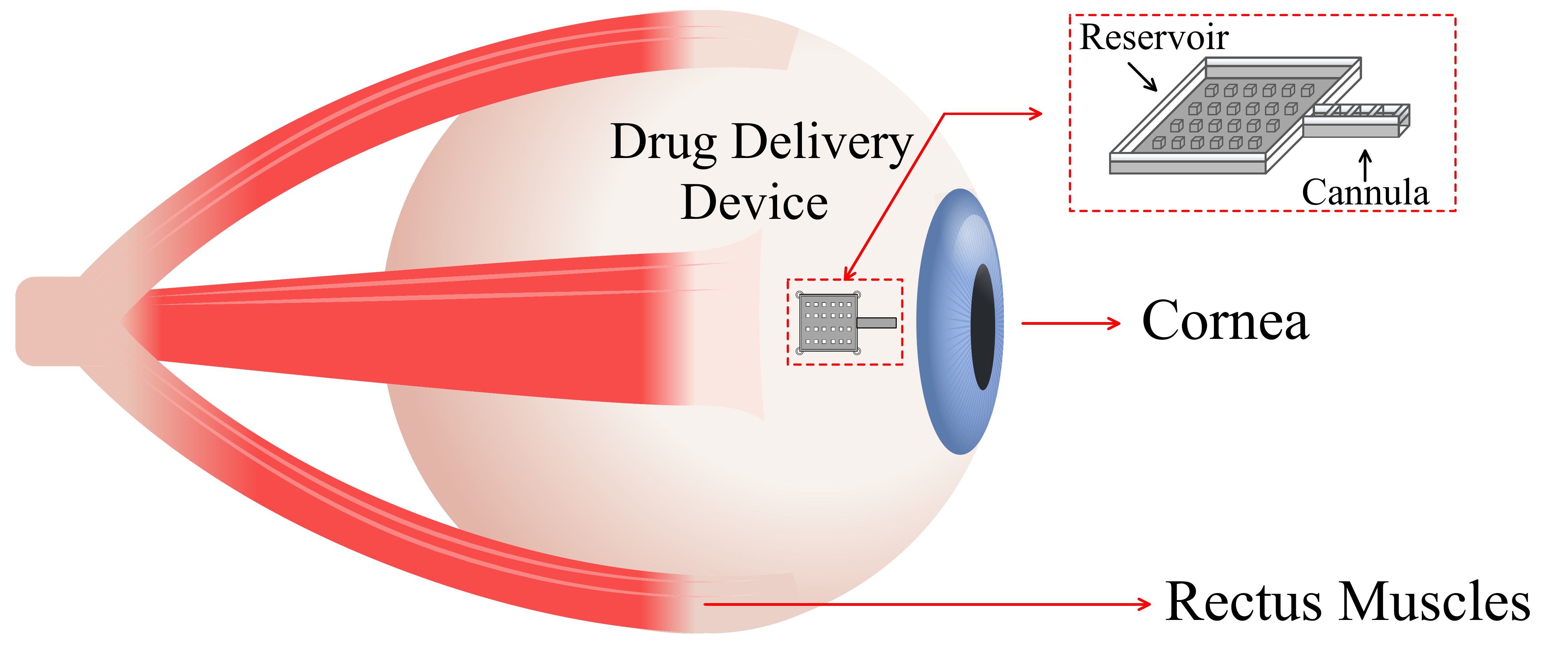}
	\caption{Schematic of a manually-actuated drug delivery device for chronic eye diseases \cite{Lo2009}.}
	\label{f5}
\end{figure}

As shown in Fig. \ref{f5}, benefiting from microfluidic systems, the drug delivery device was miniaturized to allow its placement in the limited space within the eye. The device works using \textit{\textbf{macroscale mechanical control}} (Level 3). More specifically, a pressure force, mechanically actuated by a patient's fingers, can induce the \textit{\textbf{advection-diffusion-based propagation}} (Level 1) of phenylephrine drugs from the reservoir to intraocular tissues. The released phenylephrine molecules undergo \textit{\textbf{ligand-receptor binding}} (Level 2) to adrenergic receptors and finally lead to a temporal change in pupil size, which can be measured via \textit{\textbf{in vivo imaging}} (Level 3). The observation of pupil size changes not only demonstrates successful drug injection, but also indicates that the observed output of pupil size can be manipulated by the phenylephrine concentration (i.e., \textbf{\textit{information in cellular signals}}, presented for Level 4). This device is refillable, such that only one surgical intervention and the associated pain is needed. In contrast, due to the presence of the blood-retina barrier, conventional oral medications require large doses in order to reach therapeutic levels and can have serious negative side effects. Traditional intraocular injections for chronic diseases require frequent injections, which can induce trauma in ocular tissues \cite{Lo2009}.

\subsubsection{Synthetic Biology}
Designing and engineering biological parts via synthetic biology has enabled novel therapeutic platforms to target specific pathogenic agents and pathological pathways \cite{Khalil}. Cancer involves abnormal cell growth and proliferation with the potential to invade nearby healthy tissue and  spread to other organs. 
A significant shortcoming of current cancer therapies is that cancerous cells are difficult to distinguish and remove from surrounding healthy cells. 

One potential solution is to synthetically link the invasin (\textit{inv}) gene (from \textit{Yersinia pseudotuberculosis}) with the \textit{fdhF} promoter. The reason for this synthesis is that tumor microenvironments are low in oxygen and the \textit{fdhF} promoter is strongly expressed in such an environment. Thus, the invasin proteins become controlled to only effectively express in the oxygen-deprived environment   (\textit{\textbf{gene expression}} described for Level 2). 
The invasins \textit{\textbf{diffuse}} (Level 1) within the cellular medium and can bind with the $\beta 1$-integrins distributed on the surface of cancer cells (\textit{\textbf{molecule reception}} described for Level 2), which triggers the internalization of bacteria inside cancer cells. The invasion ability can be quantified and observed by \textit{\textbf{macroscale instruments}} (Level 3) after a gentamicin protection assay. Furthermore, the authors of \cite{Anderson2006} also synthetically linked the invasion of cancer cells to bacteria density, which is achieved by placing the \textit{inv} gene under the control of a \textit{lux} quorum sensing system (we describe quorum sensing in further detail in Section \ref{sec_quorum}). Hence, bacteria invasion of cancer cells is driven by two bits of information, i.e., the oxygen level and the bacteria density (\textbf{\textit{information in cellular signals}}, presented for Level 4). This can be interpreted as an application of the \textit{\textbf{genetic AND gate}} (Level 4) that integrates multiple inputs to achieve more accurate environmental sensing. This characteristic makes invading cells ideal carriers to release therapeutic agents to enhance tumor treatment.

\subsubsection{Magnetic Field}
To prevent drug absorption or degradation before reaching the affected target sites, one efficient approach is to place the drug as close as possible to the target sites. It has been demonstrated that accurately manipulating magnetic microrobots via magnetic fields is feasible, and the human body is `transparent' to magnetic fields (i.e., in terms of biocompatibility and safety). Motivated by this, the use of magnetic microrobots for drug delivery through \textit{\textbf{macroscale magnetic control}} (discussed for Level 3) has been widely studied and applied \cite{Yesin2006,Dogangil2008,Fusco2014}. In \cite{Ullrich2013}, a microrobot was injected into the posterior area of a rabbit eye. Once an external magnetic field was applied, the injected microrobot could achieve rotational and translational mobility, thus presenting an opportunity for ocular drug delivery. In \cite{Tietze2013}, mitoxantrone-loaded magnetic nanoparticles were injected into the femoral artery. By applying an external magnetic field above the tumors implanted in the limb of rabbits, nanoparticles moved towards the tumor region via \textit{\textbf{advection-diffusion-based propagation}} (Level 1). A higher accumulation of mitoxantrone was found near the tumor region, and a clear reduction in tumor size could be observed through \textit{\textbf{in vivo imaging}} (Level 3) as mitoxantrone is able to \textbf{\textit{bind}} with the DNA of tumor cells, thus halting tumor growth and division (\textbf{\textit{molecule reception and responses}} described for Level 2). Therefore, cancer cell differentiation is controlled via the reception of mitoxantrone signals (\textbf{\textit{information in cellular signals}}, presented for Level 4).

\subsection{MC-Assisted  Applications}
\label{MC_assisted_application}
In the following, we envision an MC-enabled automatic drug delivery system, with the aim to illustrate how MC could facilitate and enhance the aforementioned biosensing and  therapeutics applications. An automatic drug delivery system largely reduces the dependency on manual operations and should be composed of a biosensor and an actuator. The biosensor senses the extracellular environment (e.g., the concentration of glucose), and could be connected with the actuator including drug reservoirs to cooperatively support the drug regulating mechanism. Nevertheless, the biosensor and the actuator, such as the microfluidic biosensor in \cite{Roggo2018} and the microfluidic actuator in \cite{Lo2009}, are often designed separately by researchers from different fields and are likely to be physically isolated. Thus, the communication between a biosensor and an actuator is of great importance because it is the only feature that enables them to work in a synchronous and cooperative manner to reach a common goal. To address this issue, MC can be used to establish a point-to-point communication link between the two of them. In this scenario, the biosensor would serve as a transmitter, and the actuator would function as the corresponding receiver. Once the biosensor detects a relevant phenomenon, it modulates this information to a chemical signal that can be received and demodulated by the actuator. As a response, the actuator releases drug molecules to a specific area of cells.

On some occasions, the actuator may be controlled by more than one bit of information (i.e., the presence or absence of a phenomenon), implying the involvement of \textit{multiple} biosensors. In this sense, the signal processing capability of an actuator should be expanded accordingly to manipulate signals received from multiple biosensors. One example is the introduction of digital logic gates, as in \cite{Wang2013} and \cite{Anderson2006}, to control the drug-regulating mechanism. Moreover, this envisioned drug delivery system can be further optimized by integrating other MC-based concepts. For example, the implementation of coding functions at Level 4 could be added to mitigate the effects of noise, thus providing a more reliable and robust communication link.

\section{End-to-End Case Studies}
\label{sec_case}

From Sections~\ref{sec_physical} to \ref{sec_applications}, we individually discussed and presented examples for each of the five levels of the proposed communication hierarchy. While we drew connections between the levels, we did not directly apply the entire hierarchy to any one example. In this section, we have selected several prominent exemplary biological systems as case studies for a complete mapping to the proposed hierarchy, as summarized in Table~\ref{table_case_studies}. In particular, we present quorum sensing by bacteria (Section~\ref{sec_quorum}), signaling within and between neurons (Section~\ref{sec_neuron}), and information encoding in DNA (Section~\ref{sec_DNA}). Quorum sensing is an example that aligns closely with diffusion-based MC. Neuron signaling includes a mix of diffusion-based and action potential wave propagation. While the propagation of DNA information has been less of a focus of study in the MC community, its implementation of the higher levels is widely known and well understood and so it provides a useful supplement. While all three of these case studies can map to the entire hierarchy as natural systems, they  also demonstrate opportunities for synthetic interactions including control.

\begin{table*}[!t]
	\centering
	\caption{Case Study Summary.}
\resizebox{\textwidth}{!}{%
	{\renewcommand{\arraystretch}{1.4}
		\begin{tabular}{p{0.1\textwidth}||p{0.18\textwidth}|p{0.18\textwidth}|p{0.18\textwidth}|p{0.18\textwidth}|p{0.18\textwidth}}
			\hline
			\thead{Case Study\\ Name \\ (Subsection)} & \thead{Section \ref{sec_physical}: Level 1\\ Signal Propagation} & \thead{Section \ref{sec_interaction}: Level 2\\ Device Interface} & \thead{Section \ref{sec_interface}: Level 3\\ Physical/Data Interface} & \thead{Section \ref{sec_data}: Level 4\\ Local Data} & \thead{Section \ref{sec_applications}: Level 5\\ Application} \\ \hline \hline
			Quorum Sensing (A)
			& Diffusion of autoinducer molecules
			& Release, capture, and detection of autoinducers
			& Measuring threshold concentration(s) (e.g., high versus low)
			& Behavior based on estimate of local population
			& Bacteria cooperation and coordination; eavesdropping and surveillance\\ \hline
			Neuronal Signaling (B)
			& Action potential along neural axon; neurotransmitters across chemical synapse &
			Membrane potential changes due to ion channels; release and capture of neurotransmitters
			& Spike timing and frequency
			& Messages to transmit nerve stimuli and motor actions
			& Functioning of nervous system \\ \hline
			Communication via DNA (C)
			& Storage in genome
			& Transcription and translation of genes; replication of DNA
			& Controlling gene regulation; modulation with 4 bases (A, C, G, T for DNA; A, C, G, U for RNA)
			& Encoding of amino acids for proteins
			& Life (e.g., cell growth, division, differentiation)\\ \hline
		\end{tabular}
	}}
	\label{table_case_studies}
\end{table*}

\subsection{Quorum Sensing}
\label{sec_quorum}

\begin{figure}[!t]
	\centering
	\subfloat[][]{\includegraphics[width=0.33\linewidth]{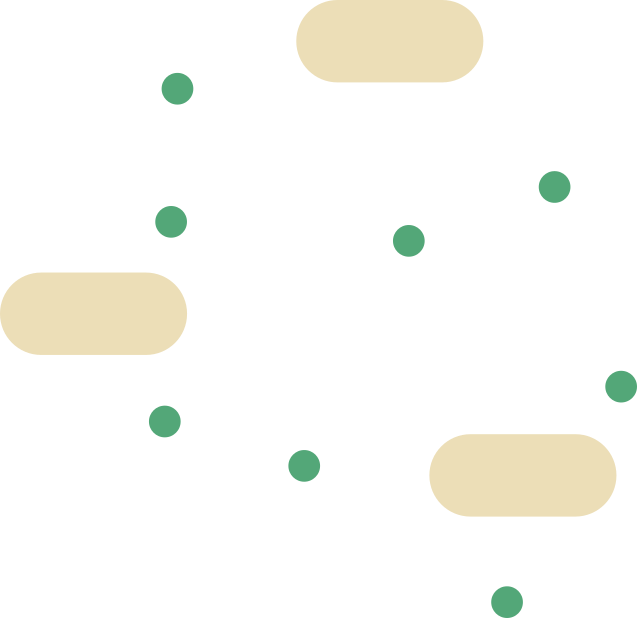}}%
	\hspace{0.05\textwidth}%
	\subfloat[][]{\includegraphics[width=0.33\linewidth]{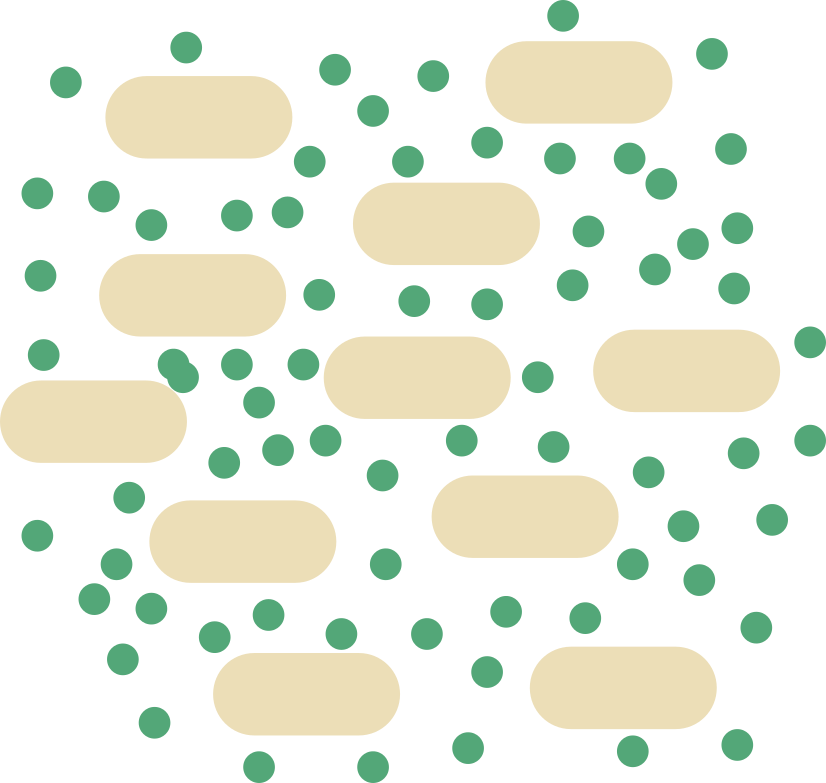}}\\
	\subfloat[][]{\includegraphics[width=0.7\linewidth]{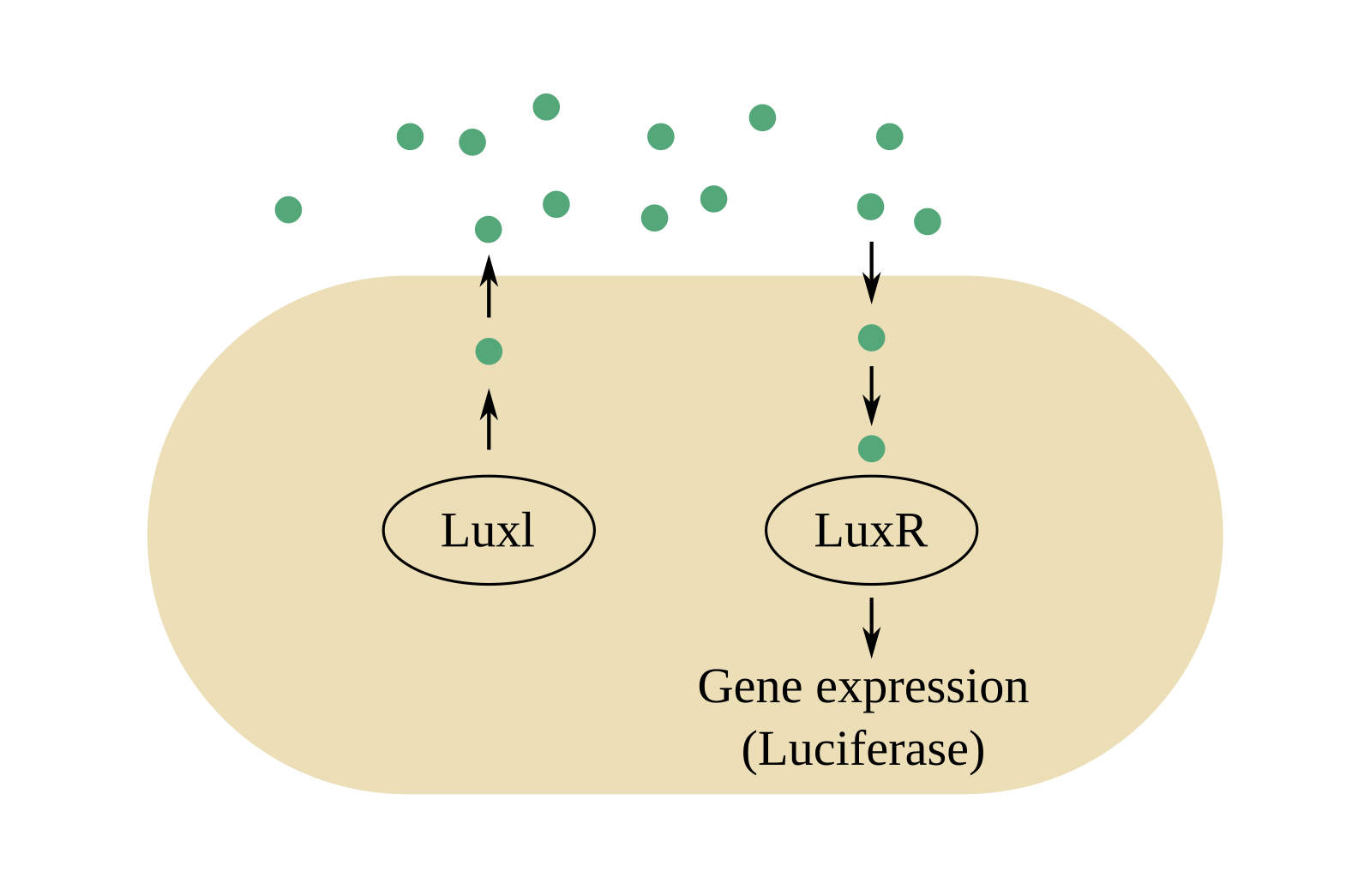}}
	\caption{Simplistic representation of quorum sensing in \textit{Vibrio fischeri}. Autoinducers (in green) produced by LuxI gene are excreted by the cell and accumulate in the environment. (a) At low cell densities, autoinducer concentration is also low. (b) Higher cell densities cause an accumulation of molecules. (c) Accumulated molecules can be sensed by the cell and activate the Luciferase genes that induce bioluminescence.}
	\label{QS}
\end{figure}

The classical view of bacteria depicts them as individual organisms that act independently as isolated entities. While this is true to the extent that a bacterium is a distinct autonomous cell, we have known for a few decades that bacteria can form groups comprised of many individuals that have been shown to exhibit coordinated behavior \cite{Bassler2006}. This includes bioluminescence (one of the first collective microbial behaviors to be characterized) \cite{Ruby1996}, biofilm formation \cite{Hammer2003}, production of virulence factors and secondary metabolites \cite{DeKievit2000}, and induction of competence for foreign DNA uptake \cite{Kleerebezem1997}. These processes are made possible by communication between bacteria via a process termed quorum sensing (QS). A recent review of QS can be found in \cite{Mukherjee2019b} and a visual summary is provided in Fig.~\ref{QS}. QS relies on the exchange of small extracellular signaling molecules called \textit{autoinducers}. Exchanging signals in this way enables bacteria to assimilate information conveyed by different types of autoinducers to control specific genes. This enables communication between the same and distinct species and even between bacteria and animal cells \cite{Papenfort2016}.

QS is very energy efficient as a communication system. Signaling molecules are based on intermediates that have a key role in the central metabolism \cite{Brameyer2015a,Fuchs2013}. Thus, investment in a specialized production chain is not required and the high affinity and selectivity of the molecules means that a very small amount is sufficient for effective communication, resulting in a very small production cost. Cost effectiveness is further improved by the fact that many QS molecules can serve multiple purposes. Examples include photopyrones, a small QS molecule in \textit{Photorhabdus luminescens} that in high concentrations can act as an insect toxin \cite{Brachmann2013}, or dialkylresorcinols (DARs) that can act as an antibiotic \cite{Joyce2008}.

Understanding biological systems is inherently complicated, with many components often serving multiple functions in highly-interconnected networks that make separation of functionality into layers a challenging task, and QS is no different. From the perspective of our proposed hierarchy, Level 1 and Level 2 are the diffusion of molecules and the mechanisms for the release and reception of autoinducers, respectively, as listed in Table~\ref{table_case_studies}. Subsequent levels are less intuitively defined. We propose concentration threshold detection as Level 3, and functions such as estimation of cell density and individuals switching behavior as Level 4. Level 5 describes features that emerge across the bacteria population using QS (e.g., coordinated behavior, cooperation, eavesdropping).

\subsubsection{Level 1} In QS communication, Level 1 is the random \textbf{\textit{diffusion}} of the autoinducers in the environment. Diffusion has been summarized mathematically in Sections~\ref{Diffusion} to~\ref{Convection-Diffusion-Reaction} of this survey. Bacteria that have formed a biofilm merit separate discussion, as fluid flow is non-existent or very restricted inside their extracellular matrices \cite{Stoodley1994}. High cell densities inside a biofilm also have a significant effect on both the diffusion distance of a molecule, and the speed at which diffusion occurs. Experimental and theoretical studies have determined the reduction of the diffusion coefficient within biofilms to be between 0.2 and 0.8 when compared with diffusion in water \cite{Stewart2003}. The distance that a molecule can cover while diffusing through a biofilm is effectively given by the dimensions of the biofilm cluster \cite{Stewart2003}.

\subsubsection{Level 2} Level 2 concerns the mechanisms of release and reception of autoinducers. QS signaling molecules differ between bacterial types. Gram-negative bacteria typically use acyl-homoserine lactones (AHL) as autoinducers. AHLs are small molecules that can freely diffuse through membranes. The system depends on two proteins, LuxI and LuxR (see Fig.~\ref{QS}(c)). LuxI helps in the synthesis of the autoinducer N-3-(oxo-hexanoyl)-homoserine lactone (30CC6HSL) family of proteins \cite{Engebrecht1984, Schaefer1996}. To describe the system here, we refer to individual proteins of a specific QS system (\textit{V. fischeri}), although we clarify that LuxI, LuxR, and 30CC6HSL are members of protein families found among all gram-negative bacteria, with each bacterial species carrying their own version. After the AHL is synthesized, it diffuses freely through the cell membrane in both directions, and its concentration rises as the microbial population increases \cite{Ng2009} (see Figs.~\ref{QS}(a), \ref{QS}(b)). LuxR is the receptor for 30CC6HSL in the cytosol, as well as the transcriptional activator of the luciferase operon \cite{Engebrecht1983, Engebrecht1984}. The role of 30CC6HSL is to stabilize LuxR (otherwise LuxR naturally degrades rapidly) and enable it to persist long enough to recognize and bind to a consensus sequence encoding for luciferase and accessory proteins \cite{Stevens1994, Meighen1991} (see also the general discussion on gene expression in Section~\ref{sec_gene_expression}). Stabilized LuxR also activates LuxI in a feed-forward loop so, when the QS system is engaged, production of autoinducers accelerates and the environment is filled with the signal molecule.

Gram-positive bacteria have a different membrane structure that is impermeable to AHLs. They instead use oligopeptides (also referred to as autoinducing peptides) as autoinducers and these are actively transported to and from the cell surface. Reception of these molecules is based on cell surface-bound receptors collectively termed \textit{two-component signaling proteins} that upon activation set up a series of reactions inside the cell (i.e., a \textit{signal cascade}) \cite{Ng2009}.

Recent findings suggest that for all types of bacteria, larger molecules such as hydrophobic AHLs that cannot pass through the membrane are instead being released in the environment via membrane \textbf{\textit{vesicles}} \cite{Mashburn-Warren2006,Morinaga2018}. The size range of these vesicles appears to be between 40 and 500\,nm \cite{Schwechheimer2015,Brown2015}. In addition to QS, these vesicles have been shown to be involved in horizontal gene transfer for the exchange of virulence factors. A review of QS vesicles can be found in \cite{Toyofuku2019}.

\subsubsection{Level 3}
In QS circuits, the default mode is the continuous expression of the gene that encodes the required behavior (e.g., luminescence, virulence). In a typical arrangement for biological systems, this gene expression is suppressed by the rapid degradation of the mRNA molecules transferring the information for the corresponding protein production. Thus, in low cell densities the product cannot be synthesized as the mRNA is not produced or is readily degraded. When autoinducers in the environment reach a critical (i.e., threshold) concentration, this suppression ceases and the mRNA are able to reach their target destination \cite{Waters2005}. Thus, the desired behavior of a QS system is digital, in the sense that QS acts as a sensing mechanism that regulates the transition from one behavior to another. By implementing a threshold mechanism, a cell is able to quantify autoinducer concentration and translate it into information to infer cell density and the presence of other species.

Function in biological systems is usually tied to physical structure, so in this discussion of threshold measurements it is also appropriate to outline the mechanisms that implement these measurements, as the way each system is implemented mirrors its behavior and suggests how it might be controlled. QS systems differ in the arrangement of their internal components, reflecting different needs in their implementation of signal quantification. For example, there are QS systems containing circuits that act in parallel or in series, others with different system components acting in opposition, and also systems that upon activation confer a permanent change to the organism \cite{Waters2005,Ng2009}. In the following, we summarize parallel and series implementations to emphasize the diversity of \textbf{\textit{microscale signal operations}} in QS.

In a parallel QS architecture, it is typical to implement more than one autoinducer in separate signal transduction cascades. Because they all have the same result (e.g., suppression of the mRNA suppressor), their signals reinforce one another. In addition, the need for the simultaneous presence of two (or more) signals for the activation of the system ensures that specific requirements are met (e.g., availability of nutrients, presence of another species), not unlike an AND gate. This parallel sensing approach might be helpful for noise reduction and to filter foreign signal-mimicking molecules \cite{Waters2005}.

Series QS circuits differ in the sense that activation of one circuit is required for the activation of the subsequent circuits. This is the mechanism \textit{P. aeruginosa} employs for virulence. Experiments have shown that unlike what happens in a  parallel system, some genes in these circuits may be expressed in response to one autoinducer only, while others respond to any of the legitimate signals, and yet others require the simultaneous presence of all signals for their expression. In addition, their activation occurs at different times during cell growth, an indication that timely ordered gene expression is very important for these organisms \cite{Schuster2003,Waters2005}.

A number of techniques are currently used to observe QS behavior, depending on the nature of the expected microbial response to QS signals. Microbial antibiotic assays can be employed for the detection of antimicrobial agents secreted by bacteria in response to the presence of other organisms. \textbf{\textit{Fluorescence microscopy}} coupled with microfluidics is a suitable tool for the observation of gene expression at the level of an individual cell \cite{Carcamo-Oyarce2015,Grote2015}. Detection and real-time tracking of autoinducers has been achieved using bacterial reporter strains \cite{Yong2009}, high performance liquid chromatography \cite{Thiel2009}, and nanosensors \cite{Zhang2014}.

\subsubsection{Level 4}
Autoinducer release, random diffusion outside the cell, and subsequent detection provide bacteria with a tool for the estimation of presence and density of microorganism populations around them, such that behavior can update once threshold conditions are observed. The QS processes enable precise regulation of a large number of genes and the fine-tuning of responses, e.g., input-output range and dynamic behavior, synchronization, and noise control \cite{Novick1995, Seed1995, Svenningsen2009,Feng2015}. In large microbial populations (particularly in biofilms), there are inevitable variations in each individual cell's local environment due to differential access to resources, accumulation of metabolic byproducts in pockets, and oxygen penetration. Thus, the QS behavior of a bacterial population is not necessarily homogeneous. For example, in a biofilm, oxygen penetration is slow, creating a gradient from the outside to the center. The cells at the periphery then sense a completely different environment than those further inside the matrix, leading to variable individual responses that are essential to maintaining the biolfilm. Recent evidence also suggest that some microbial populations can exhibit a stochastic expression of QS genes, resulting in segments of the population being in different QS modes \cite{Carcamo-Oyarce2015,Grote2015}.

\subsubsection{Level 5}
The highest level in the hierarchy corresponds to aggregate behavior, e.g., QS stimulating coordination between bacteria. This is of great interest to the scientific community as the results of coordinated microbial actions have major economic importance (e.g., biofilms, virulence, biofouling). For example, model estimates in 2013 predicted the economic burden of antibiotic resistance on global GDP to be between US\$14~billion and US\$3~trillion by 2050 \cite{Naylor2018}. Naval biofouling by barnacles is initiated by microbial mats that enable barnacles to attach to a ship hull. This adds a considerable annual cost that can reach a few million US\$ \textit{per vessel} due to the subsequent increase in drag \cite{Schultz2011}. These processes all rely on the coordination between microorganisms realized through the exchange of molecular signals. As an example of inter-species cooperation, consider the QS system used by the gram-negative bacterium \textit{Vibrio fischeri}. It is the canonical example of a QS system in gram-negative bacteria and was also the first to be described during an investigation of bioluminescence in the Hawaian Bobtail squid \textit{Eupryma scolopes} \cite{Ruby1996}. Favorable conditions inside a specific organ of the squid allow \textit{V. fischeri} to reach high cell densities, and through the activation of a QS system to induce the expression of the luciferase operon. The light produced benefits the squid host by providing protection from predators \cite{Ruby1996}.

There are several interesting communication security applications and problems that can be observed in QS systems.
Since the autoinducers released can be unique for each species using QS, different bacteria can send signals that only individuals of the same species will detect. This can establish a secure communication channel, although it can be compromised when other species are able to detect the same signals (i.e., ``eavesdropping'') \cite{Rebolleda-Gomez2019}. For example, the soil bacterium \textit{Myxococcus xanthus} (\textit{M. xanthus}) is a predatory species that actively seeks other bacteria as prey. \textit{M. xanthus} is able to detect a range of QS molecules used by different gram-negative bacteria, which enable it to infer the presence and direction of many species \cite{Lloyd2017}. Some QS signals can be deliberately detected by a number of different species to enable inter-species communication. This function in gram-negative bacteria relies on variations in the structure of the autoinducer molecules. Gram-positive bacteria exert more control on the final structure of the peptides used as signal molecules, as these are DNA-encoded, resulting in a unique genetic sequence for each organism \cite{Ng2009}.

Microbes have the ability to attach to surfaces and form biofilms \cite{Flemming2019}. Examples include plaque in teeth and rock-coating slime in water. It has become increasingly apparent over the last couple of decades that biofilm communities are the predominant form of microbial life and that they are of great importance to medicine, industry, and the environment \cite{Donlan2002, Kolter2006, Pintelon2012}. Biolfilms help microbes to engage in symbiosis with other species, avoid predators, and be shielded from antibiotic compounds. Biofilms are also very dynamic; they can have significant heterogeneity within a population and also change behavior depending on the conditions (e.g., nutrient availability) \cite{Stewart2008,Grote2015}. In the canonical example of \textit{V. fischeri}, it is reported that although QS signals are flooding the environment, there can be a significant variation in the level of bioluminescence between individual microbes \cite{Perez2010}. QS plays a key role in biofilm formation, and bacterial species can have diverse biofilm-forming strategies. For example, \textit{Pseudomonas aeruginosa} creates biofilms when cell density is high, while \textit{Vibrio cholerae} and \textit{Staphylococcus aureus} form biofilms at low cell density \cite{DeKievit2000, Bronesky2016}. In the latter cases, autoinducer accumulation \textit{suppresses} the excretion of biofilm molecules.

Another example of physiological activity regulated by QS is the releasing of virulence factors to destroy tissues in target host cells during the initiation of an infection \cite{Psarras}. The synthesis and secretion of virulence factors are expensive and they are needed in a large quantity to successfully attack a host. Thus, since autoinducer molecules are less expensive, QS is used to regulate the expression of virulence factors in bacteria so that they are produced only once the bacterial population density is sufficiently high.

\subsection{Neuronal Communication}
\label{sec_neuron}

\begin{figure}[!t]
	\centering
	\subfloat[][]{\includegraphics[width=0.28\linewidth]{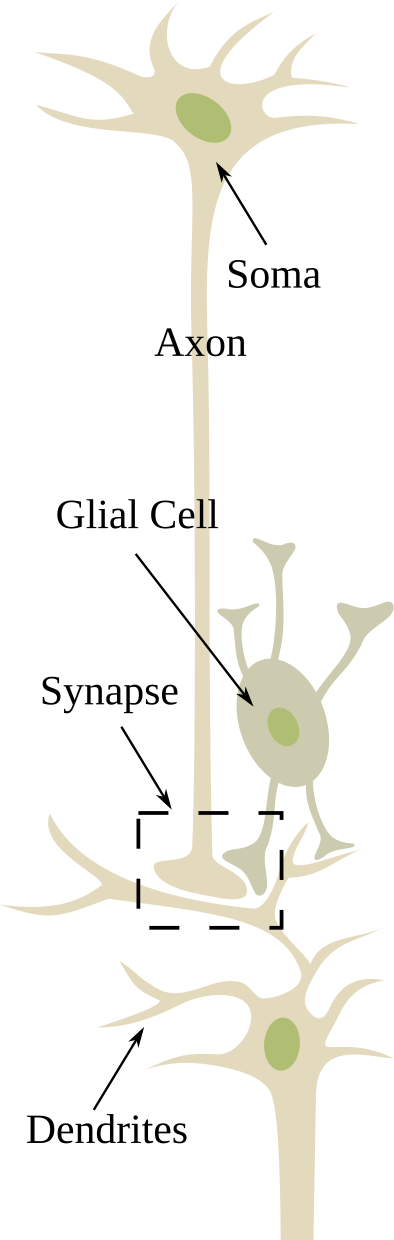}}%
	\subfloat[][]{\includegraphics[width=0.72\linewidth]{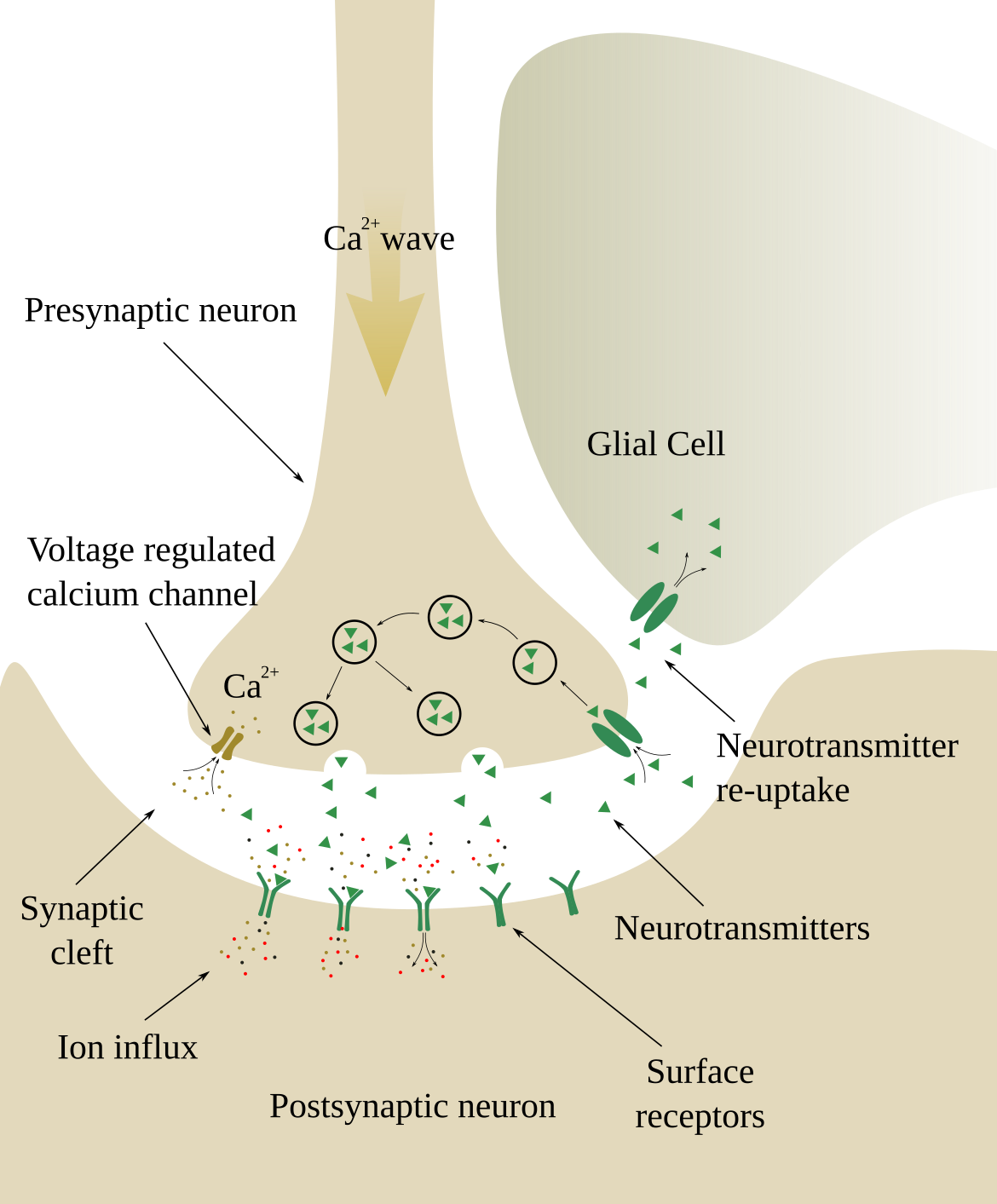}}
	\caption{Schematic drawing of a neuronal synapse and neuronal chemical signaling. (a) Main parts of a neuronal cell, showing the connection of an axon to the dendrites of another neuron. (b) A calcium wave propagating along the axon triggers the opening of calcium channels in the pre-synaptic neuron's outer membrane. Increase in Ca$^{2+}$ concentration causes the brief release of neurotransmitter molecules, before their rapid re-absorption by both the presynaptic neuron and adjacent glial cells. Released ions bind to and activate receptors in the postsynaptic neuron, leading to an influx of ions into the second neuron, and to the propagation of the signal.}
	\label{NS}
\end{figure}

Neurons are important cells for storing and processing information in most animals. In order to swiftly carry information throughout the body, they require very rapid and reliable communication mechanisms. Neuron signaling is an interesting example of \textbf{\textit{ microscale signal propagation}} because the physical dimensions and performance requirements of neurons demand a diversity of propagation techniques both within and between neurons and other cells. Since neurons can be extremely elongated (see Fig.~\ref{NS}(a)), signaling \emph{within} neurons is as important as signaling \emph{between} neurons. Neurons have branching dendrites around the cell body (\textit{soma}) to receive inputs from other neurons, and usually one long \textit{axon} to signal outputs (at terminal branches) to distant targets (including other neurons). Axons in vertebrates are typically from less than $1\,$mm to more than $1\,$m in length \cite[Ch.~11]{Alberts2015}. The contact sites between neurons are known as \textit{synapses}. The most common modality are chemical synapses \cite[Ch.~11]{Alberts2015} (see Fig.~\ref{NS}(b)), which are uni-directional, though there are also bi-directional electrical synapses and there is evidence that chemical and electrical synapses functionally interact with each other \cite{Pereda2014}.

To apply the proposed communication hierarchy, we recognize the dichotomy of signal propagation both within and between neurons to identify two distinct implementations of Level 1 and Level 2 behavior. As summarized in Table~\ref{table_case_studies}, the propagation of an action potential spike along the axon carries information within a neuron, whereas neurotransmitter molecules diffuse across a chemical synapse to carry information between neurons (or other cells connected to neurons). However, these implementations merge at Level 3, since both types of signals carry information in the timing of action potential spikes. Level 4 concerns the information in neuronal signals and Level 5 describes where and why these signals are used.

We have already discussed neuronal communication in several instances throughout this work. We have briefly mentioned neurons as examples in the context of calcium signaling in Level 1 (Section~\ref{sec_physical}), storage and release of neurotransmitters in Level 2 (Section ~\ref{sec_interaction}), and the speed of synaptic responses in Level 3 (Section~\ref{sec_interface}). The reader may re-visit the details in those sections to supplement the holistic discussion of this case study.

\subsubsection{Level 1}

The signals within neurons are changes in the local electrical potential across a neuron's plasma membrane. Active mechanisms are needed to amplify signals in larger neurons so that the signals can propagate along the axon without attenuation. The active mechanisms create a traveling wave known as an action potential, which can propagate at speeds of $100\,\mathrm{m/s}$ or more. The primary components are voltage-gated ion channels (e.g., Na$^{+}$ and K$^+$), which open with positive feedback (to trigger the opening of neighboring channels) and then close with a refractory period (to prevent repetitive firing so that the wave travels along the axon). The first contributions to quantitatively model the propagation of action potentials in neurons were by Hodgkin and Huxley \cite{Hodgkin1952}, who treated the membrane as an electrical circuit with variable conductances due to the transfer of Na$^{2+}$, K$^+$, and other ions. Thus, the propagation of an action potential spike can be modeled as an electrical transmission line using the cable equation \cite[Ch.~17]{Phillips2013}.

The signals across chemical synapses, as shown in Fig.~\ref{NS}(b), are more similar to the \textbf{\textit{reaction-diffusion}} processes that we described for Level 1 in Section~\ref{sec_physical}. The synapse has the transmitting neuron at the pre-synaptic side and the receiving neuron (or other cell, e.g., muscle) at the post-synaptic side. While there is significant diversity in the components and precise function of chemical synapses, they generally signal by releasing neurotransmitter molecules that diffuse across the synapse to the post-synaptic neuron's outer membrane. There are many different types of neurotransmitters; common ones include acetylcholine, glutamate, serotonin, glycine, and $\gamma$-aminobutyric acid \cite[Ch.~11]{Alberts2015}. Since chemical synapses are quite narrow (only $10$--$30$\,nm wide \cite[Ch.~12]{Nelson2008}), this diffusion process is fast. Nevertheless, signaling pathways within the cleft provide mechanisms to destroy neurotransmitters, recycle them via re-uptake by the pre-synaptic neuron, or remove them via re-uptake by glial cells \cite[Ch.~11]{Alberts2015}. Cleansing the synapse of neurotransmitters helps to make the synapse available for future transmissions.

\subsubsection{Level 2}

For Level 1, we explained two distinct propagation mechanisms of neuronal signaling. For Level 2, we discuss the transitions between these two mechanisms. The propagation of an action potential along the axon is controlled by the opening of voltage-gated ion channels that are at the dendrites. These ion channels are opened by external signals that can include both biological and synthetic sources. If the transmitter is another neuron, and the environment between it and the receiver neuron is a chemical synapse, then the opening of the receiver's ion channels are controlled by the binding of neurotransmitters to the receptors \cite[Ch.~11]{Alberts2015}. The distribution of ion channels (e.g., Na$^{+}$ and K$^+$) in a membrane dictates how it reacts to the synaptic inputs; neurons can be tuned to their individual roles based on where the ion channels are expressed. For example, chemical synapses can be either excitatory or inhibitory (i.e., generate or suppress action potential firing in response to stimulus), depending on the ion channels present and the current ionic conditions. Generally, the overall firing rate of an excitatory neuron is proportional to the strength of the stimulus.

Action potentials travel along the axon until they reach the axon terminal. For terminals that are connected to cells via \textit{chemical} synapses, the arrival of an action potential triggers the fusion of synaptic \textbf{\textit{vesicles}} with the pre-synaptic neuron's outer membrane \cite[Ch.~12]{Nelson2008}, as we described for Level 2 in Section~\ref{sec_interaction}. This releases neurotransmitters into the synapse, triggering ion channel activity in the following cell and the cycle continues.

Due to the diversity of neuron and synapse structures, the relative timing of Levels 1 and 2 can vary considerably. The authors of \cite{Lotter2020} showed that, depending on the size of a chemical synapse and its associated reaction rates, communication via the synapse could be either diffusion-limited or rate-limited.

\subsubsection{Level 3}

The network of neurons in the body creates an enormous number of connections to relay and store information. Broadly speaking, an individual neuron does not directly modulate and demodulate the information that it relays, but it is generally understood that information is contained in the number and timing of the action potential spikes. A sequence of such spikes is referred to as a \textit{spike train}.

Synthetic means to interface with neurons try to control the membrane potential via the ion channels, e.g., using \textbf{\textit{macroscale electric control}} and \textbf{\textit{optical control}} as described for Level 3 in Section~\ref{sec_interface}. The authors of \cite{Abbasi2018} report recent experimental work that used electrodes to transmit digital messages across anesthetized earthworms, thereby creating an artificial communication link across a biological channel where the neurons are used literally as relays. Optical stimulation has been popular in the biology community, where light-sensitive ion channel membrane proteins (called opsins) are expressed by the introduction of genes to make a cell artificially sensitive to light. When an opsin protein is activated, it opens to enable a current pass through the membrane. This approach, known as \textit{optogenetics}, has been used to control action potentials in neurons and also other cells \cite{Deisseroth2011}. Because of the directionality of light, optogenetics is a promising solution for precise control of the behavior of individual cells.

\subsubsection{Level 4}

As noted, neurons are primarily \textit{relaying} information. We do not have a complete understanding of exactly how neurons modulate information; different metrics exist to quantify the information in spike trains, e.g., measuring the number of spikes over some period of time or the rest interval between spikes \cite{Houghton2011}. There has also been recent research to quantify the information transmissible over the different stages of neuron signaling. For example, the authors of \cite{Ramezani2017a,Aghababaiyan2018} analyzed axon memory and propagation and \cite{Ramezani2018} considered the capacity of vesicles released into the synapse. The authors of \cite{Veletic2016a} measured the channel capacity of information in a chemical synapse. The authors of \cite{Aghababaiyan2019} maximized the transmissible bit rate across an axon and synapse by optimizing the action potential spike rate and the receiver decision threshold. Moreover, the authors of \cite{Lindner2016} modeled neurons as filters and considered the effect of information filtering.

\subsubsection{Level 5}

The connections between neurons can be quite extensive; one neuron can receive inputs from thousands of neurons and have synapses connecting to thousands of neurons and other cells \cite[Ch.~11]{Alberts2015}. There are about $10^{11}$ neurons in a human brain and $10^{14}$ synaptic connections. Collectively, they enable the capacity to learn from and react to external stimuli. One particular example is at the neuromuscular junction, where acetylcholine is released by a motor neuron into a chemical synapse with a skeletal muscle cell. This scenario is very well-studied due to its accessibility, unlike most synapses in the brain and spinal cord. Reception of acetylcholine leads to a rapid influx of Na$^{+}$ and triggers muscle contraction.

Due to their nature as relays of information, neurons are recognized as key junctions for having an interface between the external macroscale world and the \textit{in vivo} microscale world for biomedical applications. Besides ``high-jacking'' an organism's nervous system to build an artificial communication channel, as demonstrated with a worm in \cite{Abbasi2018}, there are many opportunities to develop technologies for brain implants and interfaces to detect and treat neurological diseases with artificial systems \cite{Veletic2019}. State-of-the-art implant technologies include electromagnetic, chemical, and optical stimulation. Optogenetics has also been proposed as part of a bridge to interface between biological systems and computer networks \cite{Jornet2019}.

\subsection{Communication via DNA}
\label{sec_DNA}

DNA is the foundation language of life. It is inherently a storage mechanism, as it contains the information required to encode proteins, but it also includes the supporting machinery to control when to produce each protein and how much. Thus, it is more useful to think of DNA communication as the sharing of information that propagates over \textit{time} instead of space. Nevertheless, as we have mentioned in earlier sections, there are also characteristics of DNA communication that include signaling over physical space.

To apply the proposed communication hierarchy to DNA, we emphasize our perspective that DNA is primarily a storage medium. As summarized in Table~\ref{table_case_studies}, Level 1 deals with the storage of genes in DNA, though there are also aspects of DNA communication that rely on spatial propagation. Level 2 covers the biochemical processes of transcription from DNA to RNA and translation from RNA to protein. Level 3 covers how genetic information is modulated and how it is controlled both locally and experimentally. Level 4 describes how proteins are encoded in DNA, and Level 5 considers what proteins are needed and how they support life.

DNA communication and processes supporting it have already been mentioned in all of the sections discussing the individual levels of the proposed hierarchy, even though genetic information is often an exception to many of the general trends of microscale MC (e.g., it supports far higher information rates than diffusive signaling). This includes the role of diffusion for DNA binding and the sharing of DNA via bacterial conjugation in Level 1 (Section~\ref{sec_physical}), biochemical pathways for gene expression in Level 2 (Section~\ref{sec_interaction}), gene regulation and macroscale control of it in Level 3 (Section~\ref{sec_interface}), the information in DNA bases and using DNA synthetically to realize storage in Level 4 (Section~\ref{sec_data}), and as a component in a microfluidic biosensing application in Level 5 (Section~\ref{sec_applications}). In this case study, we tie these ideas together with a focus on DNA's role in storing genetic information.

\subsubsection{Level 1}

The primary function of DNA is the preservation and function of life. From a communication perspective, it stores information until it is needed, which could be on-going or in response to particular life events or external signals.

While there have been limited works within the MC community to model DNA as a ``conventional'' information molecule (such as \cite{Bilgin2018} where DNA was proposed for a diffusion-based communication system), there are aspects of DNA signaling that can benefit from spatial propagation modeling. These include the \textbf{\textit{diffusion}} of proteins that travel along DNA to regulate what genes are expressed \cite{berg1993random}, the propagation of RNA out of a cell's nucleus for translation into a protein by ribosomes in the cytoplasm \cite[Ch.~6]{Alberts2015}, and the exchange of DNA by \textbf{\textit{conjugation}} when two bacteria come into contact with each other \cite{Cabezón2015}. We highlight a particular laboratory method because it has been modeled using the \textbf{\textit{advection-diffusion-reaction}} equation that we discussed for Level 1 in Section~\ref{sec_physical}. Polymerase chain reaction (PCR) modeling is an important tool for making copies of a region of DNA \cite[Ch.~8]{Alberts2015}. It includes a three-step process to 1) separate DNA into single strands; 2) bind primers to the ends of a single strand; and 3) generate the compliments of the single strands. These steps take place in different regions with different temperatures, which causes a flow that affects the movement of chemical molecules. Since all of the components are affected by circulatory flow, diffusion, and temperature-dependent chemical reactions, the advection-diffusion-reaction equation can be used to analyze PCR \cite{Yariv2005}.

\subsubsection{Level 2}

There are several distinct DNA processes that we associate with Level 2 behavior, i.e., at the interface between device and the propagation medium and the \textbf{\textit{biochemical signaling pathways}} therein. These include the steps of \textbf{\textit{gene expression}}, i.e., transcription from DNA to RNA and translation from RNA to protein using ribosomes, which we have already discussed in some detail for Level 2 in Section~\ref{sec_interaction}. Additional processes include replication of DNA and gene mutations. Our general understanding of Level 2 as discussed in Section~\ref{sec_interaction} includes the generation of information molecules. DNA is unique in this sense because the stored information is \textit{persistent}; nature obtains more DNA by copying existing DNA. We do not go into the biochemical details of DNA replication here (the reader can learn more in \cite[Chs.~4, 5]{Alberts2015}), but we draw attention to one of the profound ideas that drive evolution. The processes that maintain and repair DNA are extremely precise but imperfect. The imperfections are actually quite important, because they lead to the mutations that enable evolution.

Synthetic creation of gene mutations is referred to as gene editing. The basic idea behind gene editing is to modify a particular gene (i.e., create mutations) and then observe the effect on the organism \cite[Ch.~8]{Alberts2015}. Mutations include ``gene knockouts,'' where the gene is simply removed from the genome, and modifications where the experimenter controls the conditions under which the gene is expressed, e.g., to make the gene sensitive to a signaling molecule that can turn the gene on and off, and which we associate with Level 3 behavior. The general goal is to understand the role of a gene and the proteins that it produces. From the perspective of our proposed hierarchy, such gene editing is altering the devices themselves. We are able to artificially introduce or remove physical interfaces to the environment and therefore control communication within a cell, between cells, or with an experimenter. For example, a common modification is to fuse a gene with one that encodes a fluorescent protein. We can then monitor gene expression by measuring the level of fluorescence.

\subsubsection{Level 3}

We discussed gene regulation, i.e., the processes controlling when to activate or deactivate the expression of a particular gene, in some detail for Level 3 in Section~\ref{sec_interface}. Here, we emphasize how the bases in DNA (and RNA) correspond to sequences of information. Both DNA and RNA are composed of subunits labeled as one of four bases. These subunits are placed sequentially in a chain, such that we can read the information in a chain as a sequence of bases. Thus, DNA is usually described as a sequence of As, Cs, Gs, and Ts, and RNA is usually described as a sequence of As, Cs, Gs, and Us. Each subunit stores $\log_2 4 = 2$ bits of information.

Not surprisingly, there is considerable interest in reading DNA and RNA sequences at macroscale. Technology for doing so has been in development since the 1970s and the emergence of dideoxy (or Sanger) sequencing. Two common sequencing methods today are Illumina sequencing and ion torrent sequencing, which both rely on PCR to amplify DNA \cite[Ch.~8]{Alberts2015}. Newer methods are in development to avoid the amplification stage and read a sequence directly from individual molecules. There are also possibilities for applying operations normally expected of a word processor, i.e., cut, copy, and paste, using restriction enzymes, PCR, and ligases, respectively \cite{Little1967,Smith1970,Mullis1989,O'Driscoll2013}. Data storage in DNA is expected to be orders of magnitude more energy efficient than currently available technologies \cite{Adleman1994,O'Driscoll2013,DeSilva2016}.

A macroscale control method for DNA that was not introduced for Level 3 in Section~\ref{sec_interface} is optogenomics \cite{Jornet2019}. An optogenomic interface uses light to activate or deactivate specific genes in eukaryotic cells with subcellular resolution and high temporal accuracy. It has already been validated for the activation of cellular responses and expressing individual genes but could furthermore be applied to regulate and correct DNA structure.

\subsubsection{Level 4}

Level 4 behavior for DNA is relatively well understood, since we know how the nucleotide bases in DNA map to amino acids in protein. As we discussed in Section~\ref{sec_data}, triplets of bases encode the twenty amino acids that are commonly found in proteins. However, not all DNA maps directly to amino acids \cite[Ch.~6]{Alberts2015}. For many genes, transcription to RNA is the final step and there is no corresponding translation to protein. RNAs themselves can play important structural or catalytic roles, such as being a base for ribosomes (which conduct translation). RNAs can also be used to regulate gene expression, e.g., by degrading other target RNAs.

\subsubsection{Level 5}

The natural applications of DNA are somewhat self-evident since it is the foundation language of all life. DNA encodes proteins, which perform thousands of distinct cellular functions \cite[Ch.~2]{Alberts2015}. The propagation of genetic information over time, regulated to manifest at the right moment or in response to the right stimulus, facilitates the cell life cycle and correspondingly the function and behavior of multi-cellular organisms. For example, cell signaling, mobility, growth, mitosis (i.e., division of eukaryotic cells), and differentiation (i.e., a cell changing to a different type) are all behaviors that are driven by the biochemical actions of proteins.

As discussed in Section~\ref{sec_MicroStorage}, DNA can be a promising solution for data storage due to its information density and long ``shelf life''. Another application that shows potential is the use of DNA molecules as building blocks for nanomachines. The forces between DNA bases that define its double helical shape are well understood. By carefully selecting the sequence of base-pairs, it is possible to create synthetic double-stranded DNA molecules that self-assemble into predetermined structures (``DNA Origami") with a specific size and shape \cite{Rothemund2006,Douglas2009}. Additionally, because DNA assembly occurs due to hydrogen bonding between base-pairs, the conformation of these kinds of structures can be controlled by temperature. Heating or cooling affect the level of association-dissociation between complementary strands and change the shape of the molecule. Using this principle it is possible to construct molecular motors for the movement of nanomachines \cite{Endo2018}. Such nanomachines can be controlled in a number of ways, such as temperature, light, pH, metal ions or other external stimuli \cite{Liu2014}. Using this technology, it is possible to construct nanoscale devices that perform complicated tasks such as monitor physiological functions \cite{Surana2011} or targeted drug release \cite{Ranallo2017}.

\section{Open Problems}
\label{sec_open}

In this section, we summarize open challenges and opportunities with the support of our proposed hierarchy for signaling in cell biology. We intend for this discussion to guide further interest and research in this interdisciplinary field. Our proposed hierarchy enables us to organize these problems and gain some insight on how to effectively tackle them.

We emphasize that there are many opportunities for the application of MC theory and communications analysis in biological systems that are already relatively well studied, in addition to the design of synthetic communication networks. Natural scientists focus on describing systems end-to-end, in that they provide as much detail as possible for individual components and often omit interactions with other systems. However, given life's reliance on communication, studying these same systems from a communications engineering point of view can provide tools to inform understanding and develop methods for control. We facilitate this kind of exploration by enabling researchers to map system components and how they integrate in a communications networking sense.

In the following, we discuss problems that align with each of the five levels of the proposed hierarchy. We then describe problems that derive from the integration of the different levels. Finally, we describe opportunities associated with our end-to-end case studies of QS, neuronal communication, and communication via DNA.

\subsection{Level 1}

The existing literature on diffusion for MC focuses on theoretical descriptions of particle propagation using the mathematics of diffusion \cite{Jamali2019b}. For tractability, these models usually make simplifying assumptions about the system  (e.g., ideal propagation, homogeneous molecule characteristics, simplified channel geometry). Despite the progress that has been made, both natural and synthetic systems can be much more complex than what existing contributions can sufficiently model \cite{Kuscu2019a}. For example, molecules can participate in intricate chemical reaction networks while they are diffusing, and practical fluid flow patterns can be more spatially-varying than the models we have summarized here. Such details can make the corresponding differential equations for propagation more complex and heterogeneous. There are opportunities to identify closed-form solutions to such scenarios, or to develop robust numerical methods where closed-form solutions are not achievable.

An important related question is how realistic a model needs to be in order for it to be useful in practice, i.e., to make informed predictions or to effectively guide system design. There likely is scope to effectively apply existing MC channel modeling to biological systems. In particular, some biological communication modalities could be described as the integration of multiple communication channels. For example, neuron signals propagate as both a traveling electrical potential wave (along the axon) and as a reaction-diffusion signal (across a synapse) \cite[Ch.~11]{Alberts2015}. Other examples include gap junctions and plasmodesmata \cite{Saez2003,Brunkard2015}, which create locally-regulated parallel channels between cells with diffusion on either side. There is also a range of communication modalities that have so far received limited attention from the MC community but that could benefit from their engagement (e.g., contact-based communication, including cell conjugation \cite{Cabezón2015}).

\subsection{Level 2}
As summarized in Table \ref{table_BCs}, channel responses have been derived under ideal molecule generation and reception models. However, the physical and temporal scales of molecule generation and reception may not be sufficiently small compared with the propagation channel, which requires us to take the corresponding  biophysical and biochemical processes into account to understand the channel response, such as mRNA propagation from nucleus to ribosome and stochastic chemical reactions in gene expression. Not surprisingly, the inclusion of these practical processes will complicate the initial and boundary conditions applied to propagation equations, and thus introduce challenges when deriving channel responses \cite{Kuscu2019a}. The shift of theoretical research to more realistic models also imposes a requirement on related biological software to not only verify closed-form solutions but also provide numerical results for intractable problems. Moreover, through analytical characterization and verification via simulation, guidelines for choosing the optimal molecule generation method and reception strategy should also be developed to facilitate MC system design.

An important signaling mechanism discussed in Section~\ref{sec_interaction} is the use of extensive and interconnected transcription networks. Although the functions of some basic building blocks (e.g., the feed forward loop) have been identified, there is still a need to have a more comprehensive understanding of transcription networks, including the inputs, the outputs, and how a change of inputs influences the outputs \cite{Alon2006}. This would be helpful for controlling and synthesizing transcription networks to achieve target functions. Another question inspired by transcription networks concerns signaling \textit{complexity} in biological systems. It would be helpful to develop a categorization or quantified ``metric'' of signaling complexity in terms of the number and variety of molecules used for the signaling architecture of a given cell type \cite{Wilkinson2017,Green2018}. Such a metric might provide a rule of thumb that enables us, given data of well-understood cells, to predict what is unknown about other cells that are sufficiently similar (i.e., same kingdom, similar size, etc.).

\subsection{Level 3}
Although existing experimental tools make the control and observation of microscale phenomena possible, the operation of some devices (e.g., optical microscopy \cite{Hao2013-micro}) requires skills and interdisciplinary technical knowledge that may impede their adoption in communications-focused research. 
Moreover, as stated in Section~\ref{sec_macro_obs}, no single existing technique can inspect biological signaling processes across all spatial and temporal scales simultaneously. Thus, new experimental tools that span multiple spatial and temporal resolutions would be incredibly useful for microscale systems. These must satisfy the requirements of biocompatibility, non-invasiveness, and miniaturization. It would also be helpful for the new experimental tools to facilitate communication analysis, e.g., to capture the probability distributions needed to determine communication capacity and bit error rate. In addition, it is worthwhile to seek unique combinations of existing macroscale control and observation tools (e.g., see Fig. \ref{fig:Level3}) for targeted applications. One example is the guidance of drug carriers to target areas via \textit{in vivo} imaging and the subsequent release of drug molecules via macroscale magnetic control \cite{Jeon}.

With the vision of the Internet of Bio-NanoThings (IoBNT) \cite{Akyildiz2015}, more attention is needed to develop feasible interfaces to connect the microscale world with macroscale wireless networks. In particular, we can draw from the expertise of other related domains to support efforts to have an effective micro-macro interface, such as image processing, machine learning, and optical physics.

\subsection{Level 4}
Quantifying the limits of molecular signaling using information-theoretic approaches provides a way to study the potential of cell signaling. Although specific cell types and signaling pathways have been studied, such as the \textit{E. coli} bacterium strain K-12 MG1655 \cite{Pierobon2016} and the JAK-STAT pathway in eukaryotic cells \cite{Sakkaff2018}, it remains to be seen whether the obtained results and research methods can be generalized to other pathways and cell types. Moreover, the relationship between  communication capacity and cell behavior needs an accessible interpretation so that scientists from different fields can easily understand and apply each other's research results. 

Various chemical circuits are introduced in Sections \ref{sec:synthetic_circuits} and \ref{function_realization} with the aim to realize computation and communication functionalities \cite{cook2009,jiang2013,Bi2020a,Unluturk2015}, but we highlight that most contributions have been theoretical works that have not yet been validated with physical experiments, mainly due to the tedious, laborious, and expensive nature of wet lab experimentation. However, it is essential to develop a robust testing framework for validation and to optimize design parameters.

Communications engineers try to minimize the complexity of a system design in terms of different components or variations in the types of signals. We might also assume that this is true for natural systems, since every extra component or function has an associated cost, e.g., metabolic or fitness. However, in nature, we see potential over-engineering in the construction of signaling systems. For example, cells use multiple pathways and a variety of different molecules for the activation of the same gene \cite{Han1989,Chen2005,Madhusudan2006}. It is not entirely clear whether this fulfills a need for robustness via redundancy or whether designs constitute a locally optimal (but globally sub-optimal) solution obtained by evolutionary optimization. We believe that it is important to ask whether this redundancy (if any) can be identified by comparing predictions with observations. Such knowledge might lead to better understanding of minimal sets of required system components for particular functions, or increased robustness in synthetic systems. 

\subsection{Level 5}

While significant progress has been made over the last two decades in applying communications ideas to biological and synthetic systems, most of this work has taken the form of theoretical models proposed for MC schemes or exploring their limits \cite{Jamali2019b,Farsad2016,Akyildiz2019a,Kuscu2019a}. For MC to progress as a field, validation of these schemes is necessary through proof-of-concept applications that demonstrate their feasibility and usefulness. Additionally, as a large proportion of MC work concerns low- to mid-level interactions (e.g., diffusion or modulation) \cite{Akyildiz2008,Jamali2019b,Akan2017,Nakano2019}, there is scope for applying communications network theory to larger systems such as interconnected populations of cells or nanomachines. Such modeling could enable prediction and observation of the emerging behavioral dynamics of these systems, but requires the development of suitable algorithmic or computational tools to do this efficiently. Inspiration can be taken from agent-based modeling in synthetic biology, which is an approach that can simulate large networks of cells \cite{Gorochowski2016}.

\subsection{Multi-Level Problems}

The proposed hierarchy provides insight by separating system components and tasks into levels, but one of the hierarchy's salient features is to help articulate, understand, and solve problems that span multiple levels. For example, there is a direct mathematical link between Levels 1 and 2, since Level 2 provides the boundary conditions that are needed to solve the differential propagation equations at Level 1. Thus, Levels 1 and 2 are both necessary to determine a cell's signaling channel impulse response, which can then be used to determine a receiving device's observed signal given what was transmitted. Significant research efforts have been made to determine impulse responses for diffusion-based MC channels \cite{Jamali2019b}, but this survey can assist to identify important scenarios that have not received such analysis. For example, models for gene expression could be expanded to include the propagation of RNA out of the nucleus.

An important problem is to understand how the constraints and limitations of one level impact the design and performance of other levels. In particular, the biophysical and biochemical activities at Levels 1 and 2 are inherently noisy; molecule propagation and chemical kinetics are both modeled at microscale as stochastic processes. These features impact the reliability of cell signaling channels, the rate and quantity of information they carry, and how life evolved to accommodate them. There are many questions that can be posed regarding the impact of biophysical and biochemical noise on higher levels, e.g., on gene regulation and other microscale signal operations (Level 3), on our ability to experimentally observe and control cell signaling systems (Level 3), on how accurately cells can infer information about their surrounding environment (Level 4), on how robust synthetically designed MC devices can be (Level 4), and on how heterogeneous behavior emerges in large cellular populations (Level 5). Levels 1 and 2 also impose constraints on the overall communication speed. The approach to analyze the communication speed of chemical synapses in \cite{Lotter2020} might be generalized to other communication systems to discern bottlenecks and their impact on higher levels, e.g., device sensitivity and responsiveness to environmental changes.

Given the scalability of our proposed hierarchy, interesting problems can arise when deciding at what scale to define a communication channel. For example, gap junctions form channels between adjacent cells. However, signals passing through gap junctions can be relayed and reach cells at large distances from the initial transmitter. We propose to investigate whether it is suitable to describe the communication link to a distant cell as a \textit{single} aggregated channel. If so, then we can determine the reliability of this channel as the receiving cell is placed further from the source.

\subsection{Case Studies}

Finally, we highlight several open problems associated with our case studies. Concerning QS, the majority of work today is concerned with the simulation and analysis of ideal systems. Studies that consider collective behavior tend to be theoretical, in part because of the inherent complexity of large cell populations \cite{Yusufaly2016, Ueda2019}. Nevertheless, simulations of a large number of cells and multiple autoinducers could span all levels of our proposed hierarchy and provide numerical insights regarding the underlying causes of natural behavior and a useful testbed for the design of synthetic systems. In particular, one underdeveloped area is security in cellular signaling (e.g., secure communications using QS and eavesdropping on QS sources). Furthermore, inspired by the multi-level problems presented in the previous subsection, rigorous analysis of the impact of autoinducer propagation could extend our understanding of how QS architectures in individual cells contribute to collective behavior.

There is already a vast literature of theoretical, experimental, and computational studies of neuronal signaling \cite{Bear2016}. However, there is still scope to apply our proposed hierarchy to this field and address consequential problems. For example, neurological diseases could be modeled as application-level problems that arise from deficiencies in the propagation of neuronal signals, and could be treated via communication and control of neurons using optogenetic tools and other brain implants \cite{Veletic2019}. Another scenario is synaptic plasticity, which refers to biophysical processes that change synaptic strengths over time \cite[Ch.~25]{Bear2016}, and could be studied as a dynamic communication link that affects learning and memory.

We have already mentioned several open problems for signaling via DNA in our discussion of multi-level problems, e.g., the impact of RNA propagation on gene expression. Similarly, all of the biochemical pathways involved in the gene transcription and translation processes can be impaired by noise, even in a managed scenario such as PCR, which has an impact on the resulting protein production levels. Models following our proposed hierarchy to include this stochasticity could be used to determine the distribution of protein productivity and predict the reliability of DNA-based storage.

\section{Conclusion}
\label{sec_conclusion}
Unleashing the potential of MC for interdisciplinary applications requires substantial efforts from diverse scientific communities. However, the distinct approaches to articulate and study research problems gives rise to a mismatch between the different disciplines. To bridge this mismatch, in this survey, we proposed a novel communication hierarchy to describe signaling in cell biology. The proposed hierarchy is comprised of five levels: 1) \textbf{Physical Signal Propagation}; 2) \textbf{Physical and Chemical Signal Interaction}; 3) \textbf{Signal-Data Interface}; 4) \textbf{Local Data Abstraction}; and 5) \textbf{Application}. While the nominal communicating ``device'' is assumed to be an individual cell, the hierarchy readily describes communication between any devices using or observing chemical signals in a biological system, including cellular organelles and macroscale experimental equipment.

Our proposed hierarchy enabled us to map communication concepts to infrastructure and activities in biological signaling. Specifically, we started with the \textbf{Physical Signal Propagation} level (Level 1) to discuss the fundamental mechanisms of molecular propagation. This level focused on mathematical formulations of diffusion-based phenomena, and also detailed cargo-based propagation and contact-based transport. For the \textbf{Physical and Chemical Signal Interaction} level (Level 2), we reviewed physical signal generation and reception mechanisms and the associated biochemical and biophysical signaling pathways. In addition, we provided a mathematical characterization of different release and reception strategies, corresponding to different initial and boundary conditions (and hence channel responses), and mathematically described gene expression pathways. For the \textbf{Signal-Data Interface} level (Level 3), we described the mathematical quantification of the physical signals that are released and received, including the conversion between quantification and data, i.e., modulation and demodulation in communication networks. This discussion also included a survey of methods for macroscale observation and control of cell signaling behavior. For the \textbf{Local Data Abstraction} level (Level 4), we considered the significance of information in individual cells, limits on how much information is carried in natural MC signals, and how synthetic devices (including chemical and genetic designs) might be realized to represent, store, and process information. For the top of the proposed hierarchy, i.e., the \textbf{Application} level (Level 5), we selected biosensing and therapeutics as exemplary applications to show how they might benefit from the integration of natural and synthetic systems at lower levels to realize their potential.

To further demonstrate the utility and flexibility of our proposed hierarchy, we mapped all of the levels to case studies of QS, neuronal signaling, and communication via DNA. Finally, we identified a selection of open problems associated with each level and in the integration of multiple levels. We anticipate that our proposed hierarchy provides researchers from different fields with language to interpret and understand results on MC signaling from other disciplines, while simultaneously realizing the potential of opportunities for interdisciplinary collaboration. Ultimately, we intend for this survey to support the advancement of interdisciplinary cell signaling applications.

\appendix
\section{Appendix: Glossary}

In Tables~\ref{appendix:Biological_terms} and \ref{appendix:Communication_terms}, we define common biological and communication terms that appear throughout the survey, respectively.

\begin{table*}[!ht]
    \renewcommand{\arraystretch}{1.5}
	\caption{Glossary of Biological Terms.}
	\label{appendix:Biological_terms}
    \begin{tabular}{@{} l p{14cm} @{}}
    \toprule
    Term & Description\\
    \midrule
    Action potential & Rapid and transient change in electric potential across a membrane\\
    ATP & Adenosine triphosphate. Molecule able to store and transfer chemical energy within a cell\\
    Autoinducers & Diffusible signal molecules produced by cells to monitor local population changes. They can also have additional functions (e.g., act as antibiotics or toxins)\\
    Cytoplasm & The gel-like contents of a cell between the outer membrane and the nucleus. Comprised mainly of water, proteins, and salts\\
    Cytoskeleton & Complex, dynamic network of filaments spanning the entire cell\\
    Cytosol & The aqueous part of the cytoplasm\\
    DNA & Deoxyribonucleic acid. Carrier of the hereditary information for the building and maintenance of organisms\\
    Endoplasmic reticulum (ER) & Continuous membrane system connected to the nucleus. Involved in folding, modification, and transport of proteins\\
    Endosomes & Membrane-bound vesicles formed around molecules to facilitate their transport into the cell from the extracellular space\\
    Enzyme & A biologically relevant molecule acting as a catalyst, making chemical reactions possible or greatly increasing their rate\\
    Exocytosis & Active transport of material out of the cell via membrane vescicles\\
    Extracellular matrix & Complex and dynamic extracellular network of macromolecules providing structural and chemical support to cells\\
    Golgi apparatus & Large organelle of eukaryotic cells responsible for modification, packaging, and transportation of proteins\\
    G-protein-coupled receptors & Large group of cell surface receptor proteins\\
    Hydrolysis & The chemical breakdown of compounds by water\\
    Macromolecule & A molecule that consists of a large number of atoms (proteins, nucleic acids, synthetic polymers)\\
    Messenger RNA (mRNA) & Single-stranded RNA molecule, carrying the information for protein production outside the cell nucleus\\
    Microtubule & Dynamic, hollow tubes formed by protein polymers. Part of the cytoskeleton\\
    MicroRNA (miRNA) & Small non-coding RNA molecules involved in gene regulation\\
    Neurotransmitters & Chemical messengers that are released into a chemical synapse to convey a message between neurons\\
    Organelle & Cellular structure with specialized functions (e.g., endoplasmic reticulum, mitochondria, golgi apparatus)\\
    PCR & Polymerase Chain Reaction. Method of rapid replication of a given DNA sample into large numbers\\
    Plasmid & Genetic structure that can replicate independent of the host's chromosome\\
    Post-transcriptional modification & The modification of an mRNA molecule directly after transcription to produce a mature mRNA for protein production\\
    Post-translational modification & The modification of proteins after their production in ribosomes\\
    Promoter sequence & Small DNA sequence preceding a gene that marks where transcription should start\\
    Protein & Organic compound comprised of one or more macromolecules. Integral to most cellular processes\\
    Protein phosphorylation & The addition of a phosphate group to an amino acid of a protein. Reversible process crucial for cell signaling\\
    Redox reactions & Oxidation-reduction reactions involving the transfer of electrons between two chemical species\\
    Ribosomes & Protein-synthesizing factories, comprised of ribosomal RNA and associated helper proteins\\
    RNA & Ribonucleic acid. A single-stranded biopolymer that is essential for protein production by carrying sequence information from DNA to ribosomes\\
    RNA polymerase & Enzyme that can bind and follow a strand of DNA, replicating its sequence\\
    Second messengers & Small intracellular molecules relaying information received from first messengers, i.e., cell surface receptors\\
    Signaling pathway & A chain of cell components and molecules working in succession to transfer a signal\\
    Synapse & Small contact site that chemically or electrically links a neuron to other cells\\
    T-cell & Type of white blood cell (leukocyte), part of the immune system\\
    Transcription factor & A protein that can bind a specific DNA sequence, controlling the expression of a gene\\
    Transfer RNA (tRNA) & Special RNA molecule involved in protein production within ribosomes. It matches a loose amino acid to mRNA sequence\\
    \bottomrule
    \end{tabular}
\end{table*}

\begin{table*}[!ht]
    \renewcommand{\arraystretch}{1.5}
	\caption{Glossary of Communications Terms.}
	\label{appendix:Communication_terms}
    \begin{tabular}{@{} l p{13.2cm} @{}}
    \toprule
    Term & Description\\
    \midrule
    Binary CSK (BCSK) & Information is represented by two concentration levels of a chemical signal\\
    Bit & A basic unit of information in communication systems that represents a logical state with one of two possible values (usually $0$ or $1$)\\
    Bit error rate (BER) & Ratio of the number of bit errors and the total number of transmitted bits\\
    Channel & The physical medium or logical connection through which a signal propagates\\
    Channel capacity & The tight upper bound on the rate at which information can be reliably transmitted over a communication channel and is the maximum of the mutual information of the input and output of the channel\\
    Concentration shift keying (CSK) & Information is represented by different concentration levels of a chemical signal\\
    Decoding & Retrieval of original information symbols from a signal\\
    Demodulation & The process that uses the observed signals to attempt recovery of the transmitted information symbols\\
    Encoding & Conversion of information symbols into a form that can be superimposed onto a signal\\
    Frequency shift keying (FSK) & Information is represented by different frequencies of a signal\\
    Low-density parity-check (LDPC) code & A powerful linear error correction code\\
    Modulation & The process that varies one or more properties (e.g., amplitude, frequency, phase) of the physical carrier signal to transmit information\\
    Molecule shift keying (MoSK) & Information is represented by different types of messenger molecules\\
    Mutual information & Measures the mutual dependence between two random variables\\
    Quadruple CSK (QCSK) & Information is represented by four concentration levels of a chemical signal\\
    Reaction shift keying (RSK) & Information is represented by different emission patterns \\
    Receiver & A device or biological entity capable of observing a signal\\
    Single parity-check (SPC) code & A single bit is appended to the end of each transmitted frame. The parity bit is 1 if the data portion of the frame has an odd number of 1's; otherwise, it is 0\\
    Symbol & Used to represent a letter, a waveform, or a state in communication systems. Each symbol may encode one or several bits. For example, a symbol encodes a single bit in BCSK and two bits in QCSK\\
    Transmitter & A device or biological entity that physically produces a signal\\
    \bottomrule
    \end{tabular}
\end{table*}












\end{document}